\documentclass[preprint,1p,12pt,authoryear]{elsarticle}
\pdfoutput=1
\usepackage{t1enc,graphicx,amssymb,amsmath,latexsym,fullpage,float,color,enumerate}
\usepackage{tikz}
\usepackage{ulem,dsfont,algorithm,algpseudocode,amsmath,afterpage,textcomp,array}
\usepackage[english]{babel}
\usepackage{pdfsync}
\usepackage{natbib}
\usepackage{epstopdf}
\usepackage{xargs}
\usepackage{lineno}
\usepackage{hyperref}
\modulolinenumbers[5]
\usepackage{aliascnt}
\usepackage{amsthm}

\newtheorem{prop}{Proposition}
\usepackage{calrsfs}
\usepackage{tabularx}
\usepackage{lipsum}
\usepackage{array}
\usepackage{calc}
\DeclareMathAlphabet{\pazocal}{OMS}{zplm}{m}{n}
\newaliascnt{rem}{thm}
\newtheorem{rem}[rem]{Remark}
\aliascntresetthe{rem}

\theoremstyle{definition}
\newtheorem{ex}{Example}

\renewcommand*{\appendixname}{}
\def\1{\mathds{1}}

\newcommandx\ABC[2][1=]{
\ifthenelse{\equal{#1}{}}
{
\ifthenelse{\equal{#2}{}}{(\tparam,U)}{(\tparam^{(#2)},U^{(#2)})}
}
{
\ifthenelse{\equal{#2}{}}{(\tparam^{#2}_{#1},U^{#2}_{#1})}{(\tparam^{#2}_{#1},U^{#2}_{#1})}
}
}

\def\borel{\mathcal{B}}

\def\bS{\bar{S}}
\def\bs{\bar{s}}

\newcommandx\coord[3][1=1, 3=n]{(#2_{#1},\ldots,#2_{#3})}
\newcommandx\coor[3][1=1, 3=n]{#2_{#1},\ldots,#2_{#3}}

\def\dom{\lambda}

\def\esp{\mathbb{E}}
\def\eg{\textit{e.g.}\,}
\def\eps{\epsilon}

\def\fam{\mathsf{F}}

\def\halpha{\hat{\alpha}}

\newcommand{\KL}[2]{\text{KL}\left(#1\|#2\right)}

\def\leb{\lambda}

\def\ie{\textit{i.e}\;}
\def\iid{\textit{i.i.d.}\,}
\def\Id{\text{Id}}

\newcommandx\MH[2][1=]{
\ifthenelse{\equal{#1}{}}
{
\ifthenelse{\equal{#2}{}}{\param}{\param^{(#2)}}
}
{
\ifthenelse{\equal{#2}{}}{\param^{#2}_{#1}}{\param^{(#2)}_{#1}}
}
}

\def\norm{\pazocal{N}}
\def\nset{\mathbb{N}}

\def\param{\theta}
\def\paramst{\theta^{\ast}}
\def\paramalg{\vartheta}
\def\paramset{\Theta}

\newcommand{\pscal} [2]{\left\langle #1 , #2\right\rangle}

\def\rmd{\mathrm{d}}
\def\rset{\mathbb{R}}

\def\suffset{\mathsf{S}}

\def\tparam{\tilde{\theta}}
\def\targ{\pi}

\def\tparam{\tilde{\param}}

\def\ttarg{\tilde{\pi}}

\def\ttargs{\tilde{\pi}^{\ast}}

\def\U{U}
\def\Uset{\mathsf{\U}}
\def\Usetst{\mathsf{\U}^{\star}}
\def\Ualg{\mathcal{\U}}

\def\Vset{\bar{\Uset}}

\def\x{x}

\def\Y{Y}
\def\Yset{\mathsf{\Y}}
\def\Yalg{\mathcal{\Y}}

\begin{document}

\begin{frontmatter}
\title{Light and Widely Applicable MCMC: Approximate Bayesian Inference for Large Datasets}

\author[ucd,insight]{Florian Maire\corref{corresp}}
\cortext[corresp]{Corresponding author}
\ead{florian.maire@ucd.ie}

\author[ucd,insight]{Nial Friel}
\author[ensae]{Pierre Alquier}
\address[ucd]{School of Mathematical Sciences, University College Dublin, Ireland}
\address[insight]{Insight Centre for Data Analytics, University College Dublin, Ireland}
\address[ensae]{ENSAE--CREST, Malakoff, France}

\begin{abstract}
Light and Widely Applicable (LWA-) MCMC is a novel approximation of the Metropolis--Hastings kernel targeting a posterior distribution defined on a large number of observations. Inspired by Approximate Bayesian Computation, we design a Markov chain whose transition makes use of an \textit{unknown} but \textit{fixed}, fraction of the available data, where the random choice of sub-sample is guided by the fidelity of this sub-sample to the observed data, as measured by summary (or sufficient) statistics. LWA--MCMC is a generic and flexible approach, as illustrated by the diverse set of examples which we explore. In each case LWA--MCMC yields excellent performance and in some cases a dramatic improvement compared to existing methodologies.
\end{abstract}

\end{frontmatter}

\section{Introduction}
The development of statistical methodology which scales to large datasets represents a significant research frontier in modern statistics. This paper presents a generic and flexible approach to directly address this challenge.
Given a set of observed data $\coord{Y}[N]$, a specified prior distribution $p$ and a likelihood function $f$, estimating parameters $\param\in\paramset$ of the model proceeds via exploration of the posterior
distribution $\targ$ defined on $(\paramset,\borel(\paramset))$ by
\begin{equation}
\label{eq:posterior}
\targ(\rmd\param\,|\,\coor{Y}[N])\propto f(\coor{Y}[N]\,|\,\param)p(\rmd\param)\,.
\end{equation}
Stochastic computation methods such as Monte Carlo methods allow one to estimate characteristics of $\targ$. In Bayesian inference, Markov chain Monte Carlo (MCMC) methods remain the most widely used strategy,
% mainly for their simplicity while being universally applicable and supported by an abundant literature. Essentially, these methods rely on ergodic Markov kernels, such as the Metropolis--Hastings
% algorithm \citep{metropolis1953equation,hastings1970monte}, whose asymptotic convergence properties have been established under reasonable assumptions \cite[see \eg][]{meyn2009markov}. These strong theoretical
% arguments have supported the development of a large range of MCMC applications, in several domains including Ecology \citep{larget1999markov}, Finance \citep{eraker2001MCMC}, Medicine \allowbreak \citep{cauchemez2004bayesian},
% Genetics \citep{corander2003bayesian}, Telecommunications \citep{punskaya2002bayesian} and Social Networks \citep{snijders2002markov}.
Paradoxically, improvements in data acquisition technologies together with increased storage capacities, present a new challenge for these methods.
%whose theoretical guarantees can be reached but at the cost of a (virtually) infinite computational time.
Indeed, the size of the data set $N$ (along with the dimension of each observation) can become so large, that even a routine likelihood evaluation is made prohibitively computationally intensive.
As a consequence, methods such as the Metropolis--Hastings sampler
% , which needs to compute at each iteration a probability of the type
% \begin{equation}
% \label{eq:MH_rule}
% \alpha(\param,\param')=1\wedge\frac{f(\coor{Y}[N]\,|\,\param')p(\param')q(\param',\param)}{f(\coor{Y}[N]\,|\,\param)p(\param)q(\param,\param')}\,,
% \end{equation}
% where for all $\param\in\paramset$, $q(\param,\,\cdot\,)$ is a proposal density on $(\paramset,\borel(\paramset))$,
cannot be realistically considered. This issue has recently generated a lot of research activity.

While some authors have designed \textit{exact} algorithms that tend to match the theoretical requirements of usual MCMC methods, others have considered the possibility of \textit{approximate}, or \textit{noisy},
methods while still trying to derive some quantitative error bound on the resulting approximation scheme. In this context, we refer to as \textit{exact} any method that produces --possibly dependent-- samples
from $\targ$, as opposed to \textit{noisy}, which samples from an approximation of $\targ$.
In Pseudo-marginal algorithms \citep{andrieu2009pseudo}, the evaluation of the likelihood function is substituted by an unbiased and positive estimator, while still preserving the stationary
distribution $\targ$ and thus being \textit{exact}.
%; see also \citep{beaumont2003estimation,doucet2012efficient}.
Although theoretically appealing \citep{andrieu2012convergence}, finding an unbiased and positive estimator of the likelihood turns out to be a challenging problem in itself.
Two other more recent \textit{exact} approaches \citep{banterle2014accelerating,maclaurin2014firefly} overcome the need for an unbiased and positive estimator of the likelihood.
%However both of these methods require strong assumptions rendering
However, the method
proposed in \cite{maclaurin2014firefly} requires the specification of a lower bound of the likelihood -- which is a strong assumption especially for high--dimensional problems. A poor lower bound function yields
a method whose computational complexity can even exceed that of a M--H sampler. In \cite{banterle2014accelerating}, the authors suggest to break the usual M--H ratio into $N$ independent decision
steps (each corresponding to a factor involving the likelihood of a datum) such that a proposal is globally rejected as soon as it is rejected by an elementary step. This \textit{sticky} version of the M--H sampler
is nevertheless shown to be $\targ$-stationary but yields an higher asymptotic variance by a straightforward Peskun comparison argument with the standard M--H kernel \citep{tierney1998note}.

An alternative solution consists of finding an approximation of the M--H transition kernel, that would emulate the outcome of the accept/reject step without having to compute the likelihood ratio in the M--H transition kernel.
%\eqref{eq:MH_rule}. %involving a "cheap" way to make a decision between either accepting or rejecting a proposal without computing \eqref{eq:MH_rule}.
     In Bayesian settings where the observed data are independent and identically distributed (\iid), the likelihood is expressed as
        \begin{equation*}
        f(\coor{Y}[N]\,|\,\param)\propto\prod_{k=1}^{N}f(Y_k\,|\,\param)\,.
                \end{equation*}
    The issue of the reliability of making a decision to accept or reject a move based only a subset of these factors has been recently addressed \citep{korattikara2013austerity,bardenet2014towards}. In these papers, the usual M-H acceptance decision, which can be rewritten as
\begin{equation}
\label{eq:MH_decision}
\frac{1}{N}\log\left(U\frac{p(\param)q(\param,\param')}{p(\param')q(\param',\param)}\right)\leq \frac{1}{N}\sum_{k=1}^{N}\log\frac{f(Y_k\,|\,\param')}{f(Y_k\,|\,\param)},\quad U\sim\mathcal{U}(0,1)
\end{equation}
is made using a Monte Carlo approximation
\begin{equation*}
\rho_n=\frac{1}{n}\sum_{k=1}^{n} \log\frac{f(Y_{u_k}\,|\,\param')}{f(Y_{u_k}\,|\,\param)},\quad(u_1,\ldots,u_n)\in\{1,\ldots,N\}^n,\quad (u_i\neq u_j)
\end{equation*}
of the right hand side of \eqref{eq:MH_decision}. Both methods proposed in  \cite{korattikara2013austerity} and \cite{bardenet2014towards} share the same principle that
\begin{enumerate}[(a)]
\item a decision can be made with a
certain \textit{level of confidence} as soon as $\rho_n$ becomes \textit{sufficiently far} from the left-hand side of \eqref{eq:MH_decision},
\item and if this condition is not reached, $n$ is increased.
\end{enumerate}
They nevertheless differ by the way the level of confidence is derived and by the theoretical arguments motivating the approximation. %In \cite{korattikara2013austerity}, the stopping time is derived from a confidence interval of a Student-t distribution that the test statistic is assumed to follow. In \cite{bardenet2014towards}, the authors rely on a concentration inequality for a Monte Carlo estimator of a log ratio to set a confidence threshold $\gamma>0$, such that the probability of the event $\{|\rho_n-\rho_N|<\gamma\}$ achieves a given lower bound. The latter is in turn a lower bound for the probability of making the correct decision, \ie accepting or rejecting but using  $\rho_n$ instead of $\rho_N$.
However, practically, as observed in some examples presented in \cite{korattikara2013austerity} and \cite{bardenet2014towards}, both methods tend to draw a significant portion of the data (\ie $n\to N$), in order to reach the confidence interval when the Markov chain gets closer to stationarity. Finally, note that \textit{noisy} algorithms may retain some theoretical appealing feature. This was developed in \cite{alquier2014noisy}, where the authors show that using an approximation of an unavailable exact transition kernel can still provide ergodic Markov chains. In particular, this yields a class of general noisy M--H algorithms, extending the pseudo-marginal approach \citep{andrieu2009pseudo}, while relaxing the unbiased estimator assumption.

In this paper, we propose \textit{Light and Widely Applicable MCMC (LWA--MCMC)}, a novel methodology which aims to make the best use of a computational resource available for a given computational run-time, while still preserving the celebrated simplicity of the standard M--H sampler. Our approach designs a Markov chain on an extended state space whose marginal in $\theta$ targets an approximation of $\targ$. As a result, our algorithm can be cast as a \textit{noisy} MCMC method. At each transition of the Markov chain, a new candidate is proposed and accepted/rejected through a probability that only uses a fraction, $n/N$, of the available data which is by construction --and contrary to \cite{maclaurin2014firefly}, \cite{korattikara2013austerity} and \cite{bardenet2014towards}-- held constant throughout the algorithm. Moreover, unlike most of the papers mentioned before, LWA--MCMC can be applied to virtually any model (involving \iid data or not), as it does not require any assumption on the likelihood function nor on the prior distribution.

%assumption \cite{korattikara2013austerity} and \cite{bardenet2014towards}, only a fraction of the previous batch of the sub-sampled data is refreshed.

%Although subtle, this difference yields a significant change of paradigm: our method targets an approximation of $\targ$ using a Markov chain defined on an extended state space, while \cite{korattikara2013austerity} and \cite{bardenet2014towards} use an approximation of the standard M--H kernel on the marginal state-space targeting the exact target distribution.

The original target $\targ$ is extended to model a joint distribution between the parameter of interest $\param\in\paramset$ and an auxiliary $N$-dimensional boolean vector identifying the data involved in the subset.%, an idea which was proposed in \cite{maclaurin2014firefly}.
Each possible subset of data of size $n$ is weighted according to a \textit{similarity measure} with respect to the full set of data, in the spirit of the Approximate Bayesian Computation (ABC) \cite[see \eg][]{marin2012approximate}. %We postpone the complete theoretical analysis of our method for general models to future work.
In the special case of \iid realizations from an exponential model, we prove that when the similarity measure is identical the sufficient statistics, this yields an optimal approximation, in the sense of minimizing an upper bound of the Kullback-Leibler (KL) divergence between $\targ$ and the marginal target of our method.
%However, the analysis of our method in a general setting is more involved and will be the object of a future work.
%This will support and explain the promising results observed in simulations on synthetic and real data of more sophisticated models.

Our main finding is two-fold:
\begin{itemize}
\item for a fixed computational budget, our method can achieve a better bias/variance tradeoff compared to a standard Metropolis--Hastings algorithm and the method developed in \cite{bardenet2014towards}
\item we observe in different scenarios the existence of a tradeoff between the \textit{quality} and the \textit{size} $n$ of the batch of the sub-sampled of data, highlighting an LWA--MCMC \textit{optimal} setting.
\end{itemize}

We start in Section \ref{sec:first_ex} with a striking real data example which we hope will help the reader to understand the problem we address and motivate the solution we propose, without going into further technical details at this stage. In Section \ref{sec:expo}, we provide theoretical results concerning exponential-family models, which we illustrate through a probit example. This section allows us to justify our motivations supporting the LWA--MCMC general methodology developed in Section \ref{sec:alg}: we define the transition kernel on the extended state space and show that it yields a Markov chain targeting, marginally, an approximation of $\targ$. Finally, in Section \ref{sec:sim}, our method is used to estimate parameters of a time-series model and to perform a binary classification task. In the latter example, we compare the performance of our algorithm with the SubLikelihood approach proposed in \cite{bardenet2014towards}.

\section{An introductory example}
\label{sec:first_ex}
We address the problem of estimating template shapes of handwritten digits from the MNIST database
(\href{http://yann.lecun.com/exdb/mnist/}{http://yann.lecun.com/exdb/mnist/}) by inferring a partially known deformable template model \citep{allassonniere2007towards}.
Here, a $16\times 16$ matrix represents a digit whose conditional distribution given its class ($0,1,\ldots,9$) corresponds to a random deformation of the template shape, parameterized by a $d=256$ dimensional vector $\param$, of the known digit. Assuming small deformations, we can rewrite the model as a standard regression problem:
\begin{equation}
\text{given\;} I_k=i,\qquad Y_k=\phi(\param_i)+\sigma^2\eps_k
\end{equation}
where $I_k\in\{0,\ldots,9\}$ is the class of observation $Y_k$ (regarded as a vector $\rset^{225}$), $\phi:\rset^{256}\to\rset^{225}$ is some deterministic linear mapping, $\sigma>0$ is a variance parameter and $\eps_j\sim\norm(0_{225},\Id_{225})$ is additive noise. Given a set of $N=10,000$ labeled images and defining a prior distribution for $\param=\{\param_1,\ldots,\param_9\}$, one can estimate $\param$ through its posterior distribution, for example using a standard Metropolis--Hastings algorithm. However, for two main reasons, the efficiency of such an approach can be questioned: (i) computing the $N$ likelihoods in the M--H ratio dramatically slows down each transition and (ii) the highly peaked posterior distribution hinders a quick exploration of the state space.

Based on these observations, our approach aims at working with different subsets of data which addresses issues (i) and (ii). At this stage, we do not provide precise
details of the LWA--MCMC machinery nor its accuracy but we rather provide an insight of the rationale of this approach: we design a Markov chain whose transition kernel
targets the posterior distribution of the parameter of interest $\param$ given a random subset of $n$ data ($n\ll N$). More specifically, we inject in the standard M--H transition a decision about \textit{refreshing} the subset of data, which, as a result, will change randomly over time. In this example, we use the knowledge of the observation labels to promote subsets in which the proportion of each digit is balanced.

\begin{figure}[h!]
\centering
\begin{tabular}{ccc}
\hspace{-.5cm}time & M--H & \hspace{-.7cm} LWA--MCMC \\
\hspace{-1.3cm}
\begin{tabular}{c}
\vspace{1cm}3 mins\\
\vspace{1cm}15 mins\\
\vspace{1cm}30 mins\\
\vspace{0.1cm}60 mins
\end{tabular} &\hspace{-.9cm}
\begin{tabular}{c}
\includegraphics[scale=0.41]{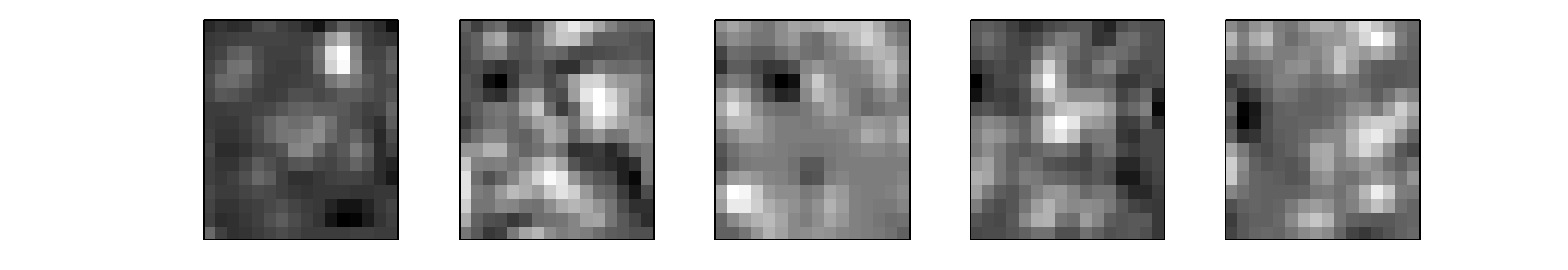}\\
\includegraphics[scale=0.41]{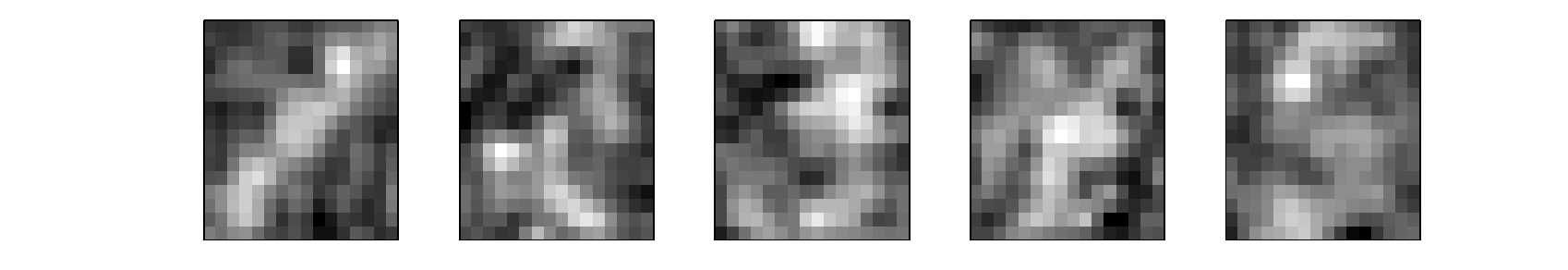}\\
\includegraphics[scale=0.41]{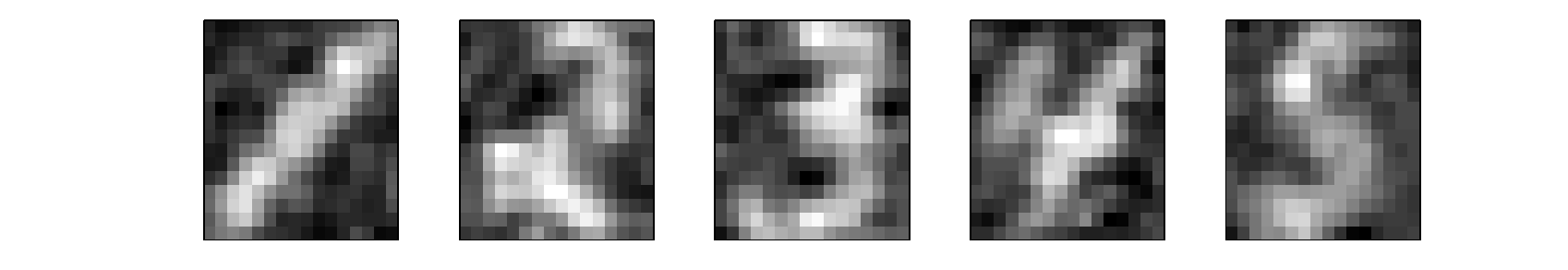}\\
\includegraphics[scale=0.41]{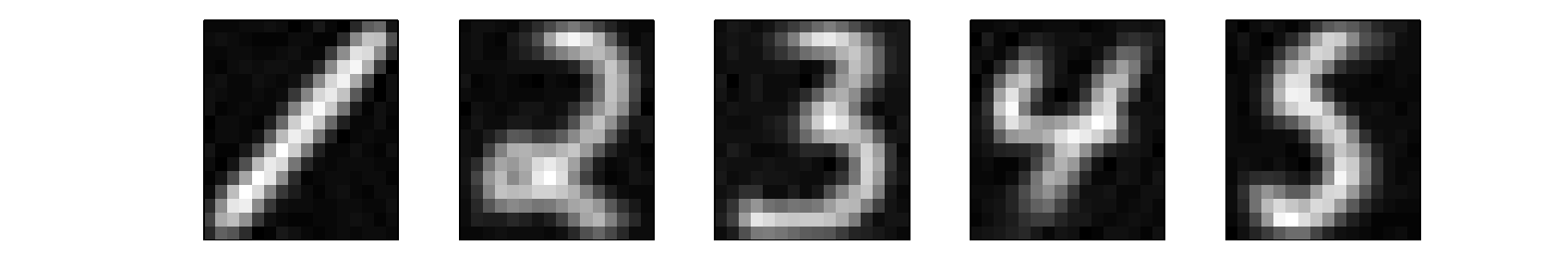}
\end{tabular}&\hspace{-1.5cm}
\begin{tabular}{c}
\includegraphics[scale=0.41]{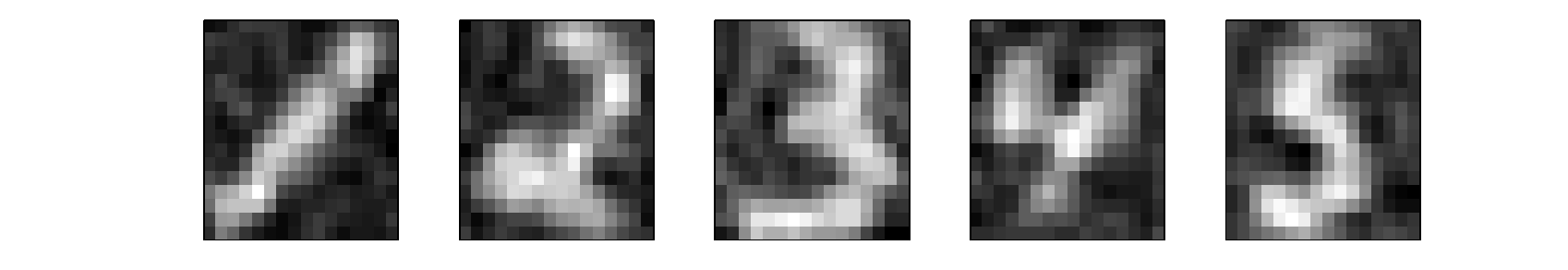}\\
\includegraphics[scale=0.41]{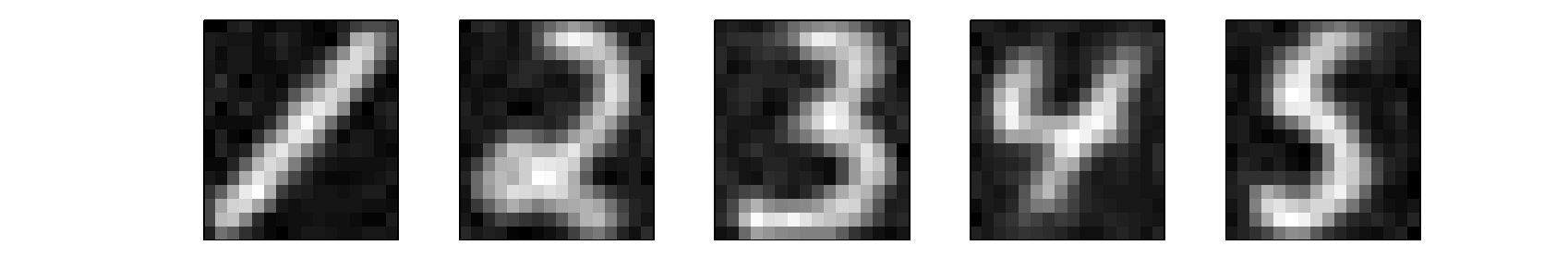}\\
\includegraphics[scale=0.41]{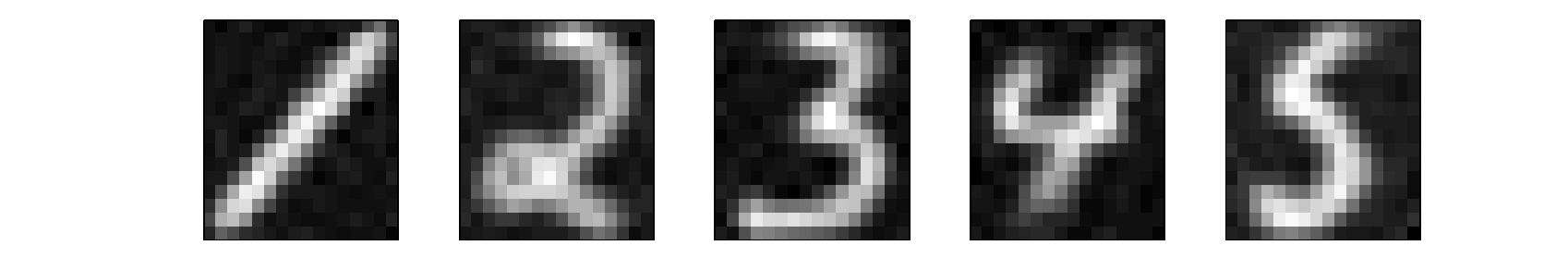}\\
\includegraphics[scale=0.41]{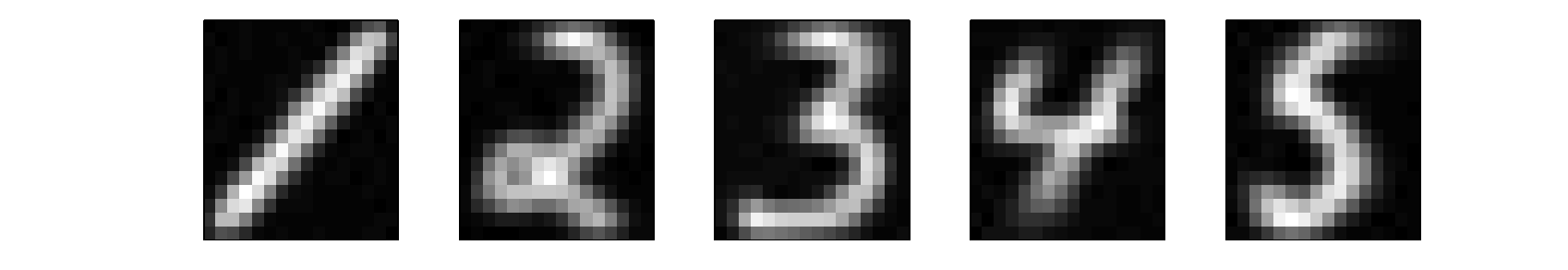}
\end{tabular}
\end{tabular}
\includegraphics[scale=0.7]{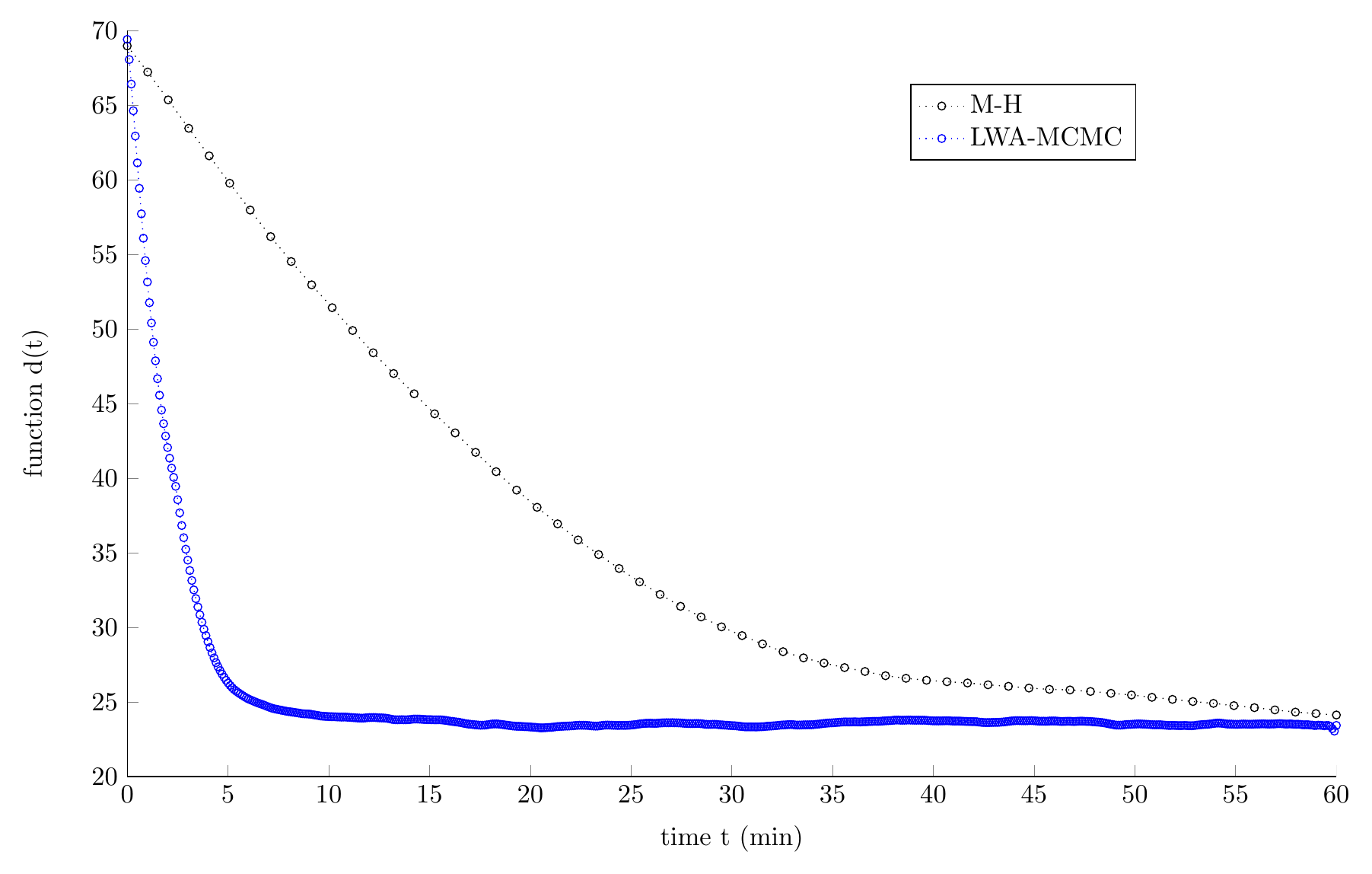}
\caption{Efficiency of template estimation through M--H and LWA--MCMC.}
\label{fig:template}
\end{figure}

We considered only five digits, $1,\ldots,5$ (for illustration purposes), subsets of size $n=100$ and a non-informative Gaussian prior for $\param$, as specified in \cite{allassonniere2007towards}. Figure \ref{fig:template} indicates a striking advantage of our method compared to a standard M--H using the $N=10,000$ data. In this scenario, we allow a given computational budget (1 hour) for both methods and compare the estimation of the mean estimate of the two Markov chains. Qualitatively, the upper part of Figure \ref{fig:template} compares the estimated template shapes of the five digits at different time steps and shows that our method allows to extract template shapes much quicker than the standard M--H, while still reaching an apparent similar graphical quality asymptotically (after one hour). This fact is confirmed quantitatively, in the lower part of Figure \ref{fig:template}, which plots, against time and for both methods, the Euclidean distance between the Markov chain mean estimate and the Maximum
Likelihood Estimate $(\param^{\ast}_1,\ldots,\param^{\ast}_5)$ obtained using a stochastic EM \citep{allassonniere2007towards}. More precisely, we compare the real valued function $\left\{d(t),\;t\in\rset\right\}$ defined as
\begin{equation*}
d(t)=\sum_{i=1}^5\|\param_i^{\ast}-\mu(\param_{i,1:\kappa(t)})\|,\qquad\text{where}
\quad
\left\{
\begin{array}{l}
\forall t\in\rset,\;\kappa(t)=\max_{k\in\nset}\{t\geq \tau_k\}\,,\\
\tau_k\;\text{is\,the\,time\,at\,the\,end\,of\,the\,}k\,\text{-th\,iteration}\,,\\
\forall k\in\nset,\;\mu(\param_{i,1:k})=(1/k)\sum_{j=1}^k\param_{i,j}\,,
\end{array}
\right.
\end{equation*}
for the two Markov chains.

LWA--MCMC provides very encouraging results for this real-data example. We formalize the method and sketch theoretical arguments in the next two sections.

\section{Approximation of the posterior distribution in exponential models: an optimality result}
\label{sec:expo}
In this section, we consider the case of $N$ independent and identically distributed (\iid) observations from an exponential model. Taking the posterior distribution given all the $N$ available data as the target distribution, which we call the \textit{full posterior}, we consider the posterior distribution of the parameter of interest given \textit{only} a subset of the $N$ observations as an approximation of the full-posterior, which we call a \textit{sub-posterior}. We investigate the influence of the choice of a subset of $n$ data on this approximation. Proposition \ref{prop:KL} shows the existence of an optimal set of possible subsets of size $n$ with respect to the Kullback-Leibler (KL) divergence between the full posterior and the sub-posterior. This result will be used in the next section to design and justify the LWA--MCMC methodology, extending this approach to non-\iid observations from general likelihood models.

\subsection{Notation}
Let $\coord{Y}[N]\in \Yset^{N}$ be a set of \iid observed data $(\Yset\subseteq\rset^m,\, m>0)$ and define
\begin{itemize}
\item $Y_{i:j}=(Y_i,\ldots,Y_j)$ if $1\leq i \leq j \leq N$ with the convention that $Y_{i:j}=\{\emptyset\}$, otherwise.
\item $Y_{-k}=(Y_1,\ldots,Y_{k-1},Y_{k+1},\ldots,Y_N)$ for all $k\in\{1,\ldots,N\}$.
\item $Y_U=\{Y_k,\,k\in U\}$, where $U\subseteq \{1,\ldots,N\}$.
\end{itemize}

In this section, we assume that the likelihood model $f$ belongs to the exponential family and is fully specified by a vector of parameters $\param\in\paramset$, ($\paramset\subseteq \rset^d,\,d>0$) and a sufficient statistic mapping $S:\Yset\to\suffset$ ($\suffset\subseteq\rset^s,s>0$) such that
$$
f(y\,|\,\param)=\exp\pscal{g(\param)}{S(y)}\bigg\slash L(\param)
$$
is the density of the likelihood distribution with respect to the Lebesgue measure $\leb(\rmd y)$. Here, the symbol $\pscal{\cdot}{\cdot}$ denotes the canonical inner product in $\suffset$, $g:\paramset\to\suffset $ is a model-specific mapping and $L(\param)$ is the likelihood normalizing constant.

The posterior distribution $\targ$ is defined on the measurable space $(\paramset,\paramalg)$ by its density function
\begin{equation}
\label{eq:posterior_full}
\targ(\param\,|\,Y_{1:N})= p(\param) \prod_{k=1}^N f(Y_k\,|\,\param)\bigg \slash \int p(\rmd \param)\prod_{k=1}^N f(Y_k\,|\,\param)\,,
\end{equation}
with respect to the Lebesgue measure on $(\paramset,\paramalg)$. $p$ is a prior distribution defined on $(\paramset,\paramalg)$ and with some abuse of notation, $p$ denotes also the probability density function (pdf) on $\paramset$ accordingly $(p(\rmd\param)=p(\param)\dom(\rmd\param))$.

For all $n\leq N$, we define $\Uset_n$ as the set of the possible combinations of $n$ different integer numbers less than or equal to $N$ and $\Ualg_n=2^{\Uset_n}$ as the powerset of $\Uset_n$. Finally, let $\Uset=\bigcup_{n\leq N} \Uset_n$ and $\Ualg=2^{\Uset}$. In the sequel, we set $n$ as a constant and wish to compare the full-posterior distribution \eqref{eq:posterior_full} with any of the sub-posterior distributions from the family $\fam_n=\{\ttarg(\,\cdot\,|\,Y_{1:N},U_n),\,U_n\in\Uset_n\}$, where for all $U_n\in\Uset_n$, we have defined
\begin{equation}
\label{eq:subposterior}
\ttarg(\param\,|\,Y_{1:N},U_n)=\ttarg(\param\,|\,Y_{U_n})=p(\param) \prod_{k\in U_n} f(Y_k\,|\,\param)\bigg \slash \int p(\rmd \param)\prod_{k\in U_n} f(Y_k\,|\,\param)\,.
\end{equation}

\subsection{Optimal subsets for the Kullback-Leibler divergence between $\targ$ and $\ttarg$}
Recall that for two measures $\targ$ and $\ttarg$ defined on the same measurable space $(\paramset,\paramalg)$, the Kullback-Leibler (KL) divergence between $\targ$ and $\ttarg$ is defined as:
\begin{equation}
\label{eq:KL}
\KL{\targ}{\ttarg}=\esp_{\targ}\left\{\log\left(\frac{\targ(\param)}{\ttarg(\param)}\right)\right\}\,.
\end{equation}
Although not a proper distance between probability measures defined on the same measurable space, $\KL{\targ}{\ttarg}$ is used as a similarity criterion between $\targ$ and $\ttarg$. It can be interpreted in information theory as a measure of the information lost when $\ttarg$ is used to approximate $\targ$, which is our primary concern here. We now state the main result of this section.

\begin{prop}
\label{prop:KL}
Consider the KL divergence between $\targ=\targ(\,\cdot\,|\,Y_{1:N})$ and $\ttarg_U=\ttarg(\,\cdot\,|\,Y_{U})\in\fam_n$, we have:
\begin{enumerate}[(i)]
\item \begin{equation}
\KL{\targ}{\ttarg_U}\leq \Psi(n,N,Y_{1:N})+B(U,n,Y_{1:N})
\end{equation}
where $\Psi$ is a deterministic function independent of $U$ and $B$ is a positive function.
\item If the set of optimal subsets $\Usetst_n\subset \Uset_n$ defined by
\begin{equation}
\label{eq:opti}
\Usetst_n:=\left\{U\in\Uset_n,\quad \frac{1}{N}\sum_{k=1}^N S(Y_k)=\frac{1}{n}\sum_{k\in U}S(Y_k)\right\}
\end{equation}
is non-empty, then for any $U\in\Usetst_n$,  $B(U,n,Y_{1:N})= 0$, hence yielding an optimal upper-bound.

\item For a given subset $U_0\in\Uset_n$ such that for all $U\in\Uset_n$
\begin{equation}
\label{eq:order1}
\left\|\frac{1}{N}\sum_{k=1}^N S(Y_k)-\frac{1}{n}\sum_{k\in U_0}S(Y_k)\right\|\leq\left\|\frac{1}{N}\sum_{k=1}^N S(Y_k)-\frac{1}{n}\sum_{k\in U}S(Y_k)\right\|\,,
\end{equation}
then we have
\begin{equation}
\label{eq:order2}
B(U_0,n,Y_{1:N})\leq B(U,n,Y_{1:N})\,.
\end{equation}
\end{enumerate}
\end{prop}

In other words, the sub-posterior distributions in $\fam_n$ with subsets having the same sufficient statistics on average as for the full dataset, will achieve an optimal approximation (with respect to upper--bounding the KL divergence). Moreover \eqref{eq:order1} defines an order on $\fam_n$ which implies an order on their relative KL divergence upper-bound \eqref{eq:order2}. The proof is outlined in Appendix \ref{app1}.

\subsection{Illustration with a probit model: effect of choice of sub-sample}
We consider a pedagogical example, based on a probit model, to illustrate Proposition \ref{prop:KL}. A probit model is used in regression problems in which a binary variable $Y_k\in\{0,1\}$ is observed through the following sequence of independent random experiments, defined for all $k\in\{1,\ldots,N\}$ as:
\begin{enumerate}[(i)]
\item Draw $X_k\sim \norm(\paramst,\gamma^2)$
\item Set $Y_k$ as follows
\begin{equation}
\label{eq:probit_model}
Y_k=
\left\{
\begin{array}{ll}
1, & \text{if}\; X_k>0,\\
0, & \text{otherwise}.
\end{array}
\right.
\end{equation}
\end{enumerate}
Observing a large number of realizations $Y_1,\ldots,Y_N$, we aim to estimate the posterior distribution of $\param$. In practice, one also estimates $\gamma$ but for illustration purpose, this parameter is considered as known here. The likelihood function can be expressed as
\begin{equation}
\label{eq:probit}
f(Y_k\,|\,\param)=\alpha(\param)^{Y_k}(1-\alpha(\param))^{(1-Y_k)}=\left(1-\alpha(\param)\right)\left(\frac{\alpha(\param)}{1-\alpha(\param)}\right)^{Y_k},
\end{equation}
where $\alpha(\param)=\int_{0}^{\infty}(2\pi\gamma^2)^{-1/2}\exp\{(1/2\gamma^2)(t-\paramst)^2\}\rmd t$ and clearly belongs to the exponential family. The full-posterior distribution can be written as
$$
\targ(\param\,|\,Y_{1:N})\propto p(\param)\left(1-\alpha(\param)\right)^{N}\left(\frac{\alpha(\param)}{1-\alpha(\param)}\right)^{\sum_{k=1}^N Y_k}\,,
$$
where $p$ is the prior density, which we assume to be non informative ($p(\param)=\norm(a,b^2)$). In this example, the mapping $\param\to\targ(\param\,|\,Y_{1:N})$ can be easily estimated for any $\param\in\paramset$, even when $N$ is extremely large, as it only requires to sum over all the binary variables $Y_1,\ldots,Y_N$. As a consequence, samples from the full-posterior distribution $\targ(\,\cdot\,|\,Y_{1:N})$ can be routinely obtained by a standard M--H algorithm and similarly for any sub-posterior distributions $\ttarg(\,\cdot\,|\,Y_{U})\in\fam_n$.

\begin{figure}[h!]
\centering
\includegraphics[scale=0.65]{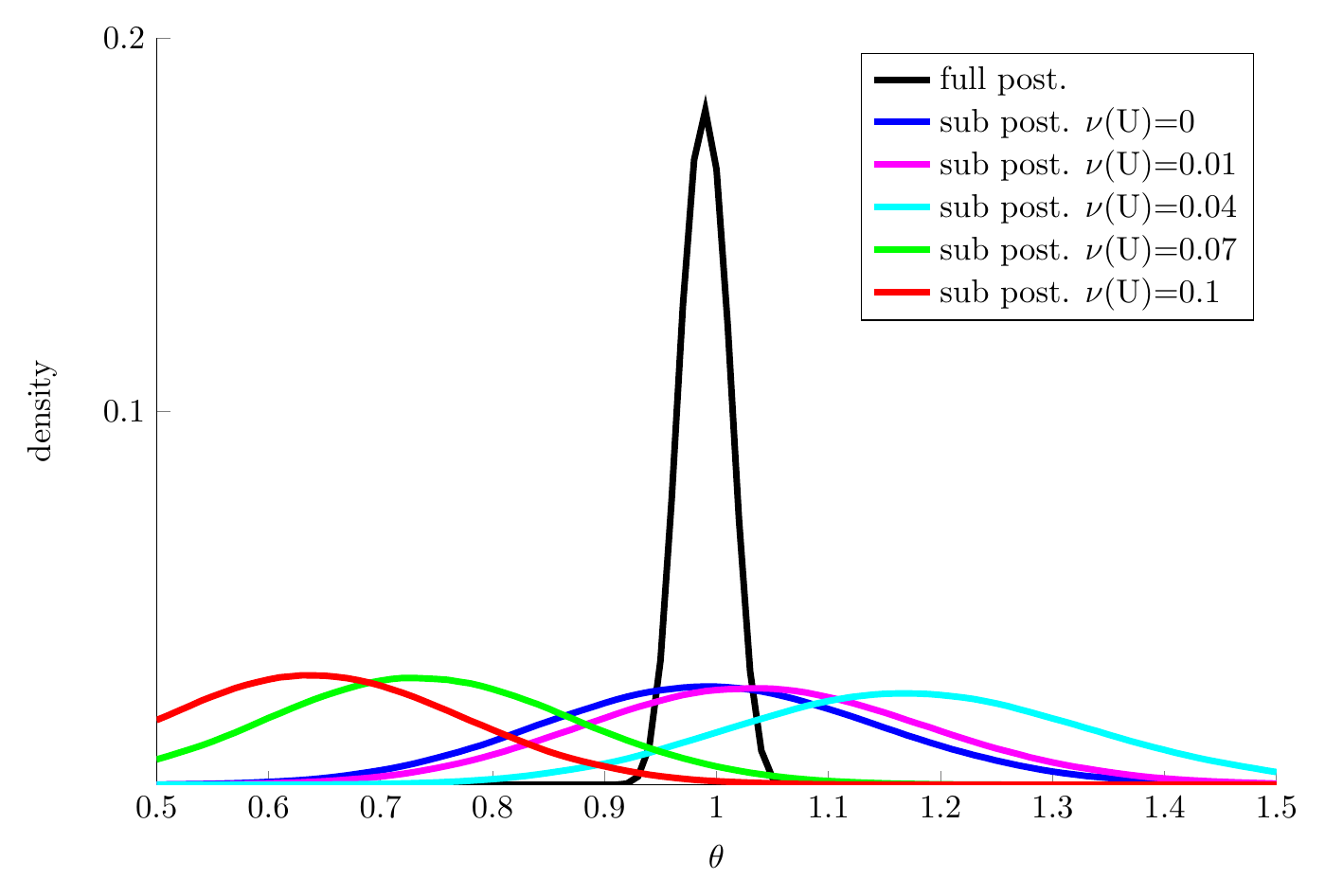}
\includegraphics[scale=0.65]{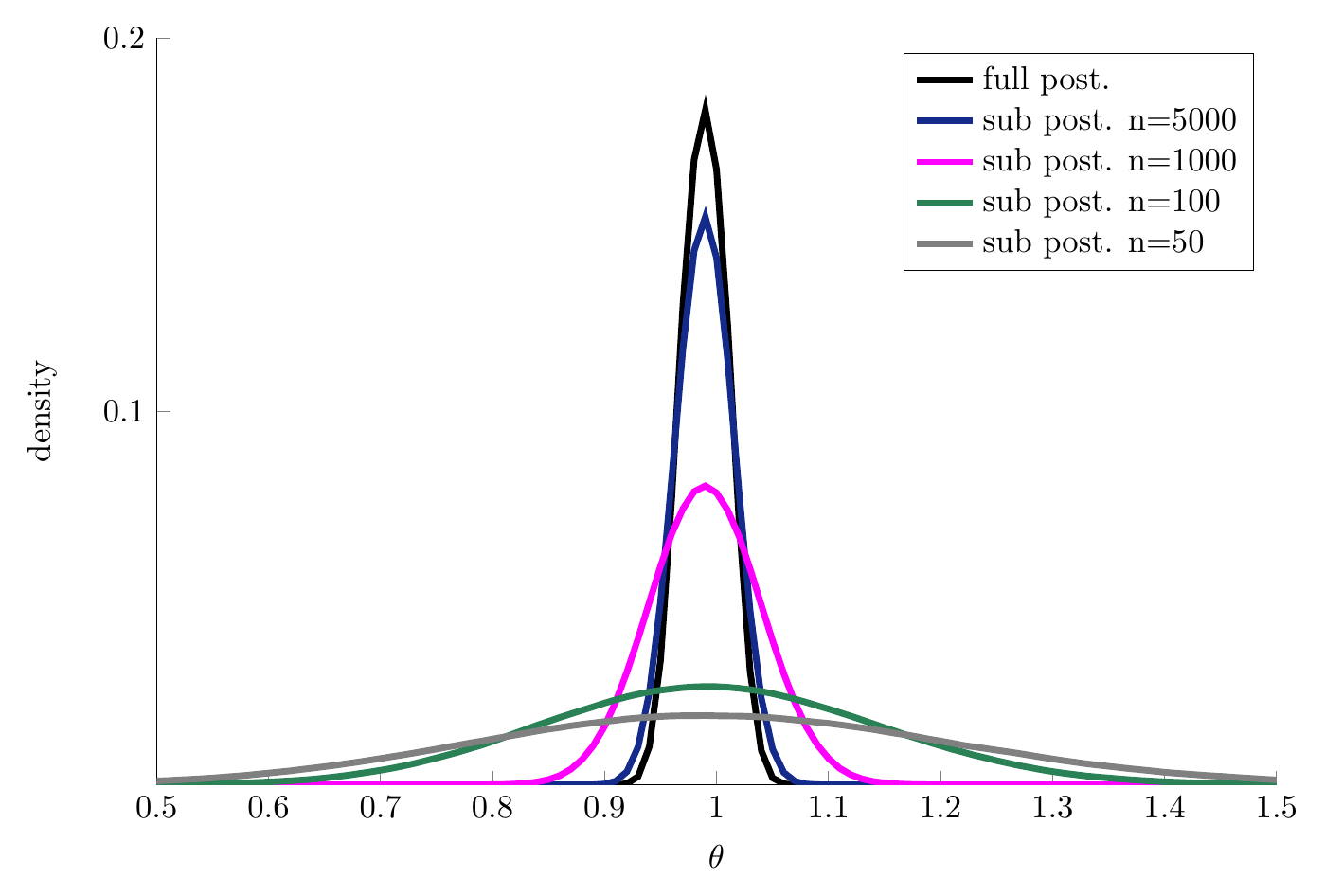}
\caption{Influence of the parameter $U\in\Uset_{100}$ on the sub-posterior distribution $\targ(\,\cdot\,|\,Y_U)$ and comparison with $\targ$ (top) -- Influence of the subsets size $n$ on the sub-posterior distribution $\targ(\,\cdot\,|\,Y_U)$, $U\in\Usetst_n$  (bottom).}
\label{fig:probit}
\end{figure}

\begin{table}
\centering
\begin{tabular}{|c|c|}
\hline
$r(U)$ & $\KL{\targ}{\ttarg_U}\slash \KL{\targ}{\ttarg_{U^{\ast}}}$\\
\hline
$0.01$ & $1.13$\\
$0.04$ & $1.39$\\
$0.07$ & $1.87$\\
$0.1$ & $2.76$\\
\hline
\end{tabular}

\vspace{.3cm}
\caption{Comparison of the KL divergence between the full-posterior and the optimal sub-posterior ($r(U)=0$) with the KL distance between the full-posterior and other sub-posterior distribution.}
\label{tab:KL}
\end{table}

We present in Figure \ref{fig:probit} some inference results for the full-posterior and several sub-posterior distributions (with different $n$ and different values of sufficient statistics) obtained with parameters $(\paramst,\gamma)=(1,1)$ and $N=10,000$ simulated data. In the upper plot, we hold $n=100$ constant and compare several sub-posterior distributions given subsets of data $U\in\Uset_{100}$ having different matches with the full data sufficient statistics $r(U)=|\bS(U)-\bS_N|$, where $\bS(U)=n^{-1}\sum_{k\in U} S(Y_k)$ and $\bS_N=N^{-1}\sum_{k=1}^N S(Y_k)$. In this probit model, $S$ is simply the identity function, implying that $r(U)$ monitories the different proportion of 1 and 0's between the full dataset and in the subset $U$. This plot, as well as the quantitative result of Table \ref{tab:KL} providing the KL divergence between the full-posterior and these different sub-posterior distributions are consistent with the statement of Proposition \ref{prop:KL}: when learning from a subset of $n$
data, one should work with a subset $U$ featuring a perfect match with the full dataset, \ie $r(U)=0$, to achieve an \textit{optimal} approximation of $\targ$. Note that the sub-posterior distributions only vary through their sufficient statistics, so obviously, in exponential models, the choice of the \textit{optimal} subset is not unique, as $\Usetst_n$ is not restricted to one element.

The lower plot of Figure \ref{fig:probit} compares the influence of $n$ on the \textit{optimal} sub-posterior distributions $\ttarg(\,\cdot\,|\,Y_U)$, $U\in\Usetst_n$ for $n\in\{50,100,1000,5000\}$. As expected, while the variance becomes wider as $n$ decreases, the expectation remains relatively constant.

\begin{rem}
\label{rem:bvm1}
The results of the lower plot of Figure \ref{fig:probit} may in some ways be related with the \textit{local asymptotic normality} (lan) theorem: Bernstein von Mises theorem \cite[see \eg][]{van2000asymptotic} states that under some mild assumptions about the likelihood and the prior distribution, the posterior distribution is asymptotically Gaussian (in $n$) in the case where the "true" parameter is an interior point of the parameter space. More precisely, the mean of the Gaussian distribution corresponds to the maximum likelihood estimate of the observed data, while the covariance is given by $H^{-1}/n$, where $H$ is the Hessian matrix at the "true" parameter value. Even though for low values of $n$, such an asymptotic result does not hold, it is nevertheless consistent with observing in Figure \ref{fig:probit} that the variance varies (in function of $n$) much more than the mean of those distributions.
\end{rem}

\begin{rem}
\label{rem:bvm2}
At this stage, one might wonder why so much effort has been put to overcome the problem of sampling from a posterior distribution given a huge amount of data, while using a local asymptotic normality theorem would virtually allow to solve this problem by sampling from a Gaussian distribution. Two main arguments actually prevent one to make use of such a result:
\begin{itemize}
\item estimation of the coefficients of the asymptotic Gaussian distribution is non-trivial, as one needs to invert the Hessian matrix at $\paramst$, which is typically unknown;
\item local asymptotic normality theorems only hold under restrictive assumptions, for example, when the observations are \iid
\end{itemize}
\end{rem}

On the basis of our analysis conducted in the case of \iid realizations from an exponential-family model, both at theoretical and experimental levels, it seems
reasonable to consider estimating the Maximum A Posteriori parameter based on a subset of data. At first glance, when one aims to estimate $\param$ using a Markov
chain targeting a sub-posterior, even optimal, $\targ(\,\cdot\,|\,Y_{U})$, $(U\in\Usetst_n)$ instead of the full-posterior may lead to a worse efficiency, iteration wise.
%{\color{red} I am happy to remove the link with ESS...} Taking for example, as an efficiency criterion, the Effective Sampling Size (ESS), defined as the number of \iid draws from the full-posterior required to reach an equivalent precision, it is clear that, because of the possible larger variance of the sub-posterior, the ESS associated to such a target will decrease as $n$ decreases. {\color{red}up to here}
However, assuming that the computational complexity of a Markov chain targeting the full posterior is prohibitively intensive, one may consider a Markov chain targeting
an optimal sub-posterior as a realistic alternative: more Markov chain iterations would be required but at a known and affordable computational cost. This yields a trade
on the subset size $n$ which will allow lighter transitions at the price of a loss in variance.

%{\color{red} Should we leave this last sentence? Although reasonable, it sounds a bit pessimistic and open the door to a quick dismiss of our idea} When estimating higher order statistics or quantiles, curvature correction techniques \citep{stoehr2015calibration} could be used in order to compensate the inherent drawback of learning from a smaller amount of data. This issue is out of the scope of the paper.

\section{Light and Widely Applicable MCMC: the general methodology}
\label{sec:alg}
In this section, we do not assume any particular correlation pattern for the sequence of observations, nor any specific likelihood model and simply write the posterior distribution $\targ$ as
\begin{equation}
\targ(\rmd\param\,|\,\coor{Y}[N])\propto p(\rmd\param) f(Y_{1:N}\,|\,\param)\,.
\end{equation}

The Light and Widely Applicable MCMC (LWA--MCMC) methodology that we describe now can be regarded as a generalization of the sub-posterior inference detailed in the previous section to non-exponential-family models with possibly dependent observations.

\subsection{Motivation of our approach}
Here, we do not assume the existence of a sufficient statistic mapping for the model under consideration. Thus, in order to allow comparison between different subsets of data, we introduce an \textit{artificial} summary statistic mapping $S_n:\Yset_n\to\suffset$ ($n\leq N$), where $\suffset\subseteq \rset^s$. The choice of the summary statistics $S_n$ is problem specific and is meant to be the counterpart of the sufficient statistic mapping for general models (hence sharing, slightly abusively, the same notation). Intuitively, a good choice of $S_n$ would capture most of the statistical features of $Y_U$ ($U\in\Uset_n$). In an attempt to derive a similar analysis to the exponential model case (Section \ref{sec:expo}) and  to reach an optimal setup in line with Proposition \ref{prop:KL}, we want to focus inference on those subsets whose summary statistics vector is close to that of the full dataset. In this context a subset $Y_U$ ($U\in\Uset_n$) is said to be more \textit{representative of the full dataset} than another $Y_{U'}$ ($U'\in\Uset_n$), if $\|\bS_n(Y_U)-\bS_N(Y_{1:N})\|\leq\|\bS_n(Y_{U'})-\bS_N(Y_{1:N})\|$, where we have set $\bS_n(Y_{U'})=S_n(Y_{U'})/n$ as the \textit{normalized} summary statistics and $\|\cdot\|$ as the Euclidean distance on $\suffset$. Since the question of specifying summary statistics also arises in ABC, one can take advantage of the abundant ABC literature on this topic to find some examples of summary statistics for usual likelihood models \cite[see \eg][]{nunes2010optimal,csillery2010approximate,marin2012approximate,fearnhead2012constructing}.

We formalise this idea by assigning a weight to any possible subset of data $U\in\Uset_n$. Let for any $n\leq N$ and $\eps>0$, $\nu_{n,\eps}$ be the distribution defined on the discrete state space $(\Uset_n,\Ualg_n)$ whose density with respect to the counting measure is:
\begin{equation}
\label{eq:ttarget_prior}
\forall\,U\in\Uset_n,\qquad
\nu_{n,\epsilon}(U)= \Phi\left(-\frac{\|\bS_n(Y_{U})-\bs_N\|}{\eps}\right)\,\bigg\slash \sum_{U'\in\Uset_n}\Phi\left(-\frac{\|\bS_n(Y_{U'})-\bs_N\|}{\eps}\right)\,.
\end{equation}
Here $\Phi:\rset\to\rset^{+}$ is a kernel function, $\bs_N=\bS_N(Y_{1:N})\in\suffset$ and the dependence of $\nu_{n,\eps}(U)$ on $Y_{1:N}$ is implicit. The parameter $\eps$ allows to control the influence of the representativeness of a subset $U\in\Uset_n$ on its overall weight $\nu_{n,\eps}(U)$: if $\eps\gg 1$, the weights tend to be uniform, whereas if $\eps\ll 1$ the weights tend to highlight the representativeness of the subsets. As a consequence, $\eps$ is a tuning parameter whose impact on the inference is significant, as we will see in Section \ref{sec:sim}. The kernel $\Phi$ allows to smooth the weights and offers protection against the possibly unbounded weights derived from Euclidean distances (situations which typically arise when weights are proportional to $\|\bS_n(U)-\bs_N\|^{-1}$ and $\Usetst_n$ \eqref{eq:opti} is not the empty set). In this paper, we let $\Phi$ be the Gaussian kernel.

\begin{rem}
\label{rem:complimentary}
Note that because the statistics used to assess the representativeness of a subset w.r.t. the full dataset are only \textit{summary} and not \textit{sufficient}, two subsets $(U,U')\in\Uset_n^2$ such that $\nu_{n,\eps}(U)=\nu_{n,\eps}(U')$ might yield two different sub-posteriors. As a consequence, should a unique optimal subset $U^{\ast}=\arg\max_{U\in\Uset_n} \nu_{n,\eps}(U)$ be available, inferring the full-posterior through this corresponding \textit{optimal} sub-posterior is likely to provide an unbalanced learning as most of the data would be simply ignored. Alternatively, when learning from a set of \textit{good} subsets, where goodness is measured by $\nu_{n,\eps}$, one can expect that each sub-posterior involved in the process will act complementarily to improve the approximation of $\targ$.
\end{rem}

At this stage, two main questions need to be addressed:
\begin{enumerate}[(i)]
\item how can a set of \textit{good} subsets be determined? Indeed, the dimension of $\Uset_n$, $|\Uset_n|=$ $N\choose n$ is prohibitively too large to estimate the normalizing constant in \eqref{eq:ttarget_prior} and therefore we reasonably assume in the following that the set of weights $\{\nu_{n,\eps}(U),\;U\in\Uset_n\}$ is unknown.
\item how to design an inference scheme that would make use of these subsets, without resorting to a set of independent Markov chains each targeting through a standard M--H transition kernel a \text{good} sub-posterior, which could become even more demanding than a standard M--H targeting $\targ$?
\end{enumerate}

Because we assume that $\{\nu_{n,\eps}(U),\;U\in\Uset_n\}$ is unknown, we consider that the subsets $U_1,U_2,\ldots$ ($U_i\in\Uset_n$) involved in the learning setup are latent variables of the model. We define the data-augmented distribution $\ttarg_{n,\eps}$ for any $A\in\paramalg$ and $U\in\Uset_n$ by
\begin{equation}
\label{eq:data_aug_dist}
\ttarg_{n,\epsilon}(A,U\,|\,Y_{1:N})=\nu_{n,\epsilon}(U)\targ(A\,|\,Y_U)\,,
\end{equation}
where
$$
\targ(\rmd\param\,|\,Y_U)={p(\param)f(Y_U\,|\,\param)\leb(\rmd \param)}\bigg\slash{\int p(\param')f(Y_U\,|\,\param')\leb(\rmd\param')}
$$
is a sub-posterior distribution.

With some abuse of notation, we write $\ttarg_{n,\eps}$ also for the density of the data-augmented distribution with respect to the product measure $\leb(\rmd\param)\1_{\Uset_n}(U)$. Integrating out the subset variable yields the marginal of interest whose density w.r.t. $\leb$ is expressed as:
\begin{equation}
\label{eq:ttarget_marg}
\ttargs_{n,\epsilon}(\param\,|\,Y_{1:N})=\sum_{U\in\Uset_n} \nu_{n,\epsilon}(U)\targ(\param\,|\,Y_U)\,.
\end{equation}
In this way, the marginal distribution $\ttargs_{n,\epsilon}$ defines an approximation of $\targ$. Straightforwardly, as $n\to N$, our approximation $\ttargs_{n,\eps}$ converges to $\targ$. %Moreover, note that the data-augmented distribution $\ttarg_{n,\eps}$ \eqref{eq:data_aug_dist} is actually a hierarchical model, where (i) a subset $U\in\Uset_n$ is drawn from $U\sim\nu_{n,\eps}$ and (ii) conditioning on $U$, a parameter $\param\in\paramset$ is drawn from $\param\,|\,U \sim\ttarg(\,\cdot\,|\,Y_U)$.
%
%\begin{rem}
%Straightforwardly, as $n\to N$, we have $\ttargs_{n,\eps}\to\targ$.
%\end{rem}
%
%\begin{rem}
%\label{rem:hier}
%The data-augmented distribution $\ttarg_{n,\eps}$ \eqref{eq:data_aug_dist} is actually a hierarchical model, where (i) a subset $U\in\Uset_n$ is drawn from $U\sim\nu_{n,\eps}$ and (ii) conditioning on $U$, a parameter $\param\in\paramset$ is drawn from $\param\,|\,U \sim\ttarg(\,\cdot\,|\,Y_U)$.
%\end{rem}

%\begin{rem}The marginal $\ttargs_{n,\epsilon}$ in \eqref{eq:ttarget_marg} can be regarded as a mixture of $\binom{N}{n}$ components each being a sub-posterior distribution of $\param$ given a subset of $n$ data. Interestingly, by construction, the weight of the components are proportional to the representativeness of the subset of data involved in the conditioning of each component. This echoes in some sense Remark \ref{rem:complimentary}: each component has a complimentary contribution to the general approximation, since each makes use of a different subset of data. This approximation also covers the exponential model case (Section \ref{sec:expo}): setting $\eps\ll 1$ will make all the weights vanish but the subset $\{\nu_{n,\eps}(U), U\in\Usetst_n\}$, if $\Usetst_n$ is non-empty, and
%$$
%\ttargs_{n,\epsilon}(\param\,|\,Y_{1:N})\approx \sum_{U\in\Usetst_n} \nu_{n,\epsilon}(U) \targ(\param\,|\,Y_U)=\frac{1}{|\Usetst_n|}\sum_{U\in\Usetst_n} \targ(\param\,|\,Y_U)=\targ(\param\,|\,Y_{U^{\star}})
%$$
%for any $U^{\star}\in\Usetst_n$, since all the sub-posteriors with subsets having identical sufficient statistics are the same.
%\end{rem}

\begin{rem}
The parameters $n$ and $\eps$ of the joint distribution $(\param,U)$, have complementary roles in the approximation of $\targ$. While $n\ll N$ lightens the full-posterior distribution, $\eps$ allows to make up for the approximation by attaching bigger weights to representative subsets in the mixture distribution $\ttarg_{n,\eps}$ \eqref{eq:data_aug_dist}. Setting $\eps$ implies a tradeoff between:
\begin{itemize}
\item $\eps \gg 1$ and $\nu_{n,\eps}$ is a flat distribution on $U$ and all subsets have the same weight in the mixture
\item $\eps \ll 1$ and $\ttarg_{n,\eps}$ has most of its probability mass on $\paramset \times \Usetst_n$
\end{itemize}
In particular, the latter scenario yields an approximation of the full-posterior based only on one or a few subsets. Depending on the relevance of the summary statistics, the choice $\eps\ll 1$ may provide a disastrous setup. The simulations detailed in Section \ref{sec:sim} clearly show the existence of an optimal $\eps$ for a given $n$ and which also depends on the likelihood model considered.
\end{rem}

\subsection{Formal description of the LWA--MCMC algorithm}

In this section, we specify LWA--MCMC which allows to sample from an approximation of $\targ$. This relies on a Markov chain whose transition kernel achieves our main target, namely having a bounded computational complexity, which can be controlled through the parameter $n$.% We start with the following preliminary remark.
%
%\begin{rem}
%The approximate target $\ttargs_{n,\eps}$ is a marginal of the data-augmented distribution $\ttarg_{n,\eps}$. It is therefore sufficient to be able to sample $(\param,U)\sim\ttarg_{n,\eps}$ to derive samples from $\ttargs_{n,\eps}$. Remark \ref{rem:hier} suggests that sampling (i) $U\sim\nu_{n,\eps}$ and then (ii) $\param\,|\,U\sim \targ(\,\cdot\,|\,Y_U)$ clearly yields independent samples from $\ttarg_{n,\eps}$. However, the assumption that drawing from $\nu_{n,\eps}$ is infeasible for all subsets prevents one to proceed this way.
%\end{rem}

\subsubsection{LWA--MCMC}

LWA--MCMC makes use of a Markov transition kernel operating on an extended state space $(\paramset\times\Uset_n,\paramalg\otimes \Ualg_n)$. Algorithm \ref{alg:lwamcmc} depicts a Markov transition of the LWA--MCMC algorithm. This embeds two successive decisions: the first one allows to \textit{refresh} the subset variable $U$ while the second one updates the parameter $\param$. $R(U,\cdot)$ and $Q(\param,\cdot)$ are proposal kernels respectively on $\Uset\times \mathcal{U}$ and $\paramset\times\borel(\paramset)$.
%More precisely, assume that we have $\param_k=\param$ and $U_k=U$. A transition $(\param_{k},U_k)\to (\param_{k+1},U_{k+1})$ involves two steps:
%\begin{enumerate}[S-1]
%\item propose $U'\sim R(U,\cdot)$ and define $U_{k+1}=U'$ with probability
%\begin{equation}
%\label{eq:acc1}
%\rho_{n,\eps}(U,U')=1\wedge\frac{\nu_{n,\eps}(U')R(U',U)}{\nu_{n,\eps}(U)R(U,U')}\,,
%\end{equation}
%and $U_{k+1}=U$, otherwise.
%\item given the following events,
%\begin{itemize}
%\item $\{\text{no subset refresh}\}$:
%propose $\param'\sim Q(\param;\,\cdot\,)$ and set $\param_{k+1}=\param'$ with probability
%\begin{equation}
%\label{eq:cacc2}
%\halpha(\param,\param'\,|\,U)=1\wedge\frac{\targ(\param'\,|\,Y_U) Q(\param';\param)}
%{\targ(\param\,|\,Y_U) Q(\param;\param')}\,,
%\end{equation}
%and $\param_{k+1}=\param$, otherwise.
%\item $\{\text{subset refresh}\}$:
%given $(\param,U_{k+1})$, draw $\param_{k+1}\sim K_{U_{k+1}}(\param;\,\cdot)$, where $K_{U_{k+1}}(\param;\,\cdot)$ corresponds to the M--H step with acceptance $\halpha(\param,\param'\,|\,U_{k+1})$ \eqref{eq:cacc2} iterated $L$ times $(L\geq 1)$.
%\end{itemize}
%\end{enumerate}

\begin{algorithm}
\caption{LWA--MCMC transition $(\param_k,U_k)\to(\param_{k+1},U_{k+1})$}
\label{alg:lwamcmc}
\begin{algorithmic}[1]
\State {\bf{Input: current state $(\param_k,U_k)$ and summary statistics of current subset $S_k=\bar{S}_n(Y_{U_k})$}}
\State propose to refresh the subset $U'\sim R(U_k,\cdot)$
\State compute the summary statistics $S'=\bar{S}_n(Y_{U'})$
\State set $U_{k+1}=U'$ with probability
\begin{equation}
\label{eq:acc1}
\rho_{n,\eps}(U_k,U')=1\wedge\frac{R(U',U_k)\Phi\left(-{\|S'-\bs_N\|}\slash{\eps}\right)}{R(U_k,U')\Phi\left(-{\|S_k-\bs_N\|}\slash{\eps}\right)}\,,
\end{equation}
\State leave $U_{k+1}=U_k$ otherwise
\If{the subset has not been refreshed}
     \State propose a change of parameter $\param'\sim Q(\param_k;\,\cdot\,)$
     \State set $\param_{k+1}=\param'$ with probability
\begin{equation}
\label{eq:cacc2}
\halpha(\param_k,\param'\,|\,U_k)=1\wedge\frac{\targ(\param'\,|\,Y_{U_k}) Q(\param';\param_k)}
{\targ(\param_k\,|\,Y_{U_k}) Q(\param_k;\param')}\,,
\end{equation}
\State leave $\param_{k+1}=\param_k$ otherwise
    \Else
    \State set $\param_{k+1}\sim T_{U_{k+1}}(\param_k;\,\cdot)$ where $T_{U_{k+1}}(\param_k;\cdot)$ is the M--H transition kernel corresponding to steps 7, 8 and 9 iterated $L$ times $(L>1)$
\EndIf
\State {\bf{return: $(\param_{k+1},U_{k+1})$ and $S'$ if $U_{k+1}\neq U_k$ and $S_k$ otherwise}}
\end{algorithmic}
\end{algorithm}

\begin{figure}[h!]
\centering
\begin{tikzpicture}
\node  at (-1.5,1) {$U_0$};
\draw [->](-1.15,1)--(-.6,0.2);
\draw [->](-1.15,-1)--(-.6,-0.2);
\node  at (-.05,0) {($\ldots$)};
\draw [->](.5,0.2)--(1.25,1);
\draw [->](.5,-0.2)--(1.25,-1);
\node  at (1.6,1) {$U_k$};
\draw [->](1.5,0.8)--(1.5,-0.8);
\draw [->](2,1)--(3.5,1);
\draw [->](2,0.8)--(3.5,-0.8);
\node  at (4,1) {$U_{k+1}$};
\draw [->](4.5,1)--(6,1);
\draw [->](4,0.8)--(4,-0.8);
\draw [->](4.5,0.8)--(6,-0.8);
\node  at (6.5,1) {$U_{k+2}$};
\draw [->](7,1)--(8.5,1);
\draw [->](6.5,0.8)--(6.5,-0.8);
%\draw [->](4.5,0.8)--(7.5,-0.8);
\node  at (9,1) {$\ldots$};
\draw [->](7,0.8)--(8.5,-0.8);

\node  at (-1.5,-1) {$\param_0$};
\node  at (1.6,-1) {$\param_k$};
\draw [->](2,-1)--(3.5,-1);
\node  at (4,-1) {$\param_{k+1}$};
\draw [->](4.5,-1)--(6,-1);
\node  at (6.5,-1) {$\param_{k+2}$};
\draw [->](7,-1)--(8.5,-1);
\node  at (9,-1) {$\ldots$};
\end{tikzpicture}
\caption{Intertwined structure of the LWA--MCMC Markov chain.}
\label{fig:dependenceLWA}
\end{figure}
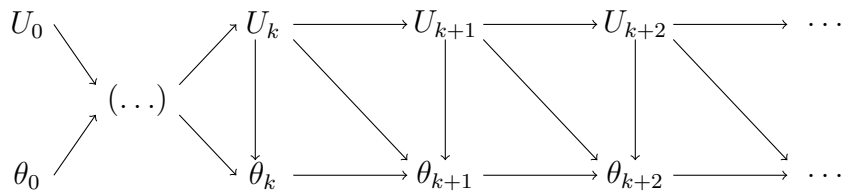

The dependence structure within the sequence of variables $\{\param_k,\,k\in\nset\}$ and $\{U_k,\,k\in\nset\}$ produced by LWA--MCMC is displayed in Figure \ref{fig:dependenceLWA}. This yields an \textit{intertwined} design in which $\{U_k,\,k\in\nset\}$ is itself a Markov chain targeting $\nu_{n,\eps}$, such that $U_k$ is independent from the past sequence of parameters $(\param_1,\param_2,\ldots,\param_k)$. This is an important feature of our proposal as allowing $U_{k+1}$ to depend on $\param_k$ could potentially lead to a local optimisation of the parameter $\param$ with respect to a given subset $Y_U$. Indeed, a parameter $\param_k$ may happen to fit particularly well the subset of data $Y_{U_k}$ and therefore the $U$ component of the chain may get stuck at $U_k$ for a large number of iterations. % This well known issue of poor mixing for MCMC targeting mixture models is usually addressed by resorting to Carlin and Chib \citep{carlin1995bayesian} or Reversible Jump MCMC \citep{green1995reversible} approaches. However, here again, these methods are ruled out due to the computational burden they generate.
We claim that LWA--MCMC allows to move freely throughout $\Uset_n$ at a limited computational cost, a fact which is empirically confirmed by the simulations. Moreover, the two distinctive accept/reject steps for $U$ and $\param$ allows the parameter $\param$ not to be hampered by a lack of move in the $U$ direction.

To summarize, the LWA--MCMC sampling scheme is appealing as it allows:
\begin{enumerate}[(i)]
\item to update the subset $U_k$ and the parameter $\param_k$ through two different decisions,
\item to make the subset update independent of $\param_k$,
\item to control the computational complexity of a transition,
\item to be applied in any inference problem where the likelihood function is tractable.
\end{enumerate}

The following remark shows that standard Markov chain methods cannot be used to sample from $\ttarg_{n,\eps}$, hence justifying the sampling machinery of LWA--MCMC from another perspective.

\begin{rem}
A block update M--H or a Metropolis-within-Gibbs algorithm targeting $\ttarg_{n,\eps}$ cannot be implemented, as both involve at some stage an intractable acceptance ratio (the normalizing constant of the sub-posteriors does not cancel anymore because of the different subsets of data involved).
\end{rem}

%\begin{rem}
%\label{rem:IBIS}
%Given that a Markov chain $\{U_k,\,k\in\nset\}$ can be sampled easily (through \eqref{eq:acc1}) and in a pre-processing step, realizations of $\ttargs_{n,\eps}$ can be obtained provided that a sequence of random variables $\{\param_k\sim\targ(\,\cdot\,|\,Y_{U_k}),\,k\in\nset\}$ is accessible, conditionally on $\{U_k,\,k\in\nset\}$. In this perspective, sampling from $\ttargs_{n,\eps}$ is equivalent to a problem of targeting a time-evolving probability distribution, which can be elegantly addressed with particles methods such as the Iterated Batch Importance Sampling particle filter \citep{chopin2002sequential}. Even though this methods is appealing since it circumvents the normalizing constant issue of \eqref{eq:MH_RNderivative}, we do not pursue it in this paper, as we want to stick closely to the M--H methodology. %Note that this approach has recently been revisited in \cite{strathmann2015unbiased} to derive an unbiased estimator whose computational complexity is sub-linear in the size of the dataset.
%\end{rem}

\subsubsection{Stability of LWA--MCMC}
Let us assume that $(\param_k,U_k)\sim \ttarg_{n,\eps}$. First, note that marginally
\begin{equation}
\label{eq:nu_marginal}
(\param_k,U_k)\sim\ttarg_{n,\eps}\Rightarrow U_k\sim \nu_{n,\eps}\Rightarrow U_{k+1}\sim \nu_{n,\eps}\,,
\end{equation}
where the later implication holds as $\{U_k,\,k\in\nset\}$ is in itself a M--H Markov chain with target distribution $\nu_{n,\eps}$ (See Figure \ref{fig:dependenceLWA} and Algorithm \ref{alg:lwamcmc} steps 2--5). On the one hand, given the event $\{\text{no subset refresh}\}$, $(\param_{k+1},U_{k+1})\sim \ttarg_{n,\eps}$ simply derives from the fact that $\param_{k+1}\,|\,U_{k+1}=U_k\sim \targ(\,\cdot\,|\,Y_{U_k})$ which holds since steps 7--9 in Algorithm \ref{alg:lwamcmc} consist in a standard Metropolis-within-Gibbs update for $\param$. On the other hand, the event $\{\text{subset refresh}\}$ disturbs the stationarity: $(\param_{k+1},U_{k+1})\not\sim \ttarg_{n,\eps}$. Indeed, in order to remain invariant, the transition kernel should produce a sample $\param_{k+1}$ such that  $\param_{k+1}\,|\,U_{k+1}\sim \targ(\,\cdot\,|\,U_{k+1})$. However, this is not achievable in a single Markov transition since the target distribution $\targ(\,\cdot\,|\,U_k)$ becomes instantaneously $\targ(\,\cdot\,|\,U_{k+1})$. Instead, the sample $\param_k$ can be regarded as the initial state of a Markov chain having $\targ(\,\cdot\,|\,U_{k+1})$ as stationary distribution. Taking $L\gg 1$ provides a state $\param_{k+1}$ which is approximately distributed under $\targ(\,\cdot\,|\,U_{k+1})$. Therefore, the LWA--MCMC transition kernel is not, stricto-sensu, stationary with respect to $\ttarg_{n,\eps}$, as when the subset is refreshed, $\param_{k+1}$ is a sample from the full conditional asymptotically in $L$.
At this point we make the following remarks.
%{\color{red} Should not we chop this point as it sounds very pessimistic}
%At this stage, one might legitimately argue that resorting to the transition kernel $T_U(\param,\cdot)$ (i) burdens the transition kernel and (ii) disturbs the main Markov chain convergence. However, we claim that:
\begin{itemize}
\item Since the sequence of distributions $\{\targ(\cdot\,|\,Y_{U_k}),\,k\in\nset\}$ are likely to be close to each other, one can use a very limited number of intermediate steps, typically $L=1$ was used in all the simulations of Section \ref{sec:sim}. Indeed, depending on the number of data refreshed by $R$, two consecutive subsets $Y_{U_k}$ and  $Y_{U_{k+1}}$ may only differ through very few observations, hence advocating setting $L=1$ does not adversely effect the convergence of the Markov chain, a statement which is supported by our experiments.

    %\vspace{1cm}
\item Tuning the prior distribution with $\eps\ll 1$ leads to a set of weights $\{\nu_{n,\eps}(U),\,U\in\Uset_n\}$ with high discrepancy. Therefore, when the marginal chain $\{U_k,\,k\in\nset\}$ reaches stationarity, the subset samples yield similar summary statistics and makes the distributions $\targ(\,\cdot\,|\,U_{k})$ and $\targ(\,\cdot\,|\,U_{k+1})$ even closer. In addition, this makes refreshing $U$ an unlikely event and a transition is therefore most of the time valid.
\end{itemize}
%{\color{red} to here}

Although, we do not carry out further theoretical analysis in this paper and leave the proof of stability of the marginal Markov chain $\{\param_k,\,k\in\nset\}$ as the central question of a forthcoming paper, we outline two considerations that are likely to help proving the ergodicity of the LWA--MCMC Markov chain.

\begin{itemize}
\item \underline{Considering $\ttarg_{n,\eps}$ as an intermediate target distribution}

The ergodicity of the data-augmented Markov chain relies on the ability of the sampler to absorb minor target changes. To the best of our knowledge and probably as a result of a rather unusual MCMC development, there has been very little literature on the convergence of Markov chain toward time-evolving target distributions. However, we connect this question to the one addressed in \cite{kuhn2004coupling} about the stability of a Markov chain targeting a sequence of distributions, each parameterized by a vector which is recursively updated through a Stochastic Approximation procedure \citep{robbins1951stochastic}.
\item \underline{Considering $\targ$ as the ultimate target distribution}

In another perspective, considering the marginal chain of interest $\{\param_k,\,k\in\nset\}$ and regarding $\{U_k,\,k\in\nset\}$ as an auxiliary sequence of parameters, the acceptance probability $\halpha(\param,\param'\,|\,U)$ of the LWA--MCMC can be viewed as an noisy version of the M--H acceptance ratio, which is the acceptance probability involved in the M--H transition targeting the full-posterior $\targ$. This observation combined with the main result of \cite{alquier2014noisy} \cite[see also][]{pillai2014ergodicity} stating that a M--H transition kernel with a noisy acceptance probability nevertheless allows to obtain samples from a distribution whose total variation distance to $\targ$ is bounded from above, provided that for all $\param_k\in\paramset$, $U_{k+1}\in\Uset$ and $\param'\sim Q(\param_k,\,\cdot\,)$,
$$
\esp\left|\halpha(\param_k,\param'\,|\,U_{k+1})-\alpha(\param_k,\param')\right|\leq \delta(\param_k,\param')\,,
$$
where the expectation is taken under $U_{k+1}$ and $\delta:\paramset^2\to\rset^{+}$ is a deterministic function.
\end{itemize}

\section{Illustrations of LWA--MCMC}
\label{sec:sim}
We evaluate the efficiency of LWA--MCMC in two different applications: the first is posterior inference of a time series observed at $N=10^7$ contiguous time steps. This example is particularly relevant since the observations are non \iid: as a result, this makes LWA--MCMC the only competing algorithm against M--H to infer such a model -- the other solutions \citep{korattikara2013austerity,bardenet2014towards,maclaurin2014firefly,banterle2014accelerating} being only suitable for \iid data. The second task is a Gaussian binary classification problem based on $N=10^7$ data, which we use to compare LWA--MCMC with the M--H Sub Likelihood inference method proposed in \cite{bardenet2014towards}. Finally, we provide some additional details to the handwritten digit example outlined in Section \ref{sec:first_ex}.

\subsection{Inference of an ARMA model}
An ARMA(1,1) time series $\{Y_k,\,k\leq N\}$ is defined recursively by:
\begin{equation}
\label{eq:arma}
\left\{
\begin{array}{l}
Y_0\sim \mu, \quad Z_0\sim \mathcal{N}(0,\sigma^2)\\
\\
Y_{k}=\alpha Y_{k-1}+\beta Z_{k-1} + \gamma +Z_{k}\,,\quad Z_k\sim \mathcal{N}(0,\sigma^2)\,,\qquad\forall\,k\geq 1,
\end{array}
\right.
\end{equation}
where $\mu$ is some distribution on $(\Yset,\Yalg)$ and $\param=(\alpha,\beta,\gamma)\in \mathbb{R}^3$. The likelihood of a trajectory can be written
\begin{equation}
\label{eq:arma_lkhd}
f(Y_{0:N}\,|\,\param)=\mu(Y_0)\prod_{k=1}^{N}g_k(Y_k\,|\,Y_{0:k-1},\theta)\,,
\end{equation}
such that for all $k\geq 1$,
\begin{equation}
\label{eq:arma_lkhd2}
g(Y_k\,|\,Y_{0:k-1},\param)=\Psi(Y_k; \alpha Y_{k-1}+\beta h_k(Y_{1:k-1})+\gamma ,\sigma^2)
\end{equation}
where we have defined $x\to\Psi(x;m,v)$ as the pdf of the univariate Gaussian distribution with mean $m$ and variance $v$ and $\{h_k,\,k\geq 1\}$ is a set of known deterministic mappings.

The purpose of this example is to infer the posterior distribution $\targ(\param\,|\,Y_{1:N})$ using LWA--MCMC. We sampled a time series $\{Y_k,\,k\leq N\}$ according to \eqref{eq:arma}, with $N=10^7$, $\mu=\norm(0,1)$ and using $\paramst=(0.5,0.7,0.1)$. The prior distribution on $\param$ is deliberately non-informative and taken as a Gaussian with mean $(0,0,0)$ and a diagonal covariance matrix with a large variance. We restrict the subset parameter to the the set $\Vset_n\subset \Uset_n$ involving $n$ contiguous observations:
$$
\Vset_n=\left\{Y_{0:n-1},Y_{1:n},\ldots,Y_{N-n+1:N}\right\}\,.
$$
Using such a sub-window yields a tractable likelihood \eqref{eq:arma_lkhd} as otherwise, the function $h_k$ in \eqref{eq:arma_lkhd2} is no longer explicit. We compare the efficiency of LWA--MCMC with respect to M--H in function of $n$, $\eps$ and $S$. In all this section, the simulations were achieved with a gaussian proposal kernel $Q$ with variance tuned so that the acceptance rate of the sequence $\{\param_k,\,k\in\nset\}$ is between $30\%$ and $40\%$. For LWA--MCMC, a subset $U\in\Vset_n$ is identified with its starting index and the proposal kernel $R$ consists in assigning a probability on the discrete alphabet $\{1,\ldots,N_n+1\}$ standing for all the possible starting time for subsets $U\in\Vset_n$. More precisely, $R$ can be written as
\begin{multline}
\label{eq:def_R}
R(n_0;n_0')\propto\omega\, {\exp{\left(-\lambda|n_0-n_0'|\right)}}+(1-\omega)\exp{\left(-\lambda|n'-n_0'|\right)},\\
\quad n'\sim\mathcal{U}(\{1,\ldots,N-n+1\})\,.
\end{multline}
The rationale is to propose $U'\sim R(U,\,\cdot)$ through a mixture of two distributions: the first gives higher weight to local moves and the latter allows jumps to remote sections of the time series (through the offset $n'$). This mixture allows to browse through $\Vset_n$, while avoiding to remain trapped in local optimal subsets. In this example, we have used $\omega=0.9$ and $\lambda=0.1$.

\begin{figure}
\centering
\includegraphics[scale=0.6]{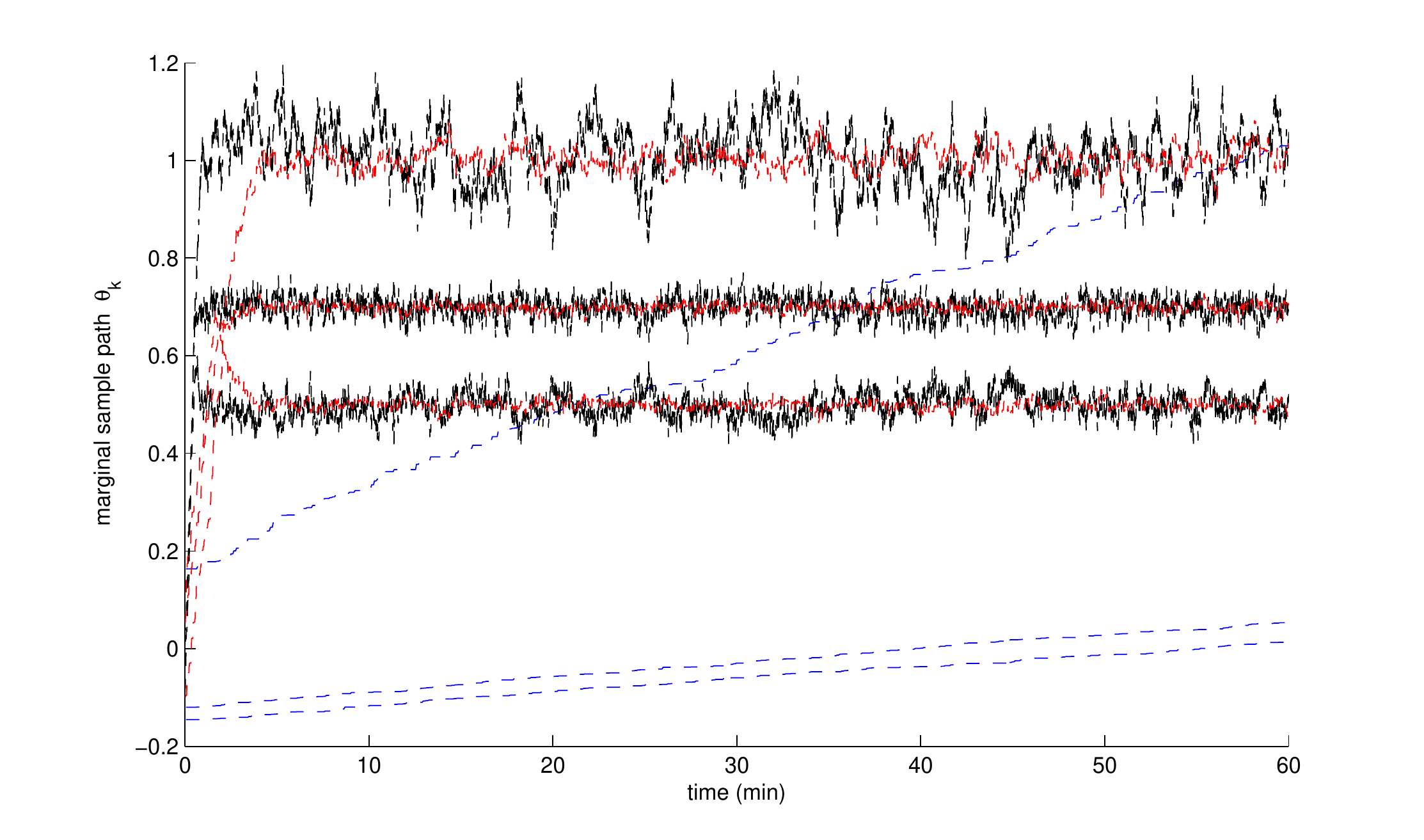}
\caption{ARMA example - M--H (dashed, blue) and LWA--MCMC (dashed, red and black) ($\eps=1$, $S=S_0$) for a fixed computational budget. \label{fig:time_norm_n}}
\end{figure}

Figure \ref{fig:time_norm_n} displays the sample path of three different Markov chains (blue, black and red) simulated during a fixed time budget (1 hour). The blue dashed sample path is the standard M--H targeting the full-posterior $\targ$ and the other two correspond to the LWA--MCMC transition kernel for $n=10,000$ and $n=1,000$, with $\eps=1$. In this setup, the summary statistics vector was defined as $S=S_0$ with

\begin{figure}
\centering
\begin{tabular}{ccc}
\includegraphics[scale=0.45]{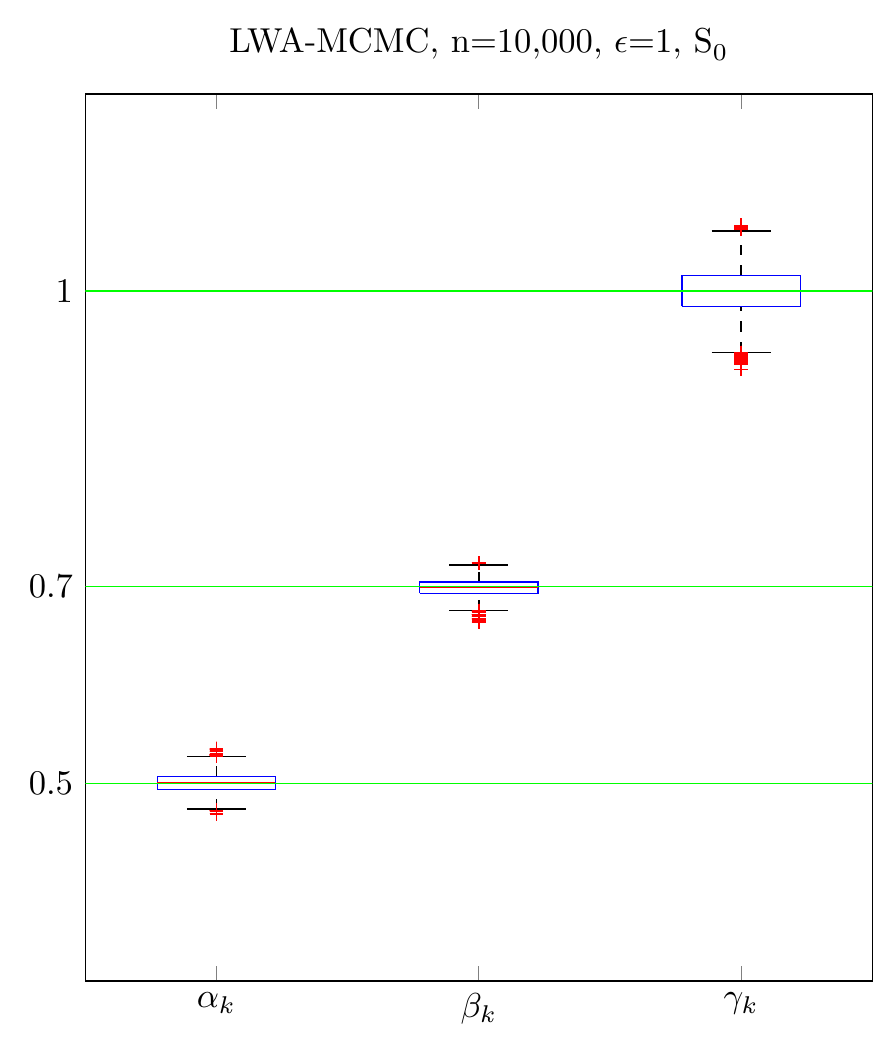}&
\includegraphics[scale=0.45]{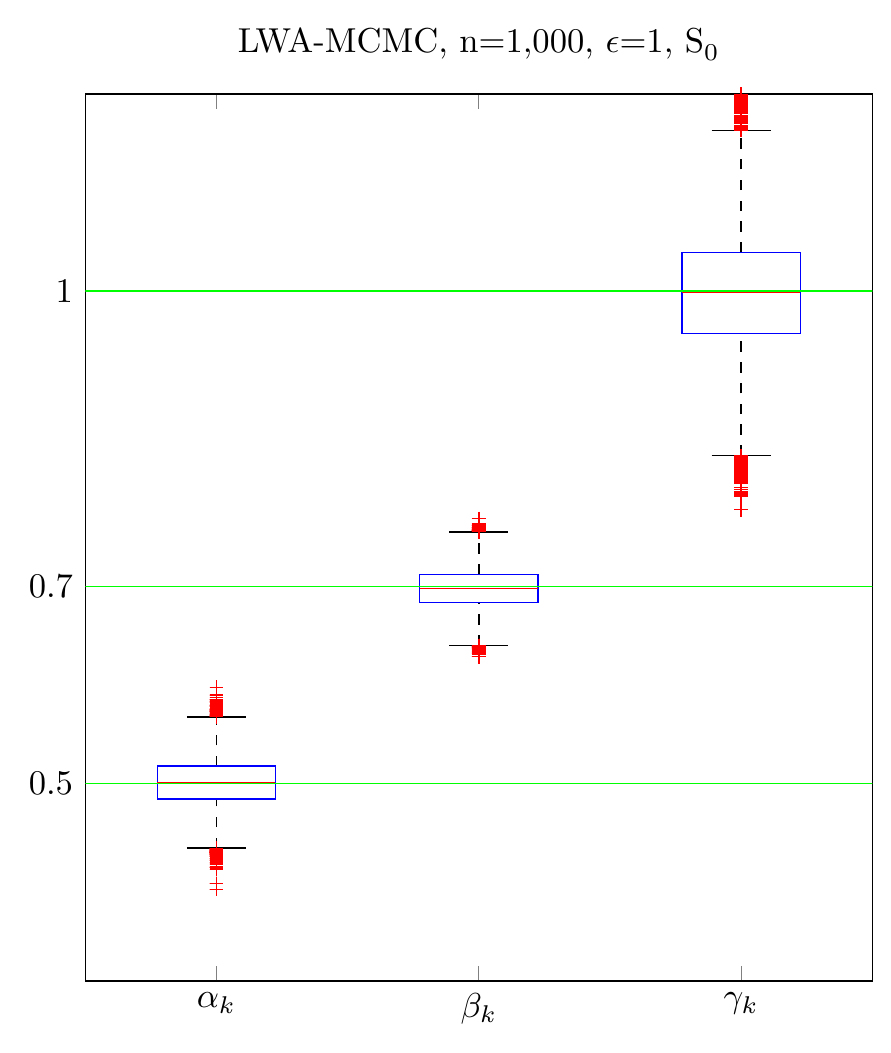}&
\includegraphics[scale=0.45]{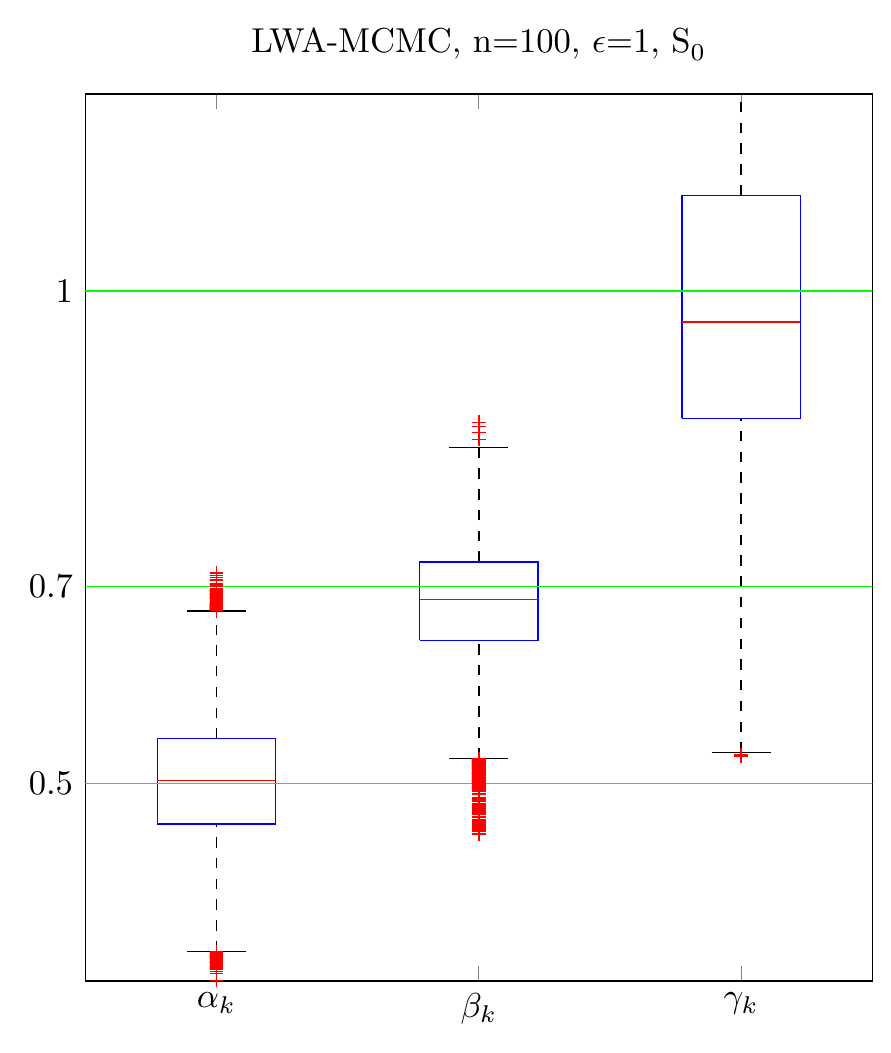}
\end{tabular}
\caption{ARMA example - Boxplot of $100$ independent LWA--MCMC sample paths for different $n$ \label{fig:lwamcmc_n} -- green lines represent the true parameter $\param^{\ast}$.}
\end{figure}

\begin{figure}
\centering
\begin{tabular}{ccc}
\includegraphics[scale=0.25]{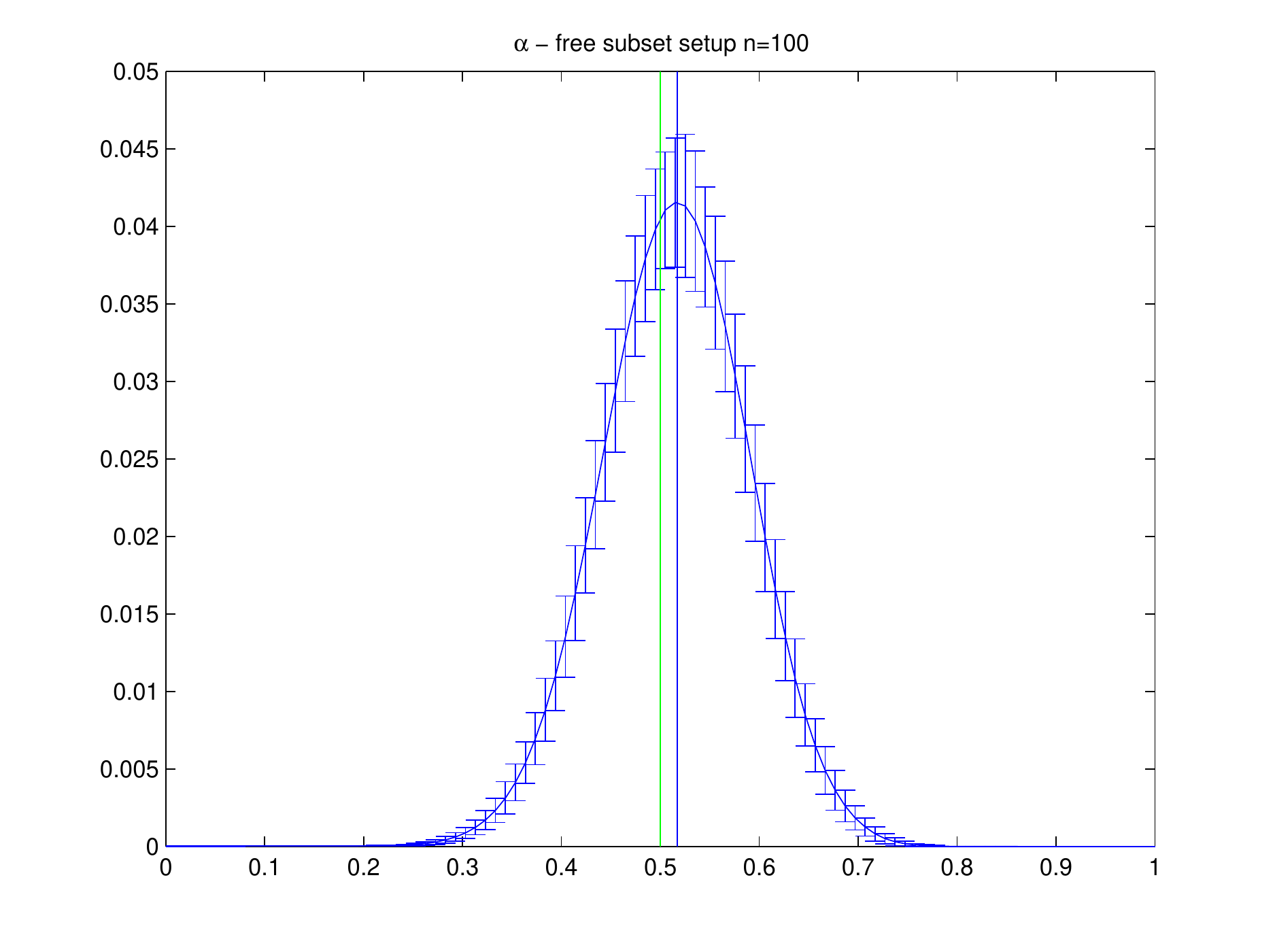}&\includegraphics[scale=0.25]{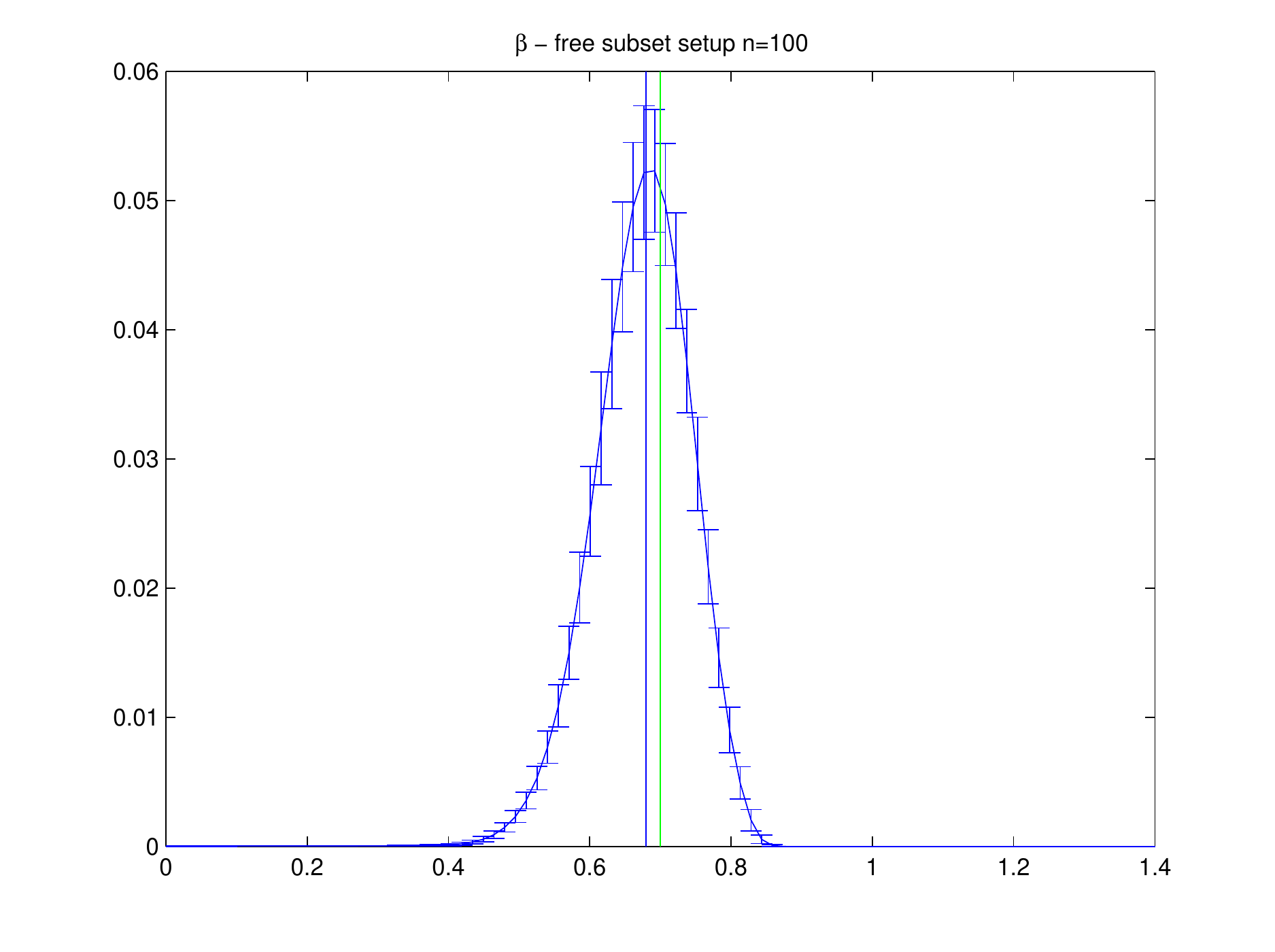}&\includegraphics[scale=0.25]{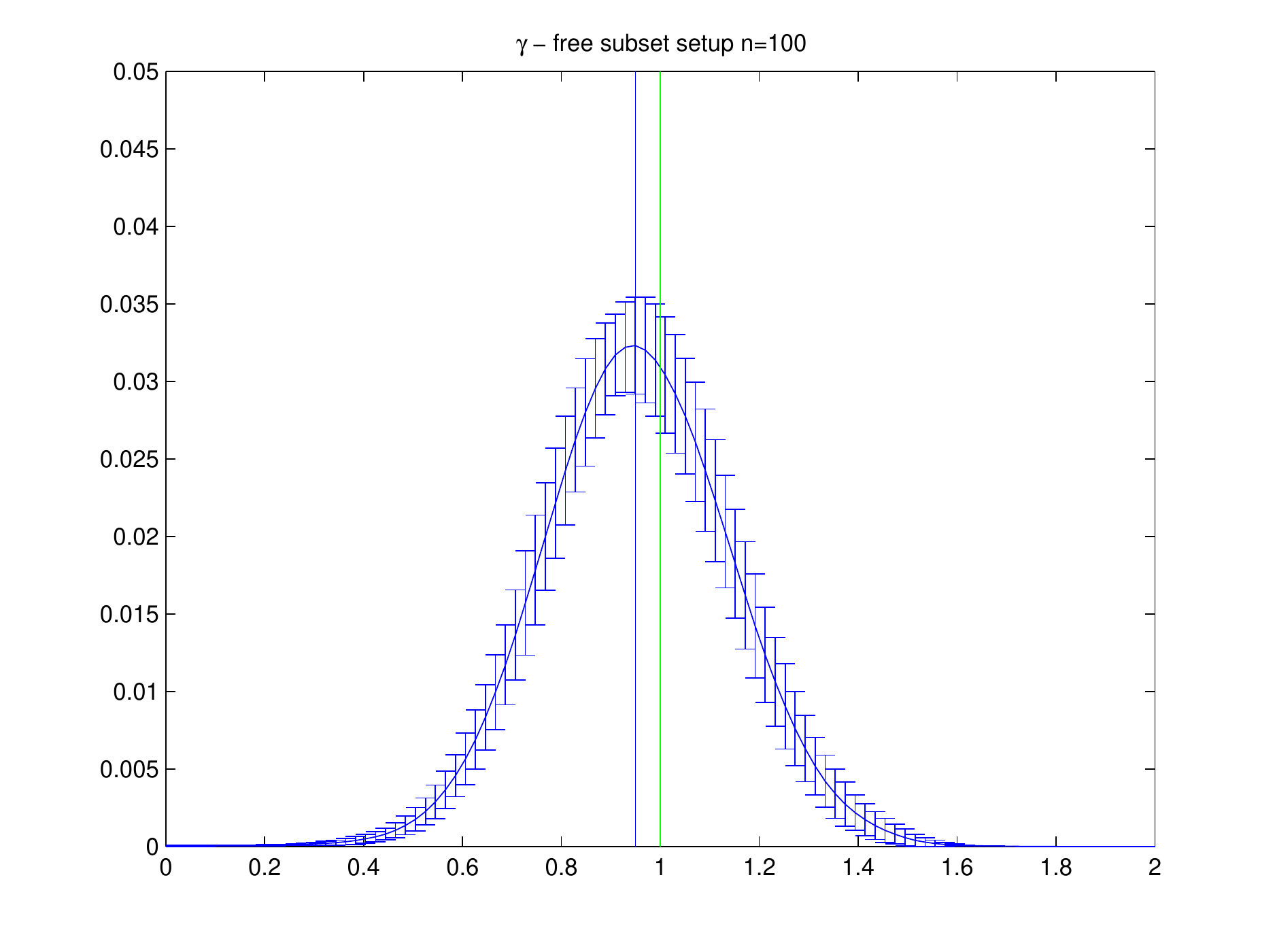}\\
\includegraphics[scale=0.25]{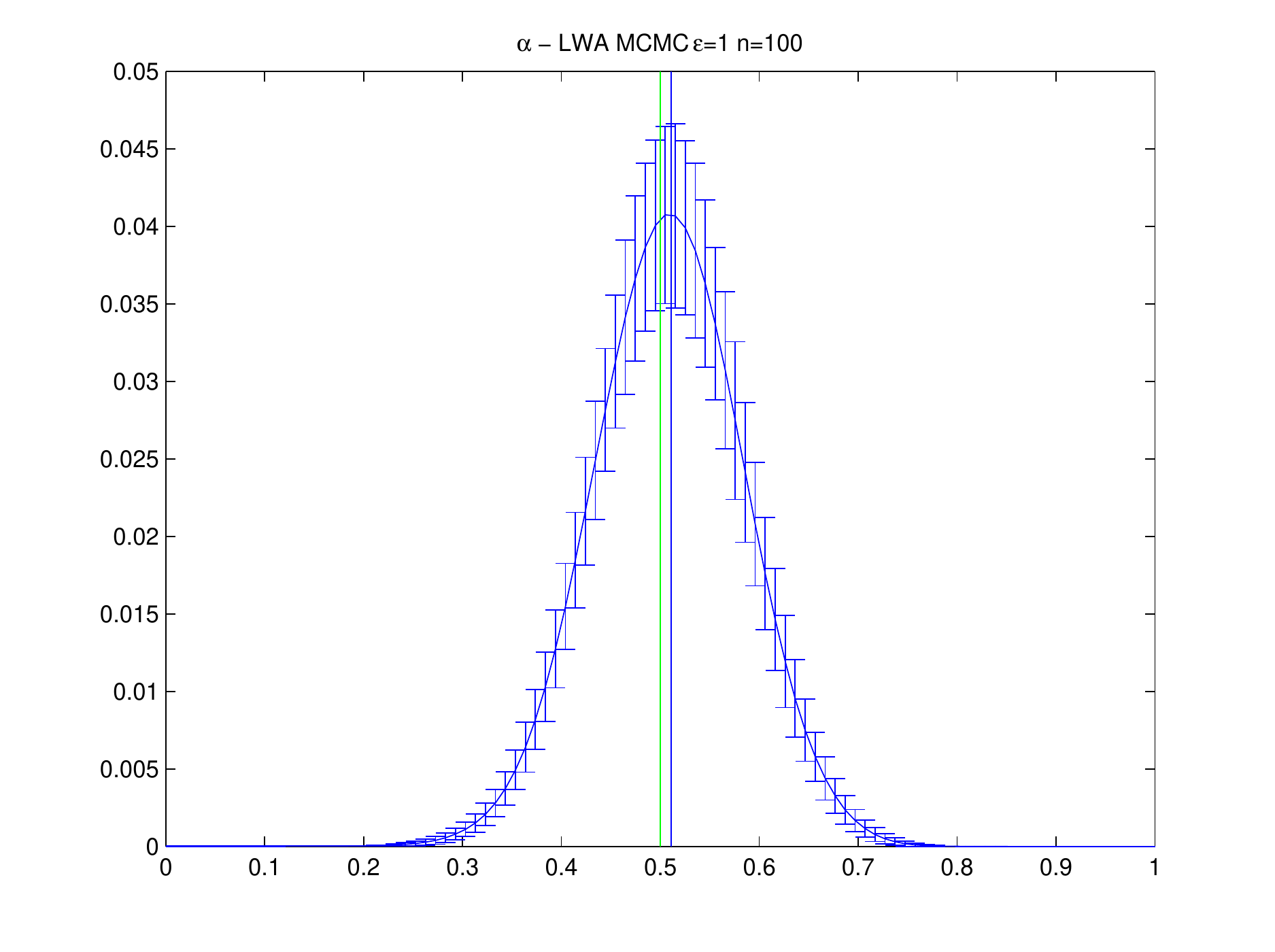}&\includegraphics[scale=0.25]{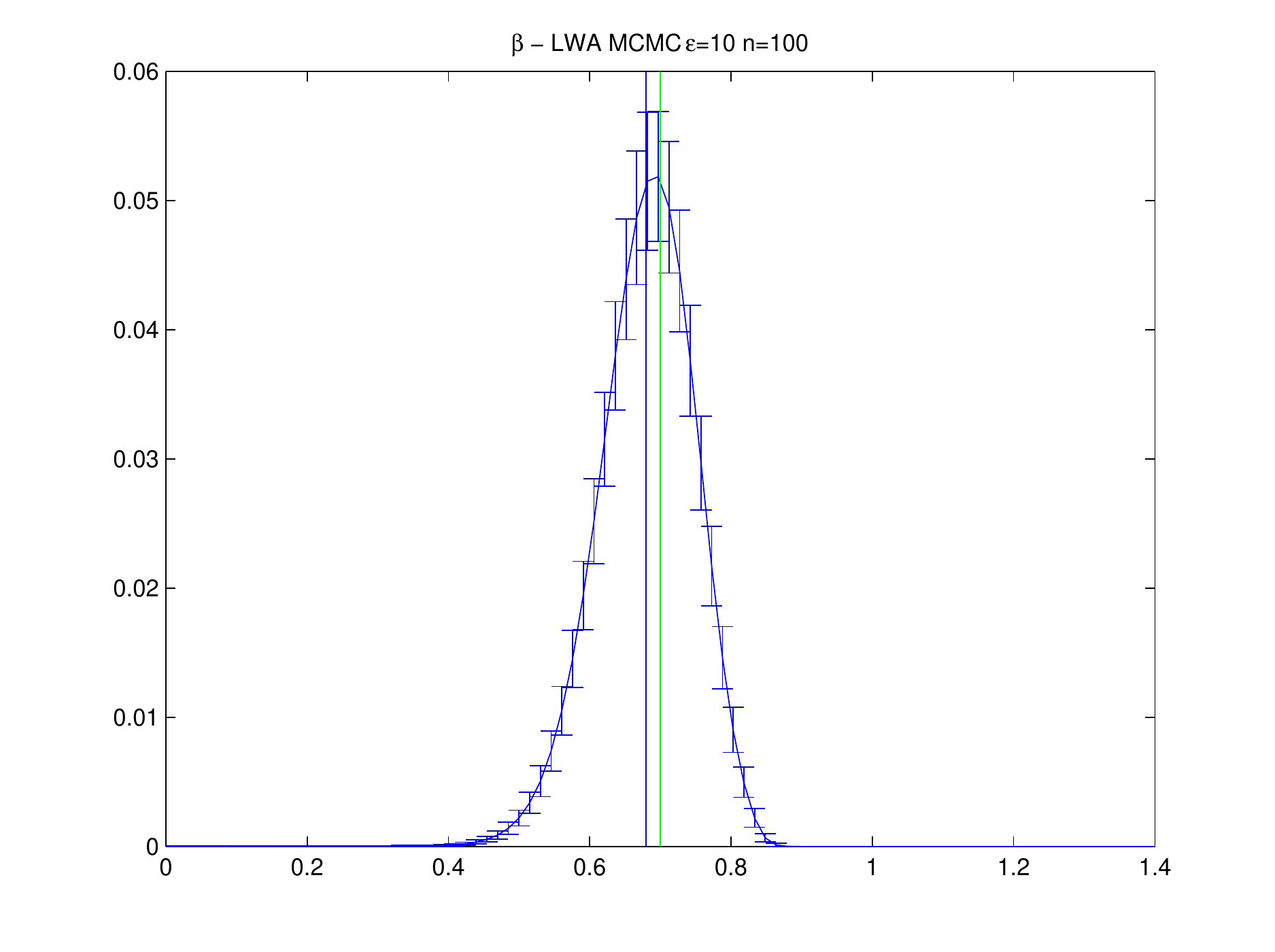}&\includegraphics[scale=0.25]{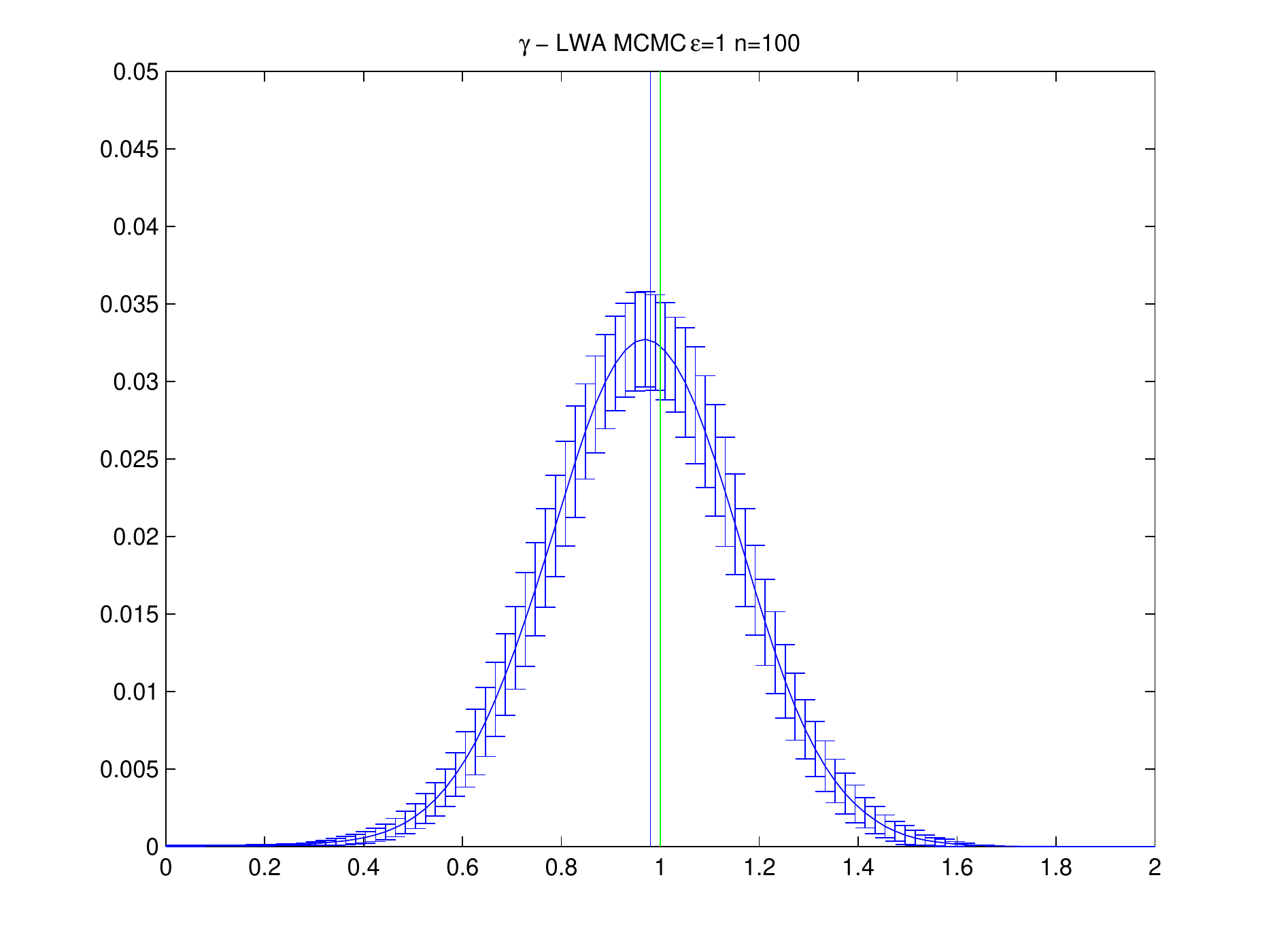}\\
\includegraphics[scale=0.25]{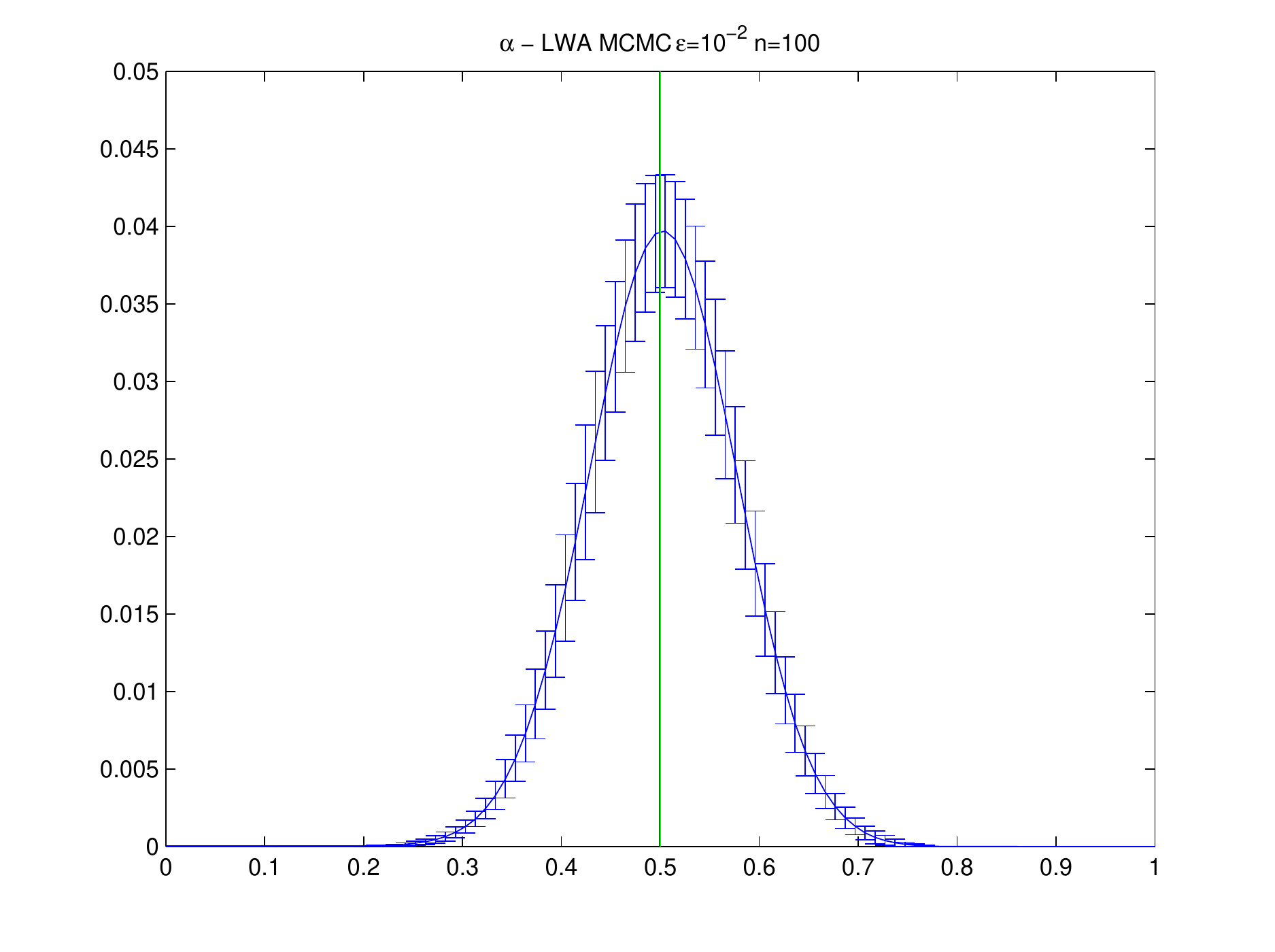}&\includegraphics[scale=0.25]{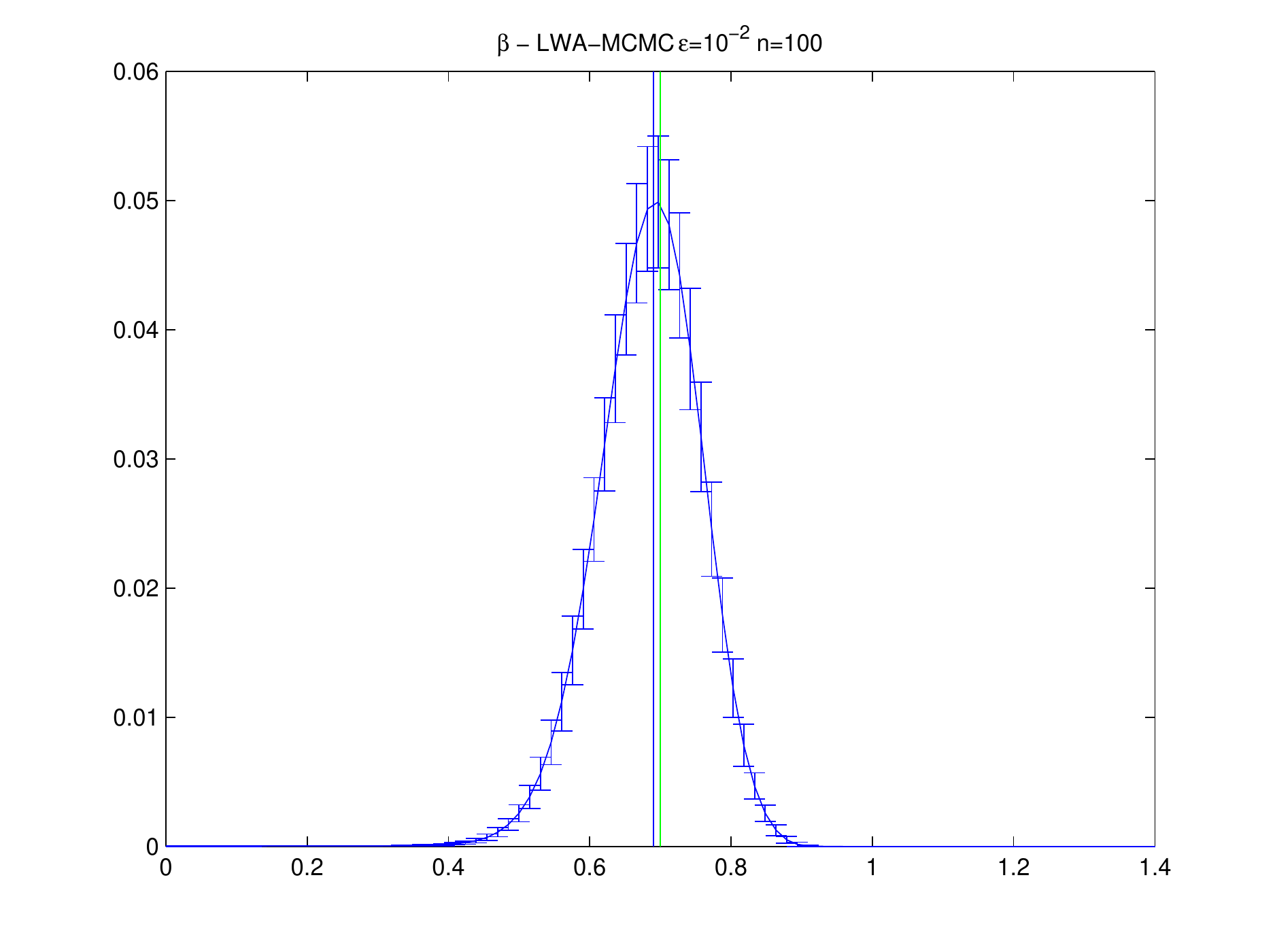}&\includegraphics[scale=0.25]{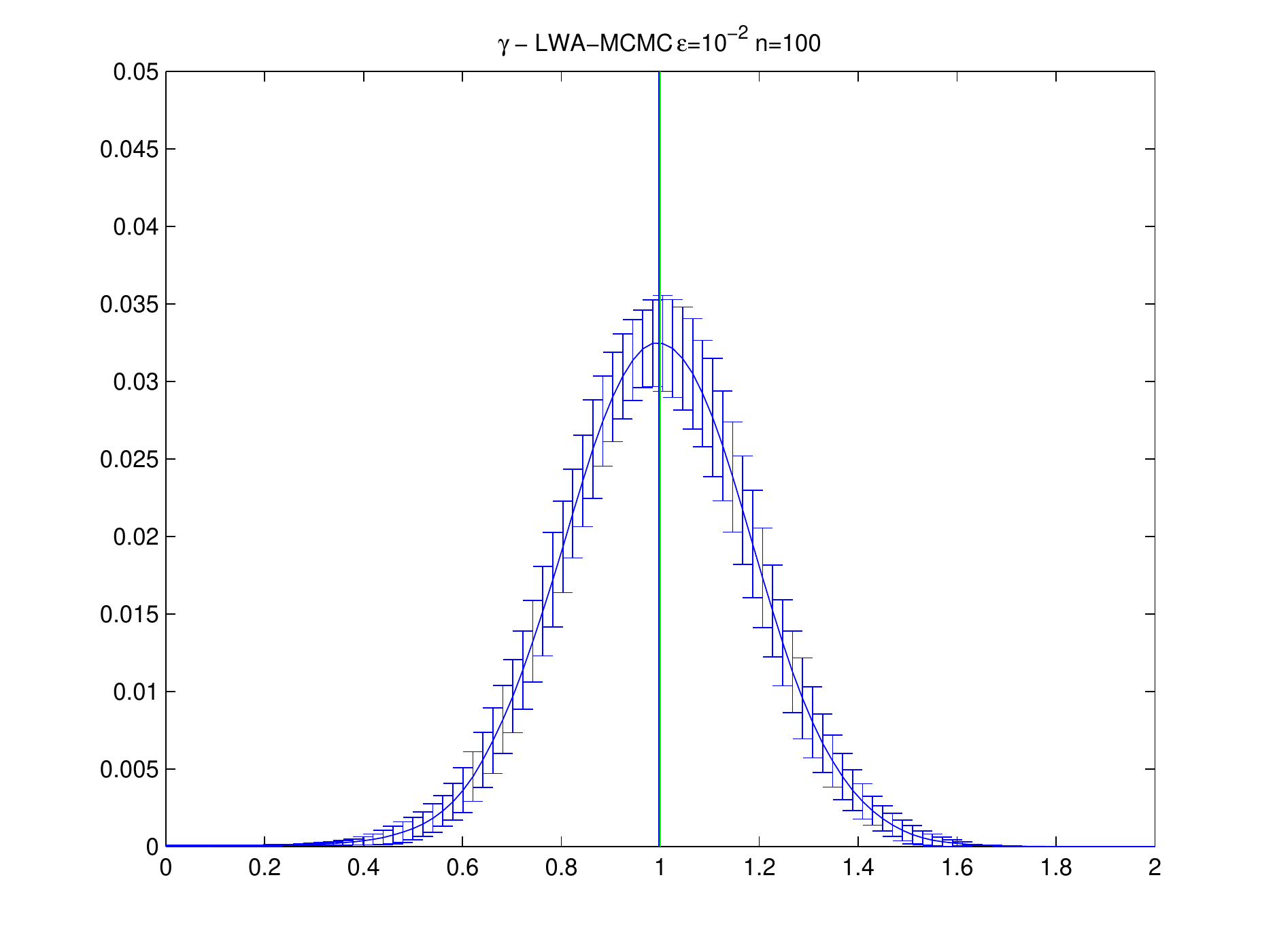}\\
\includegraphics[scale=0.25]{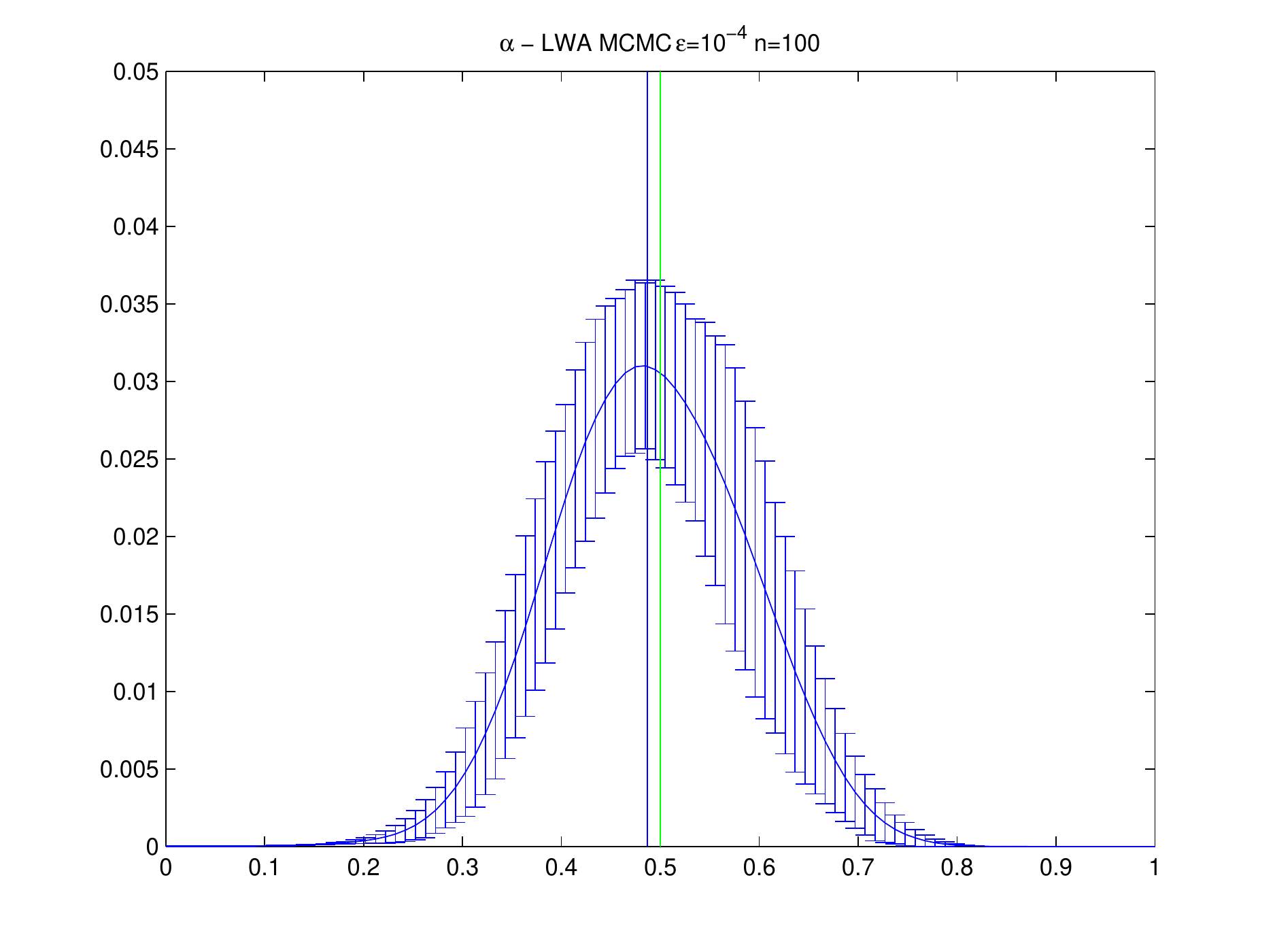}&\includegraphics[scale=0.25]{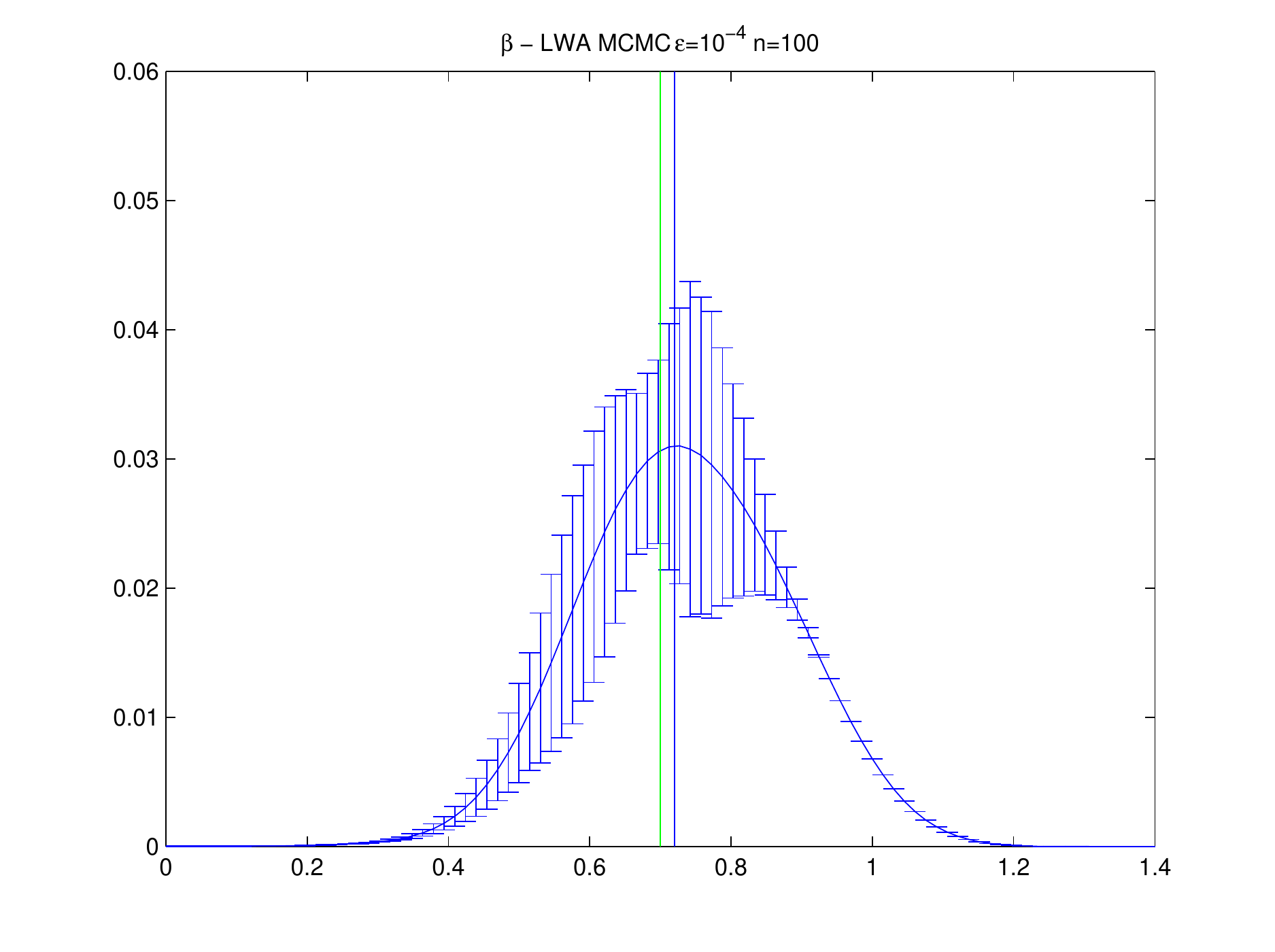}&\includegraphics[scale=0.25]{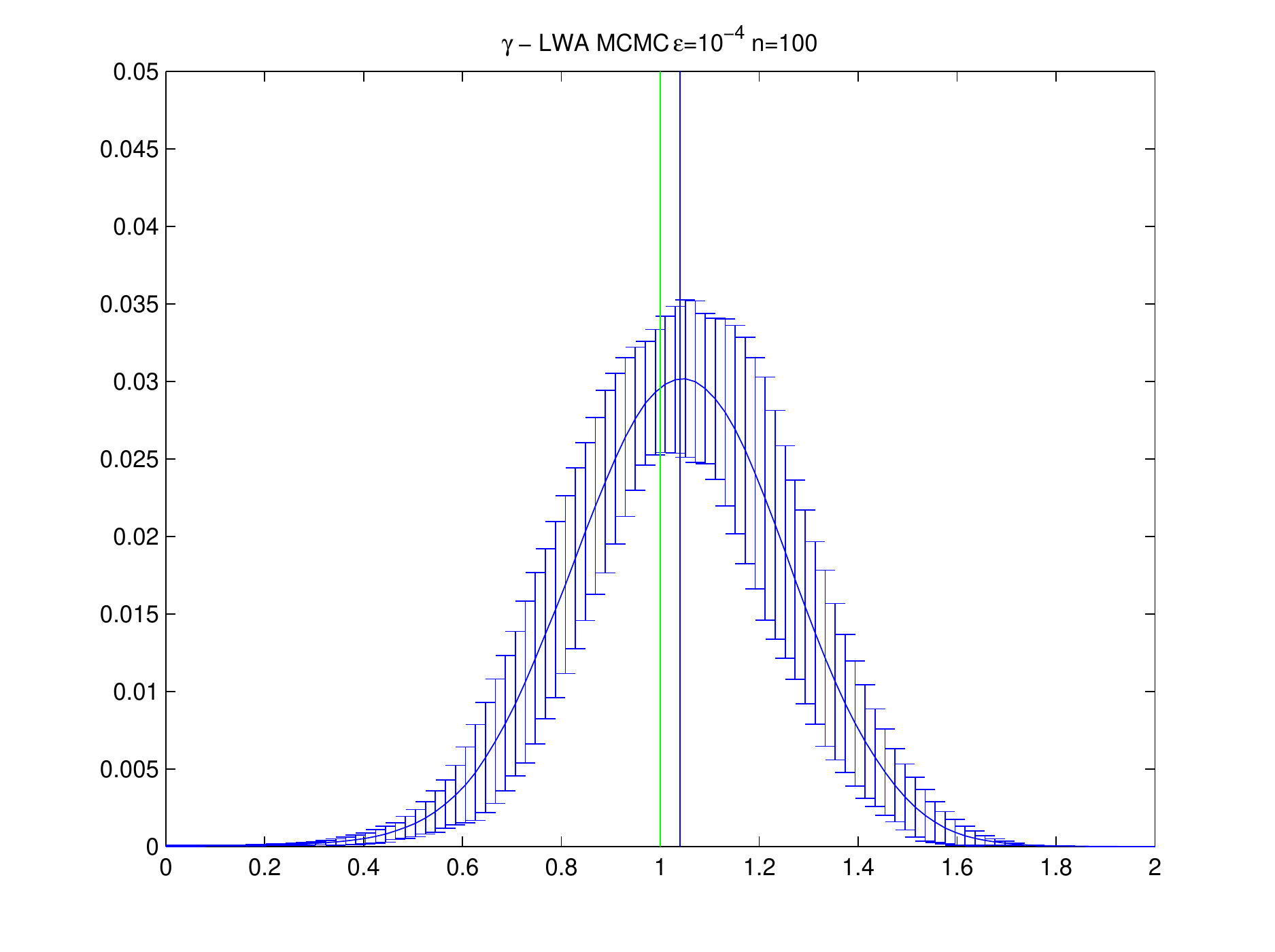}\\
\includegraphics[scale=0.25]{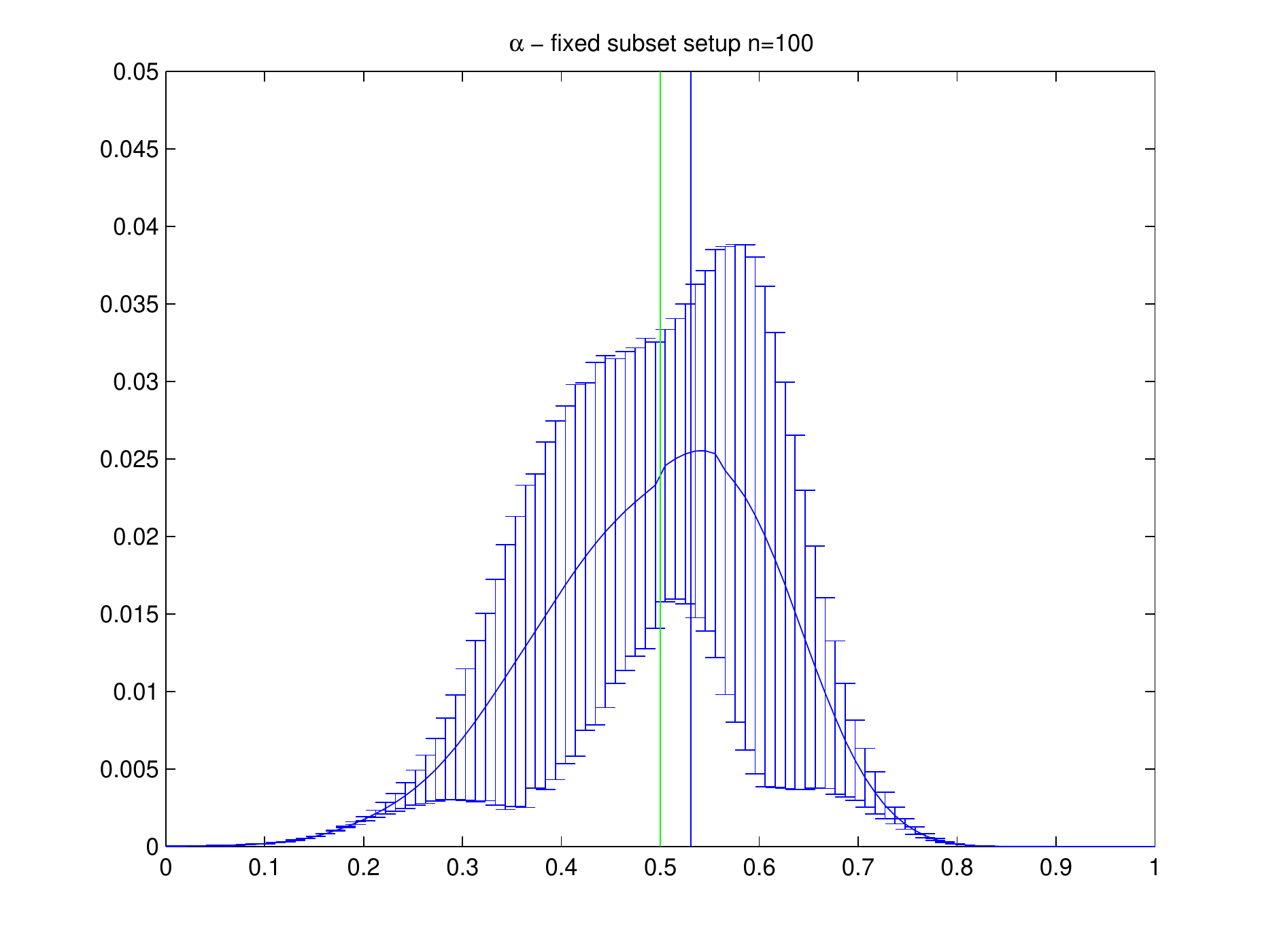}&\includegraphics[scale=0.25]{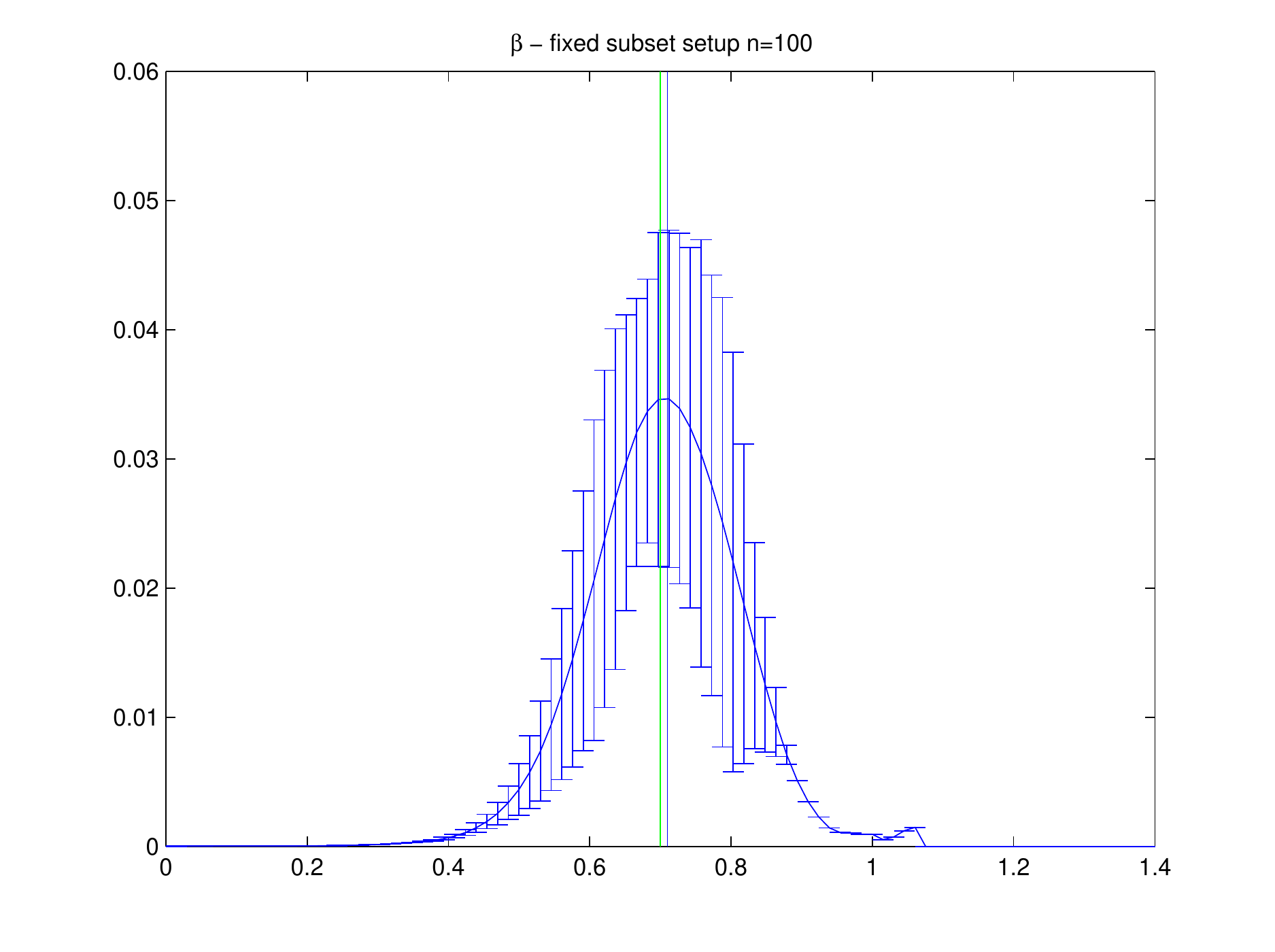}&\includegraphics[scale=0.25]{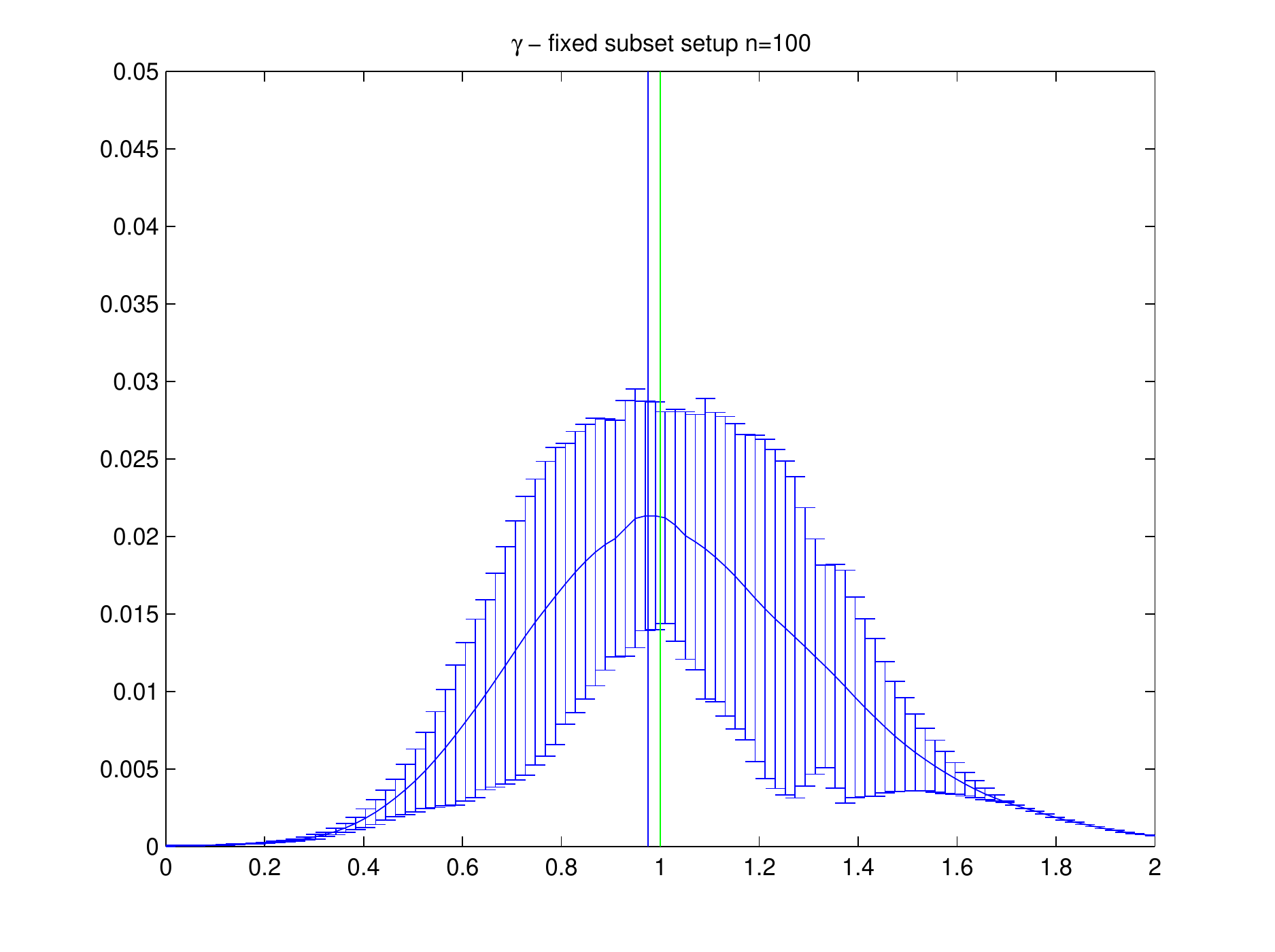}
\end{tabular}
\caption{ARMA example - Mean density and confident intervals for each parameter, obtained through 100 independent LWA--MCMC runs and for different $\epsilon$. Vertical blue lines give the mean of the mean density and vertical green lines the true parameter $\param^{\ast}$.  \label{fig:lwamcmc_eps}}
\vspace{2cm}
\end{figure}

\begin{equation}
\label{eq:sumstat}
\forall\,U\in\Vset_n \qquad S_0(Y_U)=\left(q_{.2}(Y_U),q_{.5}(Y_U),q_{.8}(Y_U),\rho_1(Y_U),\ldots,\rho_5(Y_U)\right)\,,
\end{equation}
where for any $\lambda\in(0,1)$, $q_\lambda(Y_U)$ is $Y_U$ $\lambda$-quantiles and for all $p\in\nset$, $\rho_p(Y_U)$ is the $p$-lag sample autocorrelation. As expected, the time to reach stationarity is significantly reduced when sampling through LWA--MCMC. Actually, the M--H sampler does not even reach the high density area of the support during this time frame. On the other hand, when reducing $n$ the time to reach convergence is even smaller but the stability of the chain may be affected. This is illustrated with Figure \ref{fig:lwamcmc_n} which shows the posterior distribution of the three parameters $\ttargs_{n,\eps}$ for $\eps=1$, $S=S_0$, $n=10,000$, $n=1,000$ and $n=100$ obtained through a time-normalized LWA--MCMC simulation. For the setup featuring $n=100$, the efficiency of the sampler could be criticized, since a slight bias results. Nonetheless, in an attempt to correct this behaviour, the parameter $\eps$ is decreased, making the choice of subset more critical. Here, Figure \ref{fig:lwamcmc_eps} shows that it is worthwhile to penalize subsets which are less relevant with respect to the summary statistics $S_0$. More precisely, since we are interested in the posterior distribution of the parameter we have reported the mean density observed along the confidence intervals corresponding to the $.2$-quantiles and $.8$-quantiles. Indeed, because each run is a stochastic process (a random collection of subsets is sampled) there is variability within the collection of posterior distributions provided by each run (especially when $\eps\ll 1$). These results were obtained through $100$ independent runs of LWA--MCMC for $n=100$, $S=S_0$ and $\epsilon\in\{1,10^{-2},10^{-4}\}$, each run featuring $200,000$ iterations among which the $10,000$ first were discarded for burn-in. The initial states of all chains were drawn from the prior distribution. Moreover, we have included two extra setups which can be regarded as "extreme" scenarios:
\begin{itemize}
\item \textbf{free subset} -- a standard M--H kernel targeting a sub-posterior which refreshes uniformly the $n=100$ data of the subset at each transition,
\item \textbf{fixed subset} -- a standard M--H kernel targeting a sub-posterior with a fixed subset of $n=100$ data chosen uniformly before the first iteration.
\end{itemize}
While the first one can be seen as a LWA--MCMC kernel with $\epsilon\to\infty$ since the relevance of the subset does not interfere the parameter sampling, the latter is related with a LWA--MCMC kernel which would be trap in a single subset throughout all the sampling scheme, hence $\epsilon\to 0$. The vertical green and blue lines respectively indicate the true parameter value and the expectation of the mean sub-posterior distribution targeted by the different MCMC. Clearly for $n=100$ and $S=S_0$, the parameter $\epsilon=10^{-2}$ yields an optimal setup since for all the three parameters, the expectation of the posterior distributions meets the true value. Beyond this optimal $\epsilon$, the weights $\{\nu_{n,\eps}(U),\,U\in\Vset_n\}$ become too unbalanced. As a result some relevant subsets are simply ignored and the inference is performed on a restricted set of subset. Table \ref{tab:refresh_U} shows indeed that for $\eps<10^{-2}$, the subset may be refreshed relatively infrequently.

\begin{table}
\centering
\begin{tabular}{c|c|c|c|c|c|c|c}
setup / $\eps$ & free subset & $1$ & $10^{-1}$& $10^{-2}$ & $10^{-3}$ & $10^{-4}$ & fixed subset\\
\hline
$U$ refresh rate & 1 & .81 & .34 & .05 & .001 & $10^{-4}$ & 0
\end{tabular}
\caption{ARMA example - Refresh probability of the subset $U$ when targeting $\targ_{n,\eps}$ with $n=100$ and $S=S_0$. \label{tab:refresh_U}}
\end{table}

\begin{figure}
\centering
\begin{tabular}{ccc}
\includegraphics[scale=0.45]{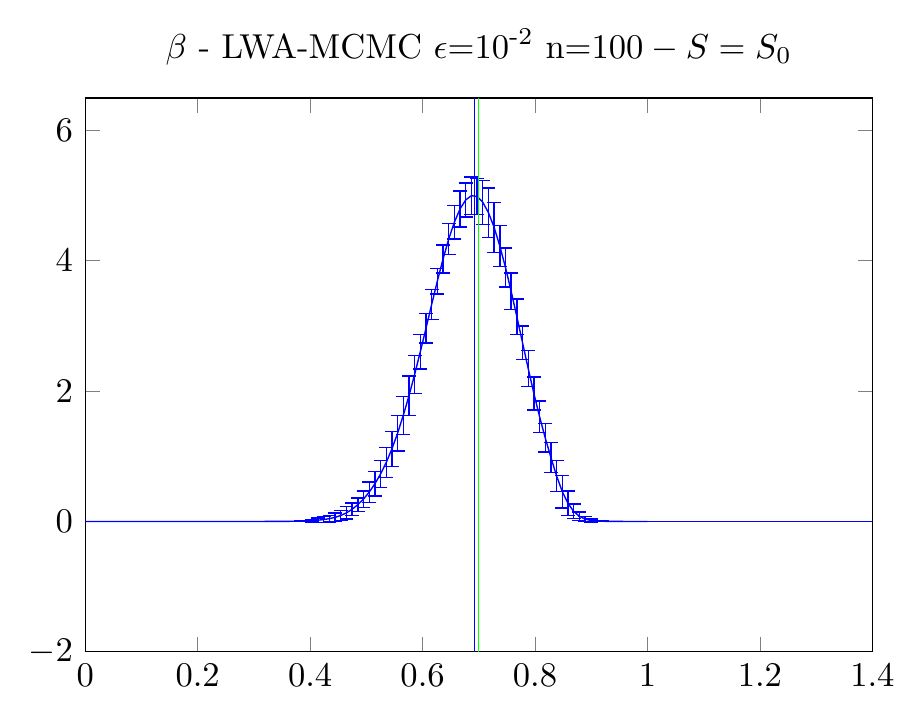}&
\includegraphics[scale=0.45]{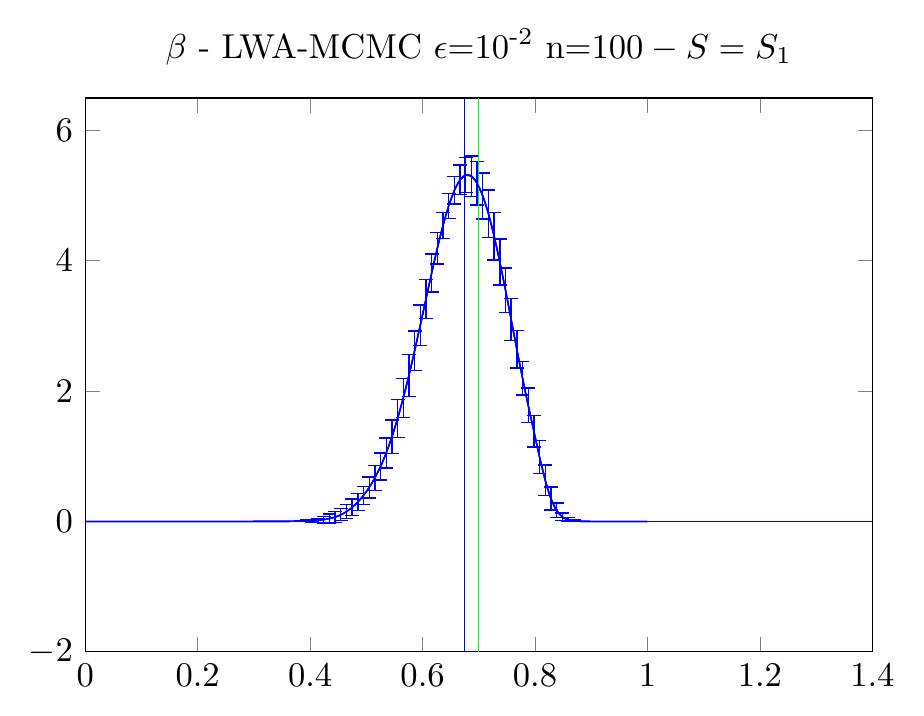}&
\includegraphics[scale=0.45]{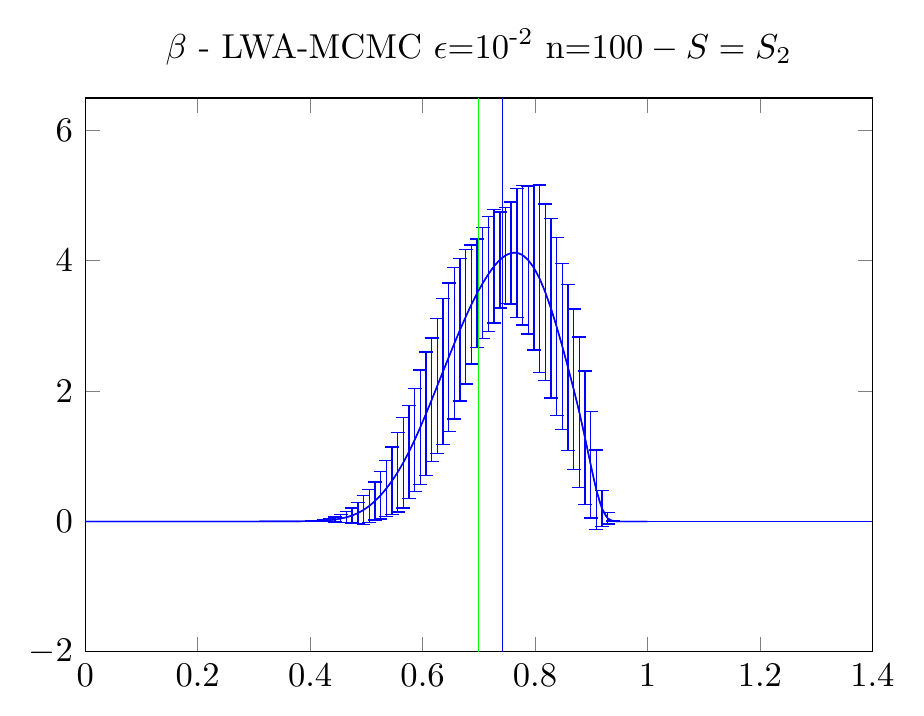}
\end{tabular}
\caption{ARMA example - Mean density and confident intervals for each parameter, obtained through 100 independent LWA--MCMC runs and for different summary statistics $S$. Vertical blue lines give the mean of the mean density and vertical green lines the true parameter $\param^{\ast}$. \label{fig:lwamcmc_S}}
\end{figure}

Finally, we consider for the setup $n=100$ and $\eps=10^{-2}$, two other possible choices for $S$:
$$
S_1(Y_U)=\left(\rho_1(Y_U),\ldots,\rho_{15}(Y_U)\right)\,,\qquad S_2(Y_U)=\left(\min(Y_U),\max(Y_U)\right)\,.
$$
Figure \ref{fig:lwamcmc_S} shows that this \textit{naive} choice of $S_2$ as summary statistics yields a poor sub-posterior target. Indeed, the minimum and the maximum of the time series does not characterize much the model parameters. Note that this setup gives worse results than the free subset setup: while the latter does not advantage any subset $U\in\Uset_n$, the LWA--MCMC setup with $n=100$, $\eps=10^{-2}$, $S=S_2$ uses more likely subsets which match the min/max statistics of the full dataset and which, as a result, may force the sampler to infer the parameters through those unrepresentative subsets. Indeed, the min/max stopping times are likely to be far from each other as they represent large deviations to the stationary process. Therefore the subsets whose min/max statistics of the $n=100$ contiguous time step are close to that of the full dataset can reasonably be labeled as "anomalies". On the other hand, when the inference is performed using $S=S_1$, the collection of subsets is only guided through the correlation between consecutive states. We see that the results are better when these relative statistics are coupled with global statistics such as the mean and the two quantiles $q_{.2}$ and $q_{.8}$, like in $S=S_0$.

\subsection{Binary classification}

\begin{figure}
\centering
\includegraphics[scale=.6]{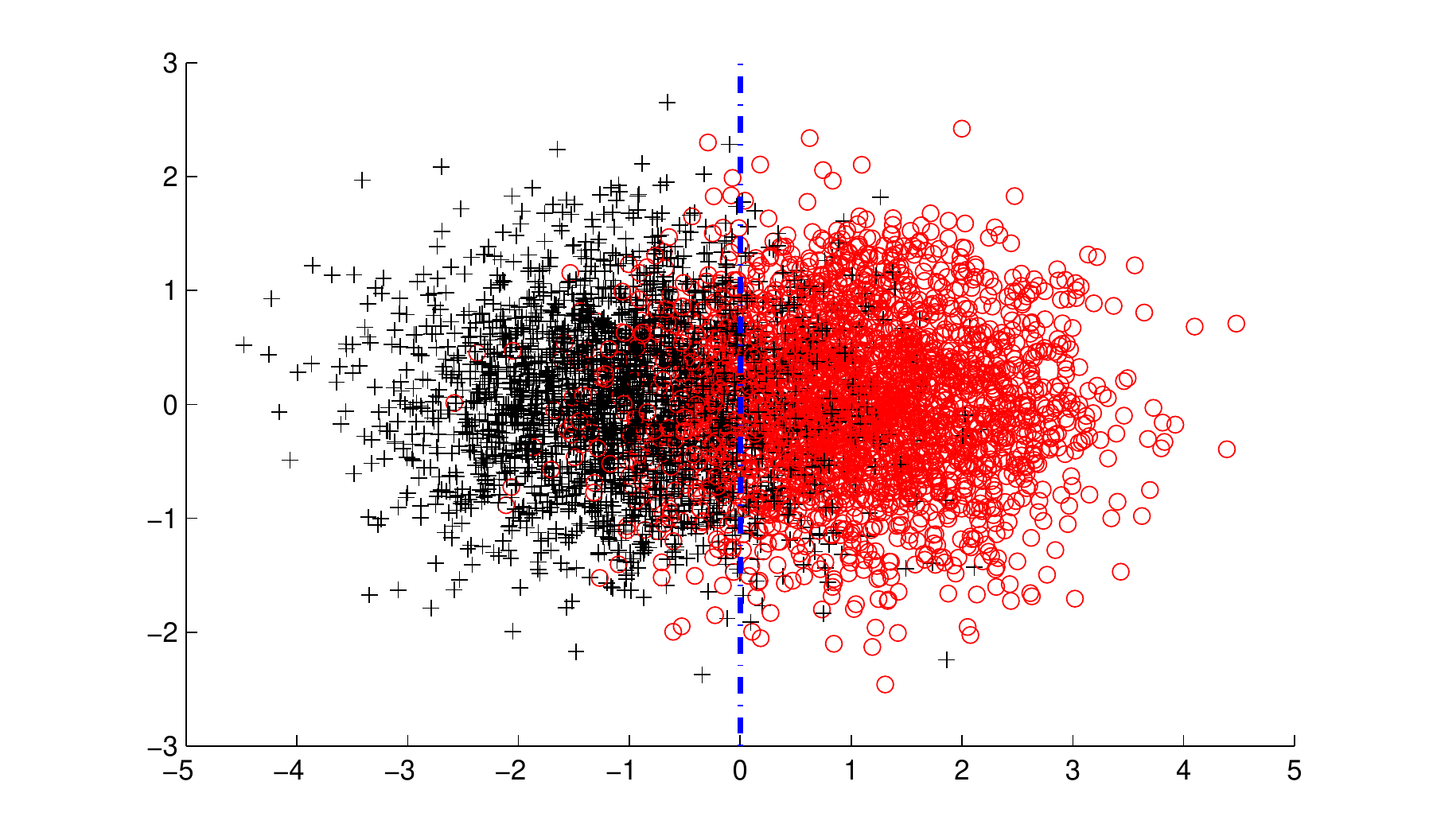}
\caption{Binary Classification example - Samples from the two-dimensional mixture of two Gaussian distributions. \label{fig:classif_obs}}
\end{figure}

In this application, we consider the binary classification example used in the MCMC SubLikelihood (MCMCSubLhd) approach \citep{bardenet2014towards}. MHSubLhd is an alternative to M--H, designed to perform Bayesian inference in large datasets contexts; see \eqref{eq:MH_decision} and the corresponding introduction section for more details. We consider here the example in Section 4.2 of \cite{bardenet2014towards} in order to compare the efficiency of LWA--MCMC and MHSubLhd. A large number ($N=10^7$) of realizations from a two-dimensional Gaussian mixture distribution with two classes were sampled with the following parameters
$$
\mu_1=[-1\,,\, 0],\; \mu_2=[1\,,\, 0],\; \Sigma_1=\Sigma(\sigma_2=.25),\;\Sigma_2=\Sigma(\sigma_2=.25),\; \Sigma(\sigma)=\text{diag}(\sigma^2;\sigma^2/2)\,,
$$
for the mean and covariance matrix of the two components respectively; see Figure \ref{fig:classif_obs}. Both LWA--MCMC and MHSubLhd provide a sequence of parameters $\{\param_{j,k}=(\mu_{j,k},\sigma_{j,k}),\allowbreak \,j\in\{1,2\},\,k\in\nset\}$ which is used to classify test-data $\{\tilde{Y}_m,\,m\leq 10^7\}$, sampled from the same model, through the following real-time maximum likelihood classifier defined at time $t>0$ by:
$$
C_t(\tilde{Y}_m)=\arg\max_{j\in\{1,2\}} f(Y_u\,|\param_{j,\kappa(t)})\,\quad\text{where}
\;
\left\{
\begin{array}{l}
\forall t\in\rset,\;\kappa(t)=\max_{k\in\nset}\{t\geq \tau_k\}\,,\\
\tau_k\;\text{is\,the\,time\,at\,the\,end\,of\,the\,}k\,\text{-th\,iteration}\,.
\end{array}
\right.
$$

On the basis of the time series simulation example, LWA--MCMC was tuned with $n=1,000$, $\eps=.01$ and $n S(U)=\left[\sum_{k\in U}\1_{\{I_k=0\}}\,;\, \sum_{k\in U}\1_{\{I_k=1\}}\right]$. The specific MHSubLhd parameters were chosen accordingly:
\begin{itemize}
\item $\delta=.1$, which states that the decision to accept/reject a proposed candidate in MHSubLhd is in accordance with the M--H decision with probability $1-\delta=.9$
\item every time a decision (accept/reject a candidate) is postponed, $n_\ell=1000\ell$ new data are added to the current subset, where $\ell\in\nset^{\ast}$ is the number of times that a decision has been postponed in the current MCMC transition (the size of the subset increments, following the guidelines provided in the Section 2.2 of \cite{bardenet2014towards}).
\end{itemize}

Finally, the same proposal kernel were used for both samplers:
\begin{eqnarray*}
\centering
&j\sim\text{Unif}(1,2),\quad(\eps_1,\eps_2)\sim\norm(0,1),
\quad\left\{
\begin{array}{l}
\mu_j'=\mu_j+\varsigma_{j}\eps_1 \\
\sigma_j'=\exp(\varrho_{j}\eps_2)\sigma_j
\end{array}
\right.,
\end{eqnarray*}
where the parameter $\{\varsigma_j,\varrho_j\}_{j\in\{1,2\}}$ were updated to maintain an acceptance rate between $.25$ and $.35$ through an adaptive Metropolis procedure \citep{haario2001adaptive}.

\begin{figure}
\centering
\begin{tabular}{cc}
\includegraphics[scale=0.25]{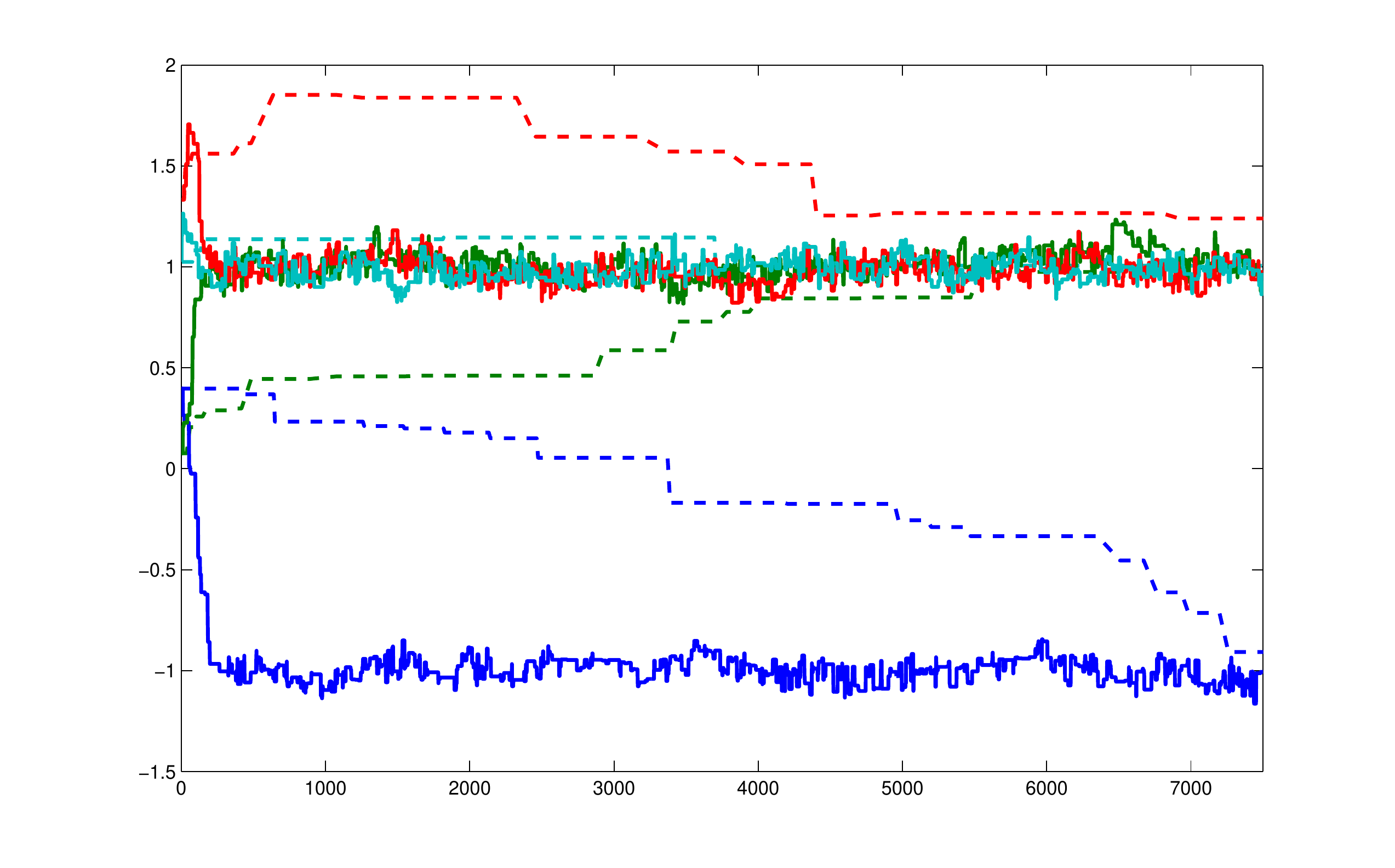}&\includegraphics[scale=0.25]{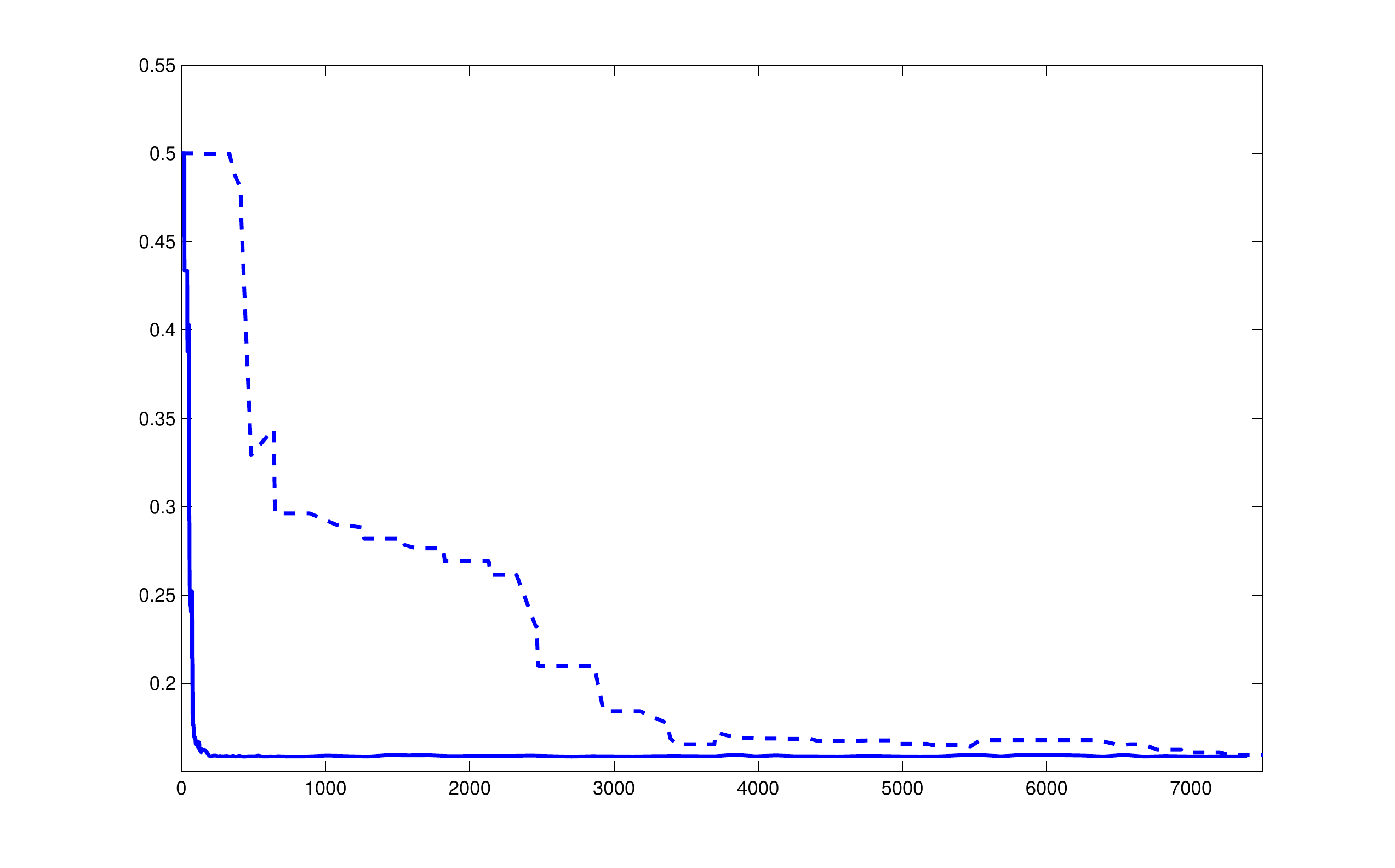}\\
\includegraphics[scale=0.25]{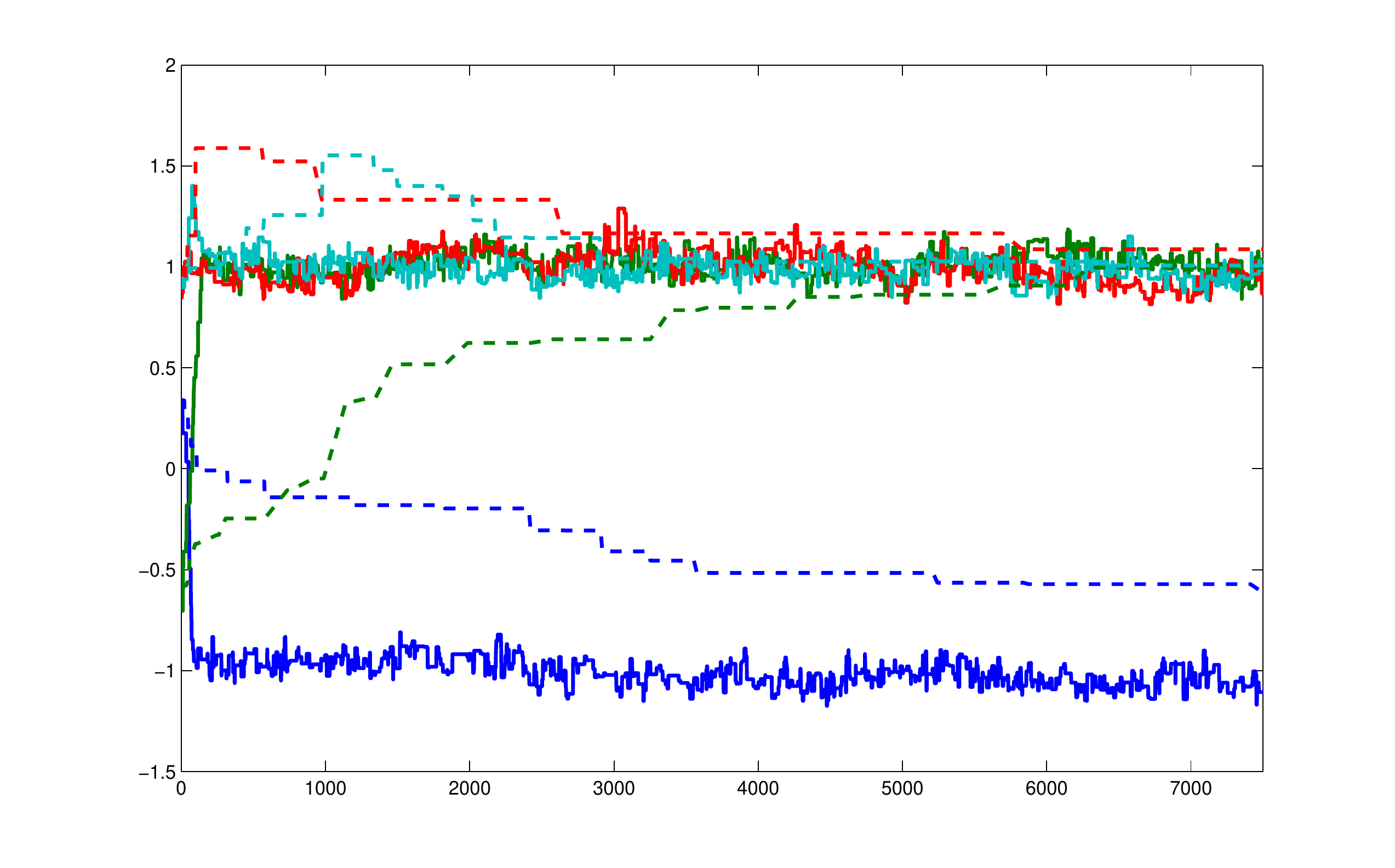}&\includegraphics[scale=0.25]{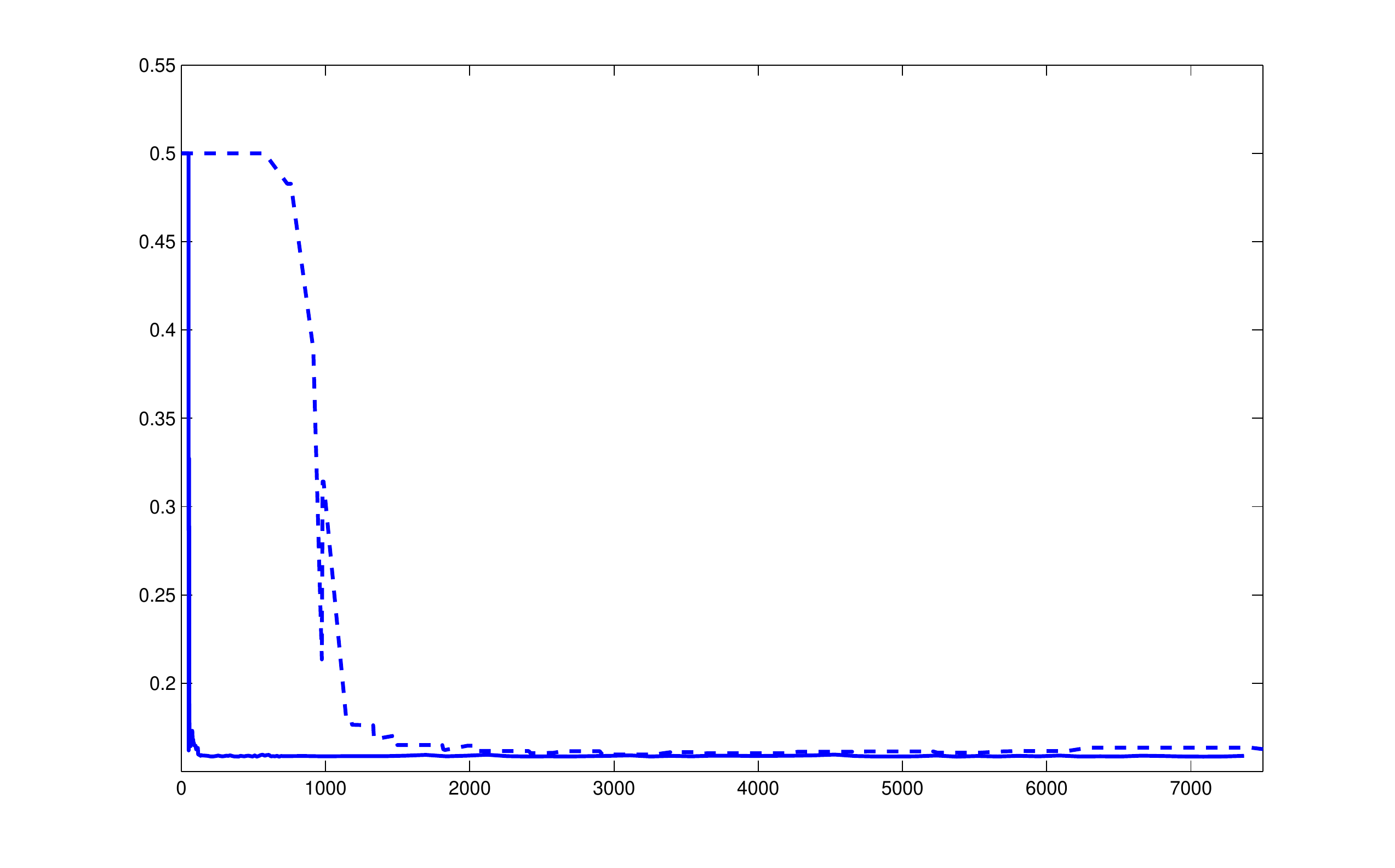}\\
\includegraphics[scale=0.25]{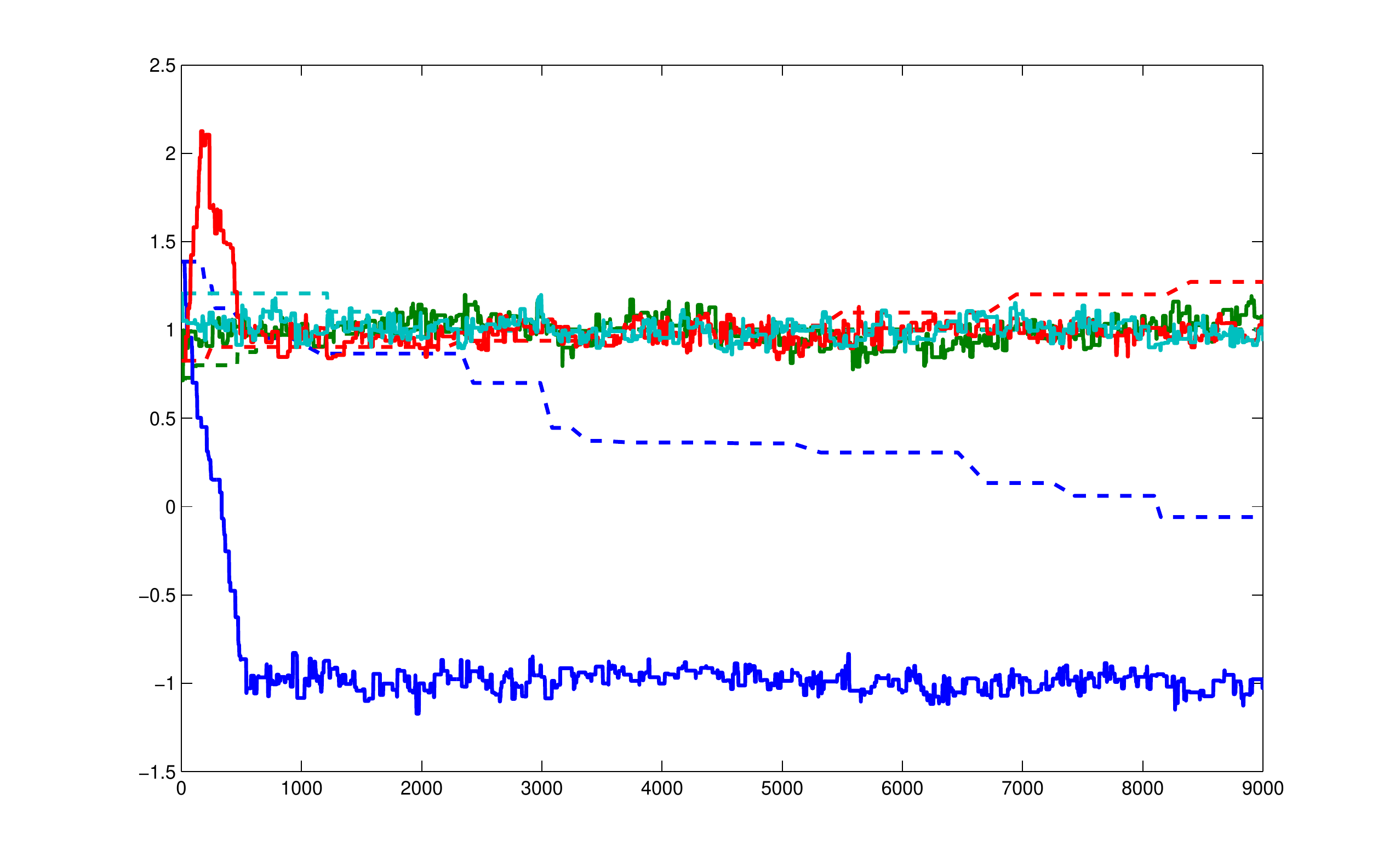}&\includegraphics[scale=0.25]{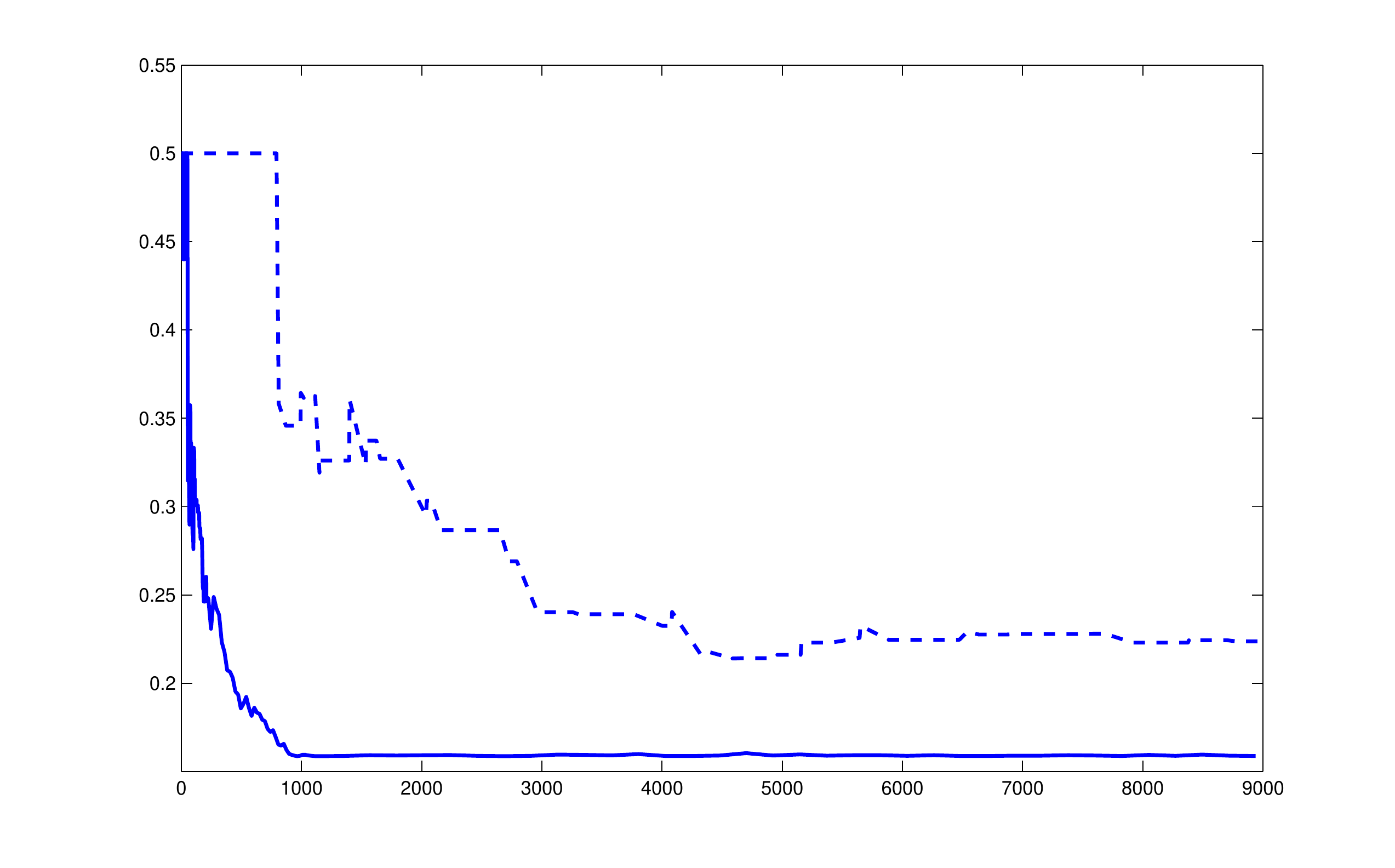}\\
\includegraphics[scale=0.25]{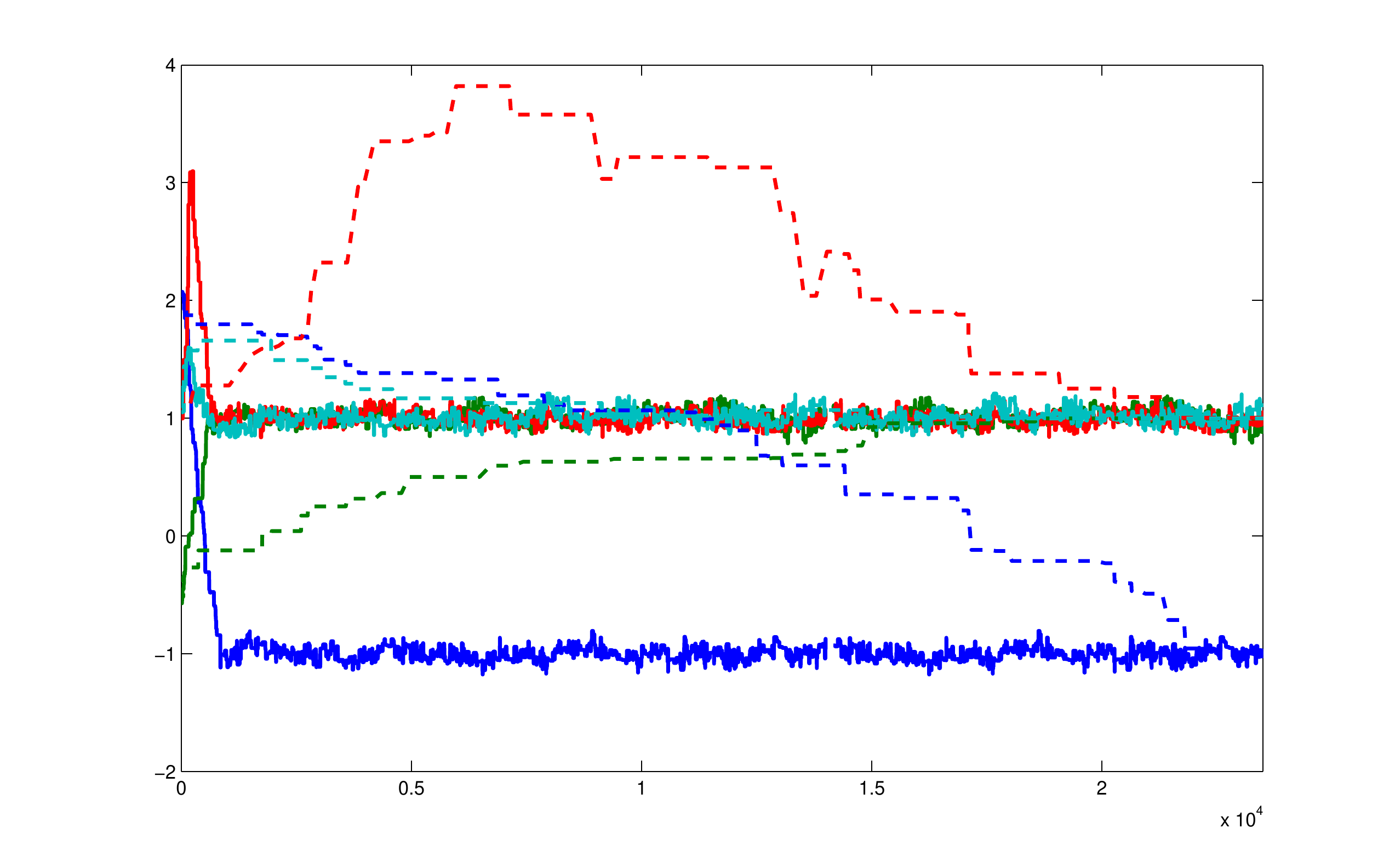}&\includegraphics[scale=0.25]{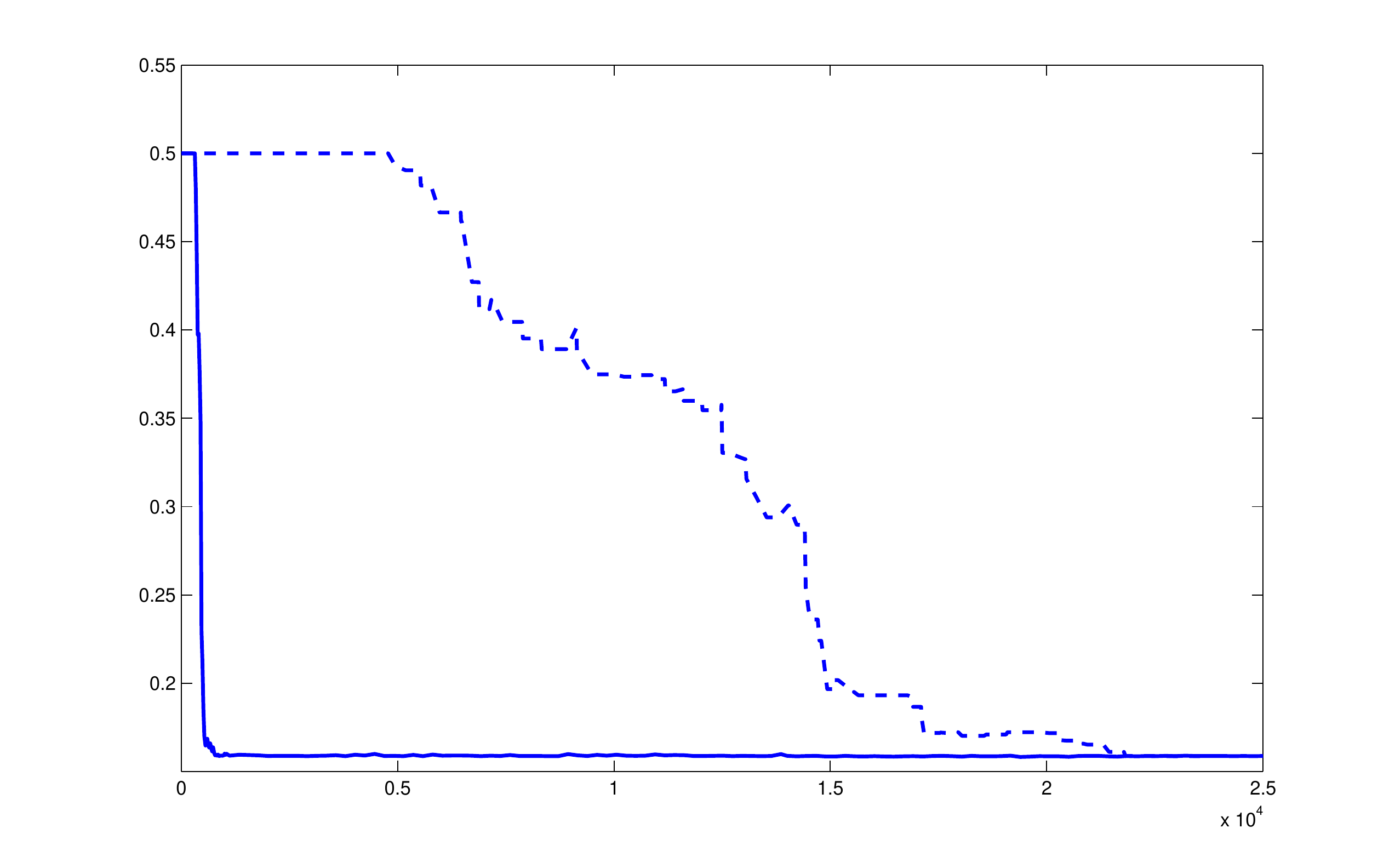}
\end{tabular}
\caption{Binary Classification example - Four independent scenarios -- left column: sample paths of the two samplers (LWA--MCMC in plain and MHSubLhd dashed) ; right column: classification error rate on the training dataset (LWA--MCMC in plain and MHSubLhd dashed).\label{fig:classif}}
\end{figure}

Figure \ref{fig:classif} shows four independent comparisons between LWA--MCMC and MHSubLhd. Both samplers starts with the same initial state (drawn from the prior) in all scenarios. The first column shows the sample path of the two Markov chains against the time (in second): plain lines are LWA--MCMC sample paths and dashed lines are MHSubLhd sample paths. We stress that the step shape of the MHSubLhd sample paths does not highlights a poor mixing chain but illustrates the fact that a single MHSubLhd transition can take a similar amount of time as a standard M--H transition. As a consequence, the chain remains at the same state for a large amount of time. Table \ref{tab:nbdata_trans} shows indeed that, on average, the MHSubLhd sampler ends up using almost $25\%$ of the full dataset at each transition. The results clearly show that, in this application, using LWA--MCMC instead of MHSubLhd results in a practically useful approach, with some spectacular convergence acceleration: in scenario 4, LWA--MCMC is 22 times faster than MHSubLhd.

\begin{table}
\centering
\begin{tabular}{c|c|c|c|c}
scenario & 1 & 2 & 3 & 4 \\
\hline
LWA--MCMC & $1,000$ & $1,000$ & $1,000$ & $1,000$ \\
MHSubLhd & $2.4\,10^{6}$  & $1.9\,10^{6}$ & $2.35\,10^{6}$ & $2.8\,10^{6}$
\end{tabular}
\caption{Binary Classification example - Average number of data used per transition. \label{tab:nbdata_trans}}
\end{table}

\subsection{Additional details for the handwritten digit example of Section \ref{sec:first_ex}}

In the handwritten digit example (see Section \ref{sec:first_ex}), we have used batches of $n=100$ data. The summary statistics were simply defined so as to promote subsets which have $20$ observations from each class. A very low bandwidth $\eps=10^{-5}$ was used in order to enforce this characteristic. Similarly to the binary classification example, the proposal kernels of LWA--MCMC and M--H were defined in the same style, namely a Random Walk kernel where at each iteration only a bloc of the template parameter of one of the 5 classes is updated ($\param\in\rset^{256}$ in this example). The variance parameter of the Random Walk is here again adapted according to the past trajectory of the chain, so as to maintain an acceptance rate of $.25$.

In contrast to the previous examples, the computational difference is not explained by the fact that the M--H acceptance probability is more expensive to compute. Indeed, both need to evaluate the function $\phi(\param_j'):\rset^{256}\to\rset^{225}$ for the updated class $j$, which represents the only heavy routine calculation. In fact, the adjusted variance of the M--H proposal turns out to be $10$ times lower than that for LWA--MCMC. Loosely speaking, on the one hand, the M--H proposal needs to provide a parameter $\param'$ which fits in about $2,000$ observations: the log of the acceptance ratio depends of $\sum_{k=1}^{N}\1_{I_k=j}\{\|Y_k-\phi(\param'_j)\|^{2}-\|Y_k-\phi(\param_j)\|^{2}\}$
for the updated class $j$. On the other hand, the LWA--MCMC proposal should match only about $20$ images through the quadratic term $\sum_{k\in U}\1_{I_k=j}\{\|Y_k-\phi(\param'_j)\|^{2}-\|Y_k-\phi(\param_j)\|^{2}\}$. As a consequence, the M--H adapted variance makes the Random Walk less efficient, which results in the Markov chain $\{\param_k,\,k\in\nset\}$ exploring the state space $\paramset$ slower.

\section{Conclusion}
The Light and Widely Applicable MCMC methodology introduced and discussed in this paper is attractive as it overcomes the critical issues encountered by the popular Metropolis--Hastings sampler in the modern development of big data inference problems. Several recent noisy M--H methods have been proposed to address these issues. However, (i) they are only valid for \iid realizations and (ii) they may use a significant portion of the dataset negating any potential computational saving. LWA--MCMC pushes the approximation one step forward to preserve the celebrated M--H simplicity. The efficiency of the sampler is illustrated in several typical Bayesian problems including parameter estimation and classification.

Our experiments have illustrated the usefulness of LWA--MCMC. Future work should extend the theory beyond the scope of exponential models.

%Apart from the special case of exponential models, the price to pay to reach this simplicity is the loss of the theoretical basis, justifying/motivating the usual MCMC samplers. We regard this as an open problem which is at the crossroad of many ongoing research question involving the ABC literature, noisy MCMC literature, doubly-intractable problems and time-evolving target simulation. In this perspective, some connection with the Bootstrap theory may also prove helpful.

\section*{Acknowledgements}
Florian Maire thanks The Insight Centre for Data Analytics for funding the Post-Doctoral Fellowship. The Insight Centre for Data Analytics is supported by Science Foundation Ireland under Grant Number SFI/12/RC/2289. Nial Friel's research was also supported by an Science Foundation Ireland grant: 12/IP/1424.

\newpage
\renewcommand\appendixname{Appendix\hspace{0.1cm}}
\appendix
\section{Proof of Proposition \ref{prop:KL}}
\label{app1}
\begin{proof}
Without loss of generality, we take $g$ as the identity on $\paramset$. By straightforward algebra,
\begin{multline}
\label{eq:pr1}
\KL{\targ}{\ttarg_U}=\esp_{\targ}\left\{\log{\frac{\targ(\param)}{\ttarg_U(\param)}}\right\}=\pscal{\esp_{\targ}\left(\param\right)}{\sum_{k=1}^{N}S(Y_k)-\sum_{k\in U}S(Y_k)}+(n-N)\esp_{\targ}\left\{\ell(\param)\right\}\,,\\
+\log{\frac{Z(Y_U)}{Z(Y_{1:N})}}\,,
\end{multline}
where
$
\ell(\param)=\log L(\param)
$.
Now, let
\begin{equation}
\label{eq:pr2}
\xi_U=\frac{1}{n}\sum_{k\in U} S(Y_k)-\frac{1}{N}\sum_{k=1}^{N}S(Y_k)
\end{equation}

and express \eqref{eq:pr1} as:
\begin{multline}
\label{eq:pr3}
\KL{\targ}{\ttarg_U}=\left(1-\frac{n}{N}\right)\pscal{\esp_{\targ}\left(\param\right)}{\sum_{k=1}^N S(Y_k)}+(n-N)\esp_{\targ}\left\{\ell(\param)\right\}-n\pscal{\xi_U}{\esp_{\targ}(\param)}\,,\\
+\log\frac{Z(Y_U)}{Z(Y_{1:N})}\,.
\end{multline}

We now want to find an upper bound of $\log Z(Y_U)/Z(Y_{1:N})$. First, note that
\begin{multline}
\label{eq:pr4}
Z(Y_U)=\int p(\rmd\param)\frac{\exp{(n/N)\pscal{\param}{\sum_{k=1}^N S(Y_k)}}}{L(\param)^n}\exp\left(n\pscal{\param}{\xi_U}\right)\\
=\int p(\rmd\param)\left\{\prod_{k=1}^N f(Y_k\,|\,\param)\right\}^{\frac{n}{N}}\exp\left(n\pscal{\param}{\xi_U}\right)\\
=\int p(\rmd\param)\prod_{k=1}^N f(Y_k\,|\,\param)\exp\left(n\pscal{\param}{\xi_U}\right)\left\{\prod_{k=1}^N f(Y_k\,|\,\param)\right\}^{\frac{n}{N}-1}
\end{multline}
Because $n/N\in(0,1)$, we can write
$$
\left\{\prod_{k=1}^N f(Y_k\,|\,\param)\right\}^{\frac{n}{N}-1}\leq \alpha(Y_{1:n})^{\frac{n}{N}-1}
$$
where $\alpha(Y_{1:n})=\inf_{\param\in\paramset}\prod_{k=1}^{N}f(Y_k\,|\,\param)$. As a consequence, we have:
\begin{multline}
\label{eq:pr5}
\frac{Z(Y_U)}{Z(Y_{1:N})}\leq \alpha(Y_{1:n})^{\frac{n}{N}-1} \int \targ(\rmd\param\,|\,Y_{1:N})\exp\left(n\pscal{\param}{\xi_U}\right)\\
=\alpha(Y_{1:n})^{\frac{n}{N}-1}\esp_{\targ}\left\{\exp\left(n\pscal{\param}{\xi_U}\right)\right\}
\end{multline}

Plugging \eqref{eq:pr5} into \eqref{eq:pr3} yields
\begin{multline}
\KL{\targ}{\ttarg_U}\leq
\underbrace{\left(1-\frac{n}{N}\right)\left\{\pscal{\esp_{\targ}\left(\param\right)}{\sum_{k=1}^N S(Y_k)}-\log\alpha(Y_{1:N})\right\}
+(n-N)\esp_{\targ}\left\{\ell(\param)\right\}}_{\Psi(n,N,Y_{1:N})}\\
\underbrace{+\log\esp_{\targ}\left\{\exp\left(n\pscal{\param}{\xi_U}\right)\right\}-n\pscal{\xi_U}{\esp_{\targ}(\param)}}_{\Omega(U,n,Y_{1:N})}\,.
\end{multline}
Now, using the Cauchy-Schwartz inequality we have:
\begin{multline*}
\Omega(U,n,Y_{1:N})=\log\esp_{\targ}\left[\exp\left\{n\pscal{\param}{\xi_U}-n\pscal{\xi_U}{\esp_{\targ}(\param)}\right\}\right]\\=
\log\esp_{\targ}\left[\exp\left\{n\pscal{\param-\esp_{\targ}(\param)}{\xi_U}\right\}\right]
\leq \underbrace{\log\esp_{\targ}\left\{\exp\left(n\|\param-\esp_{\targ}(\param)\|\cdot\|\xi_U\|\right)\right\}}_{B(U,n,Y_{1:N})}\,.
\end{multline*}
Finally we have
$$
\KL{\targ}{\ttarg_U}\leq \Psi(n,N,Y_{1:N})+ B(U,n,Y_{1:N}),\qquad\text{where}\quad B(U,n,Y_{1:N}) \geq 0
$$
such that the subset $U\in\Uset_n$ minimizing $\|\xi_U\|$ also minimize $B(U,n,Y_{1:N})$ with the special case
$$
 U\in\Usetst_n \Rightarrow B(U,n,Y_{1:N})=0\,,
$$ hence completing the proof.
\end{proof} 

%\section*{References}
\bibliographystyle{elsarticle-harv}
\bibliography{biblio}

\end{document}